\documentclass{article}
\usepackage{graphicx}
\usepackage{cite}
\usepackage{titling}
\usepackage{tikz}
\usepackage{amsmath}
\usepackage{amsthm}
\usepackage{authblk}
\usepackage{amsmath, amsfonts}
\usepackage{xspace}
\usepackage{thm-restate}
\usepackage{amsthm}
\usepackage{amssymb}
\usepackage{mathrsfs}
\usepackage{caption}
\usepackage{subcaption}
\usepackage[utf8]{inputenc}
\usepackage{booktabs}
\usepackage{array}
\usepackage{tabularx}
\usepackage{geometry}
\usepackage{float}
\usepackage{comment}
\usepackage{xcolor}
\usepackage{listings}
\definecolor{vscBg}{HTML}{FFFFFF}
\definecolor{vscText}{HTML}{000000}
\definecolor{vscBlue}{HTML}{0000FF}
\definecolor{vscPurple}{HTML}{AF00DB}
\definecolor{vscString}{HTML}{A31515}
\definecolor{vscComment}{HTML}{008000}
\definecolor{vscNumber}{HTML}{098658}
\definecolor{vscFunc}{HTML}{795E26}
\definecolor{vscType}{HTML}{267F99}
\definecolor{vscLineNo}{HTML}{237893}
\definecolor{vscFrame}{HTML}{C8C8C8}
\lstdefinestyle{sagepy}{
    language=Python,
    backgroundcolor=\color{vscBg},
    basicstyle=\ttfamily\footnotesize\color{vscText},
    deletekeywords={def,class,lambda,True,False,None,print,range,int,str,set,list,tuple,dict},
    keywordstyle=\color{vscPurple},
    keywordstyle=[2]\color{vscBlue},
    keywordstyle=[3]\color{vscType},
    keywordstyle=[4]\color{vscFunc},
    morekeywords=[2]{def,class,lambda,True,False,None,self},
    morekeywords=[3]{int,str,bool,float,set,list,tuple,dict,range,frozenset},
    morekeywords=[4]{sum,any,all,min,max,len,print,sorted,enumerate,zip,map,filter,abs,graphs},
    commentstyle=\color{vscComment}\itshape,
    stringstyle=\color{vscString},
    numbers=left,
    numbersep=10pt,
    numberstyle=\tiny\color{vscLineNo},
    showstringspaces=false,
    breaklines=true,
    frame=single,
    rulecolor=\color{vscFrame},
    framerule=0.5pt,
    columns=fullflexible,
    keepspaces=true,
    xleftmargin=14pt,
    framexleftmargin=14pt,
    captionpos=b,
    literate=%
      {0}{{\textcolor{vscNumber}{0}}}1 {1}{{\textcolor{vscNumber}{1}}}1
      {2}{{\textcolor{vscNumber}{2}}}1 {3}{{\textcolor{vscNumber}{3}}}1
      {4}{{\textcolor{vscNumber}{4}}}1 {5}{{\textcolor{vscNumber}{5}}}1
      {6}{{\textcolor{vscNumber}{6}}}1 {7}{{\textcolor{vscNumber}{7}}}1
      {8}{{\textcolor{vscNumber}{8}}}1 {9}{{\textcolor{vscNumber}{9}}}1,
}

\usepackage{tikz}
\usetikzlibrary{calc, fit, shapes, backgrounds, decorations.pathreplacing}
\tikzset{
    paper edge/.style={
        draw=black!60,
        line width=0.7pt,
        line cap=round
    },
    paper path/.style={
        paper edge,
        line width=1.4pt
    },
    paper extra edge/.style={
        draw=blue!70!black,
        line width=0.7pt,
        dashed,
        line cap=round
    },
    paper forbidden edge/.style={
        draw=red!80!black,
        line width=0.7pt,
        dashed,
        line cap=round
    },
    paper highlighted edge/.style={
        draw=red!80!black,
        line width=1.4pt,
        line cap=round
    },
    paper edge label/.style={
        font=\scriptsize,
        text=black!70,
        inner sep=0.8pt
    },
    paper highlighted label/.style={
        paper edge label,
        text=red!75!black
    },
    paper vertex/.style={
        circle,
        draw=black,
        fill=white,
        line width=0.55pt,
        inner sep=1pt,
        minimum size=11pt,
        font=\scriptsize
    },
    paper representative/.style={
        paper vertex,
        draw=red!80!black,
        fill=red!20,
        line width=0.8pt
    },
    paper panel/.style={font=\scriptsize}
}

\usepackage[ruled,vlined,linesnumbered]{algorithm2e}
\SetKwProg{Proc}{\textsc{Procedure}}{}{}%

\SetAlgoCaptionSeparator{}

\usepackage{etoolbox}
\usepackage[colorlinks=true, linkcolor=blue, citecolor=black, urlcolor=blue]{hyperref}

\newcommand{\poly}{\mathrm{poly}}
\newcommand{\id}{\mathrm{ID}}
\newcommand{\WT}{\mathrm{WT}}
\newcommand{\TC}{\texttt{tri-count}}
\newcommand{\pred}{\mathrm{pred}}
\newcommand{\rep}{\psi}

\newtheorem{theorem}{Theorem}[section]

\newtheorem{lemma}[theorem]{Lemma}

\newcommand{\clique}{\textsf{Clique}\xspace}
\newcommand{\congest} {\textsf{Congest}\xspace}

\SetCommentSty{normalfont}
\SetKwComment{tcp}{\hspace{1.5em}$\triangleright$~}{}

\usepackage[nottoc,numbib]{tocbibind}
\usepackage[most]{tcolorbox}
\definecolor{lightgrey}{gray}{0.85}

\title{A Local-Counting Approach for\\
Distributed Induced Path Detection}
\author{Keren Censor-Hillel}
\author{Julian Ewaied}
\affil{Department of Computer Science, Technion, Israel\thanks{Contacts: \texttt{ckeren@cs.technion.ac.il}, \texttt{jolian.ewaied@gmail.com}.}}

\date{}

\begin{document}

\maketitle

\begin{abstract}
A well-known application of the celebrated color-coding technique is fast detection of paths in a distributed setting with limited bandwidth, by propagating a signal along colorful paths [Even et al., DISC 2017]. However, since vertices of the same color are indistinguishable by this procedure, it cannot detect induced paths, as they appear the same as cycles.

We make the following observation. If, for a given edge, we \emph{count} the number of such signals over all permutations of the random coloring that terminate at that edge, then we can separate an induced path from a cycle: an induced path contributes one to that count, while a cycle contributes two. Refining and generalizing this observation gives  our \emph{local-counting approach}, which allows us to separate induced paths from any other subgraph.  

We establish the following end results, all improving upon the state of the art in their round complexities and refined detection variants.
\begin{itemize}
    \item An $O(1)$-round algorithm for induced $P_4$ detection.
    \item $O(\Delta)$-round algorithms for detection of induced $P_5$ and $P_6$.
    \item An $\tilde{O}(n^{1/3})$-round algorithm for detection of induced $P_5$ in induced-$C_4$ free graphs.
\end{itemize}
\end{abstract}

\tableofcontents

\section{Introduction}
Finding subgraphs within a large given graph is a fundamental problem in various computational settings, since subgraphs represent connections between elements in a data set. This work focuses on distributed algorithms for finding subgraphs, for which a survey can be found in \cite{CH21Survey}. Notably, in addition to the motivation of studying connections in a graph, it has been established that some $H$-free graphs admit faster algorithms for various tasks. In distributed settings, this has been shown for coloring \cite{PettieS15, ChungPS17, HMP24}, finding large cuts \cite{HRSS17}, detecting other subgraphs \cite{Censor-HillelFG20}, and finding an MIS \cite{KS25}.

Within this context, we consider detecting short induced paths. In a distributed setting, the typical requirement of the detection variant of finding a subgraph $H$ is that all vertices accept if the graph is $H$-free, whereas at least one vertex rejects if the graph contains a copy of $H$. For a constant $k$, detecting non-induced $k$-paths (and, more generally, any tree with $k$ vertices) is known to be possible within a constant number of rounds \cite{EvenEtAl17} using the color-coding technique \cite{AYZ95}. In a nutshell, each vertex $v$ chooses a random color $c_v\in\{1,\dots,k\}$ and sends a signal to its neighbors upon receiving a signal from a neighbor with color $c_v-1$. A vertex $v$ with $c_v=k$ that receives such a \emph{path signal} detects the $k$-path. Applying this to induced paths fails because it does not rule out the existence of chords between the vertices of the detected path. A na\"ive approach for overcoming this issue is to send the identifiers of the vertices along this path. However, since there may be many of these, this approach incurs a large overhead when the size of messages is bounded, which is the setting that we consider in this work. Specifically, we work in the synchronous \congest model, in which in every round each vertex can send an $O(\log n)$-bit message to each of its neighbors.

    It is also possible to detect non-induced $k$-paths deterministically, using \emph{representative families} \cite{KorhonenR17}. In essence, this technique shows that not all information about the $k$-hop neighborhood needs to be sent as in the na\"ive approach. Instead, a sufficiently small subset of the $k$-hop neighborhood is enough for determining whether a path exists. This  does not seem to extend to induced paths, since the need to avoid chords requires a larger subset, as well as more information for computing it.

\sloppy Prior work by \cite{miyamoto2025pk} showed algorithms for detecting induced $P_4$ and $P_5$ in \congest in $O(\log n/\log\log n)$ and $O(n\log^2 n)$ rounds, respectively. It may be hard to go below $\Omega(n^\rho)$ rounds, where $\rho \approx 0.158$ is the fast matrix multiplication exponent in the related all-to-all communication \clique model, as shown in \cite{NK21}, coinciding with known results in the centralized setting \cite{FLLL15}. To the best of our knowledge, beyond the general $k$-vertex subgraph detection $\tilde{O}(n^{2-\frac{2}{3k+1}})$-round algorithm of \cite{EFFKO19}, no non-trivial \congest algorithms are known for detecting induced $P_k$ for $k>5$. This raises the following open question:

\begin{tcolorbox}[
    colback=lightgrey,
    colframe=lightgrey,
    arc=4mm,
    boxrule=0mm
]
\begin{center}
    \textbf{Open Question.} What is the round complexity of induced $P_k$ detection in \congest ?
\end{center}

\end{tcolorbox}

We make significant progress in answering the above, by presenting detection algorithms for induced $P_4$, $P_5$, and $P_6$, whose round complexities improve upon the state of the art. Notably, our induced $P_4$ detection algorithm completes within $O(1)$ rounds. Our algorithms have further compelling features, such as promising one-sided error, and most of them promise refined variants of detection or work in the more restricted broadcast \congest model (in which the messages to all neighbors within a round are the same). Below, we elaborate on our results, and overview our local-counting approach that underlies them. 
Table~\ref{tab:induced-path-results} summarizes the known and new results.

\subsection{Our Contribution}

{
\begin{table}[t]
\small
\centering
\begin{tabularx}{\linewidth}{@{}>{\raggedright\arraybackslash}Xll>{\raggedright\arraybackslash}p{3.1cm}l@{}}
\toprule
Problem & Model & Rounds & Variant & Reference \\
\midrule
Non-induced $P_k$ & broadcast \congest & $O(1)$ & detection & \cite{EvenEtAl17} \\
Non-induced $P_k$ & \congest & $O(1)$ & detection (det.) & \cite{KorhonenR17} \\
\midrule
Induced $P_4$ & broadcast \congest & $O(\log n/\log\log n)$ & detection & \cite{miyamoto2025pk} \\
Induced $P_4$ & broadcast \congest & $O(1)$ & mem. detection & Thm.~\ref{thm:induced-p4-detection} \\
\midrule
Induced $P_5$ & \congest & $O(n\log^2 n)$ & detection & \cite{miyamoto2025pk} \\
Induced $P_5$ & \congest & $O(\Delta)$ & edge-mem. detection & Thm.~\ref{thm:induced-p5-delta} \\
Induced $P_5$ (in induced-$C_4$-free graphs) & \congest & $\tilde O(n^{1/3})$ & end-edge mem. detection & Thm.~\ref{thm:induced-p5-signal} \\
\midrule
Induced $P_6$ & \congest & $\tilde O(n^{36/19})$ & detection & \cite{EFFKO19} \\
Induced $P_6$ & broadcast \congest & $O(\Delta)$ & detection & Thm.~\ref{thm:induced-p6-signal} \\
\midrule
$k$-vertex Induced $H$ & \congest & $\tilde O(n^{2-\frac{2}{3k+1}})$ & detection & \cite{EFFKO19} \\
Induced $P_k$, $k\ge 11d$ ($d\le 2$) or $k\ge 8d$ ($d\ge 3$) & \congest & $\tilde\Omega(n^{2-1/d})$ & detection & \cite{miyamoto2025pk} \\
\bottomrule
\end{tabularx}
\caption{Known and new results for detecting paths. }
\label{tab:induced-path-results}
\end{table}
}

Our first contribution is an induced $P_4$ detection algorithm which completes in a constant number of rounds and succeeds with high probability, that is, with probability $1 - \frac{1}{n^\kappa}$ for some constant $\kappa > 0$.

\newif\ifinformalstatement
\informalstatementtrue
\begin{restatable}{theorem}{inducedpfourdetection}\label{thm:induced-p4-detection}
\ifinformalstatement
    \textup{(informal)} There exists a randomized $O(1)$-round distributed algorithm which detects induced $P_4$ with high probability.
\else
    There exists a randomized $O(1)$-round distributed algorithm that solves induced $P_4$ membership detection with high probability in broadcast \congest.
\fi
\end{restatable}

As mentioned, the previously fastest algorithm for detecting induced $P_4$ requires $O(\log n/\log\log n)$ rounds \cite{miyamoto2025pk}. It uses a sketching method that originates in the randomized multiparty simultaneous-messages protocol of~\cite{KariMRS15}, which defines the notion of twin ordering and proves that its existence implies induced-$P_4$ freeness, i.e., it is a \emph{cograph} \cite{CLB81}. This approach might generate false negatives as well as false positives. In contrast, our algorithm in Theorem \ref{thm:induced-p4-detection} improves upon the above round complexity, while promising a one-sided error. Moreover, while our algorithm still works in the broadcast \congest model as the prior state of the art, it uses only a small amount of memory beyond the list of neighbors given as input to each vertex. Further, it solves a more refined variant of subgraph detection, namely, each vertex $v$ detects whether there is an induced $P_4$ that contains $v$ itself, which is referred to as \emph{membership detection}. We prove Theorem \ref{thm:induced-p4-detection} in Section~\ref{sec:induced-p4-detection}.

~\\Our next contributions are improved algorithms for induced $P_5$ detection. The following are proven in Section \ref{sec:induced-p5-detection}. We let $\Delta$ denote the maximum degree of the input graph.

\begin{restatable}{theorem}{inducedpfivedeltadetection}\label{thm:induced-p5-delta}
\ifinformalstatement
\textup{(informal)} There exists a randomized $O(\Delta)$-round distributed algorithm which detects induced $P_5$ with high probability.
\else
There exists a randomized $O(\Delta)$-round algorithm that solves induced $P_5$ edge-membership detection with high probability in \congest.
\fi
\end{restatable}

\begin{restatable}{theorem}{inducedpfivedetection}\label{thm:induced-p5-signal}
\ifinformalstatement
    \textup{(informal)} There exists a randomized $\tilde O\left(n^{1/3}\right)$-round distributed algorithm which detects induced $P_5$ with high probability on induced-$C_4$-free graphs.
\else
    There exists a randomized $\tilde O\left(n^{1/3}\right)$-round algorithm that solves induced $P_5$ end-edge membership detection with high probability in \congest on induced-$C_4$-free graphs.
\fi
\end{restatable}

The prior state of the art for induced $P_5$ detection is the aforementioned randomized $O(n\log^2 n)$-round algorithm of~\cite{miyamoto2025pk}. Our algorithm for Theorem \ref{thm:induced-p5-delta} obtains an improved round complexity, as well as a refined detection variant, which we refer to as \emph{edge-membership detection}: for each edge, both of its endpoints detect whether it is a part of an induced $P_5$. 

Our algorithm for Theorem \ref{thm:induced-p5-signal} receives as input a graph that is induced-$C_4$-free, and detects induced $P_5$ within $\tilde{O}(n^{1/3})$ rounds. It gives an
\emph{end-edge membership detection} variant: for each edge, both of its endpoints detect whether it is an \emph{end edge} of an induced $P_5$. Exploiting $H$-freeness for going below the known complexity of detection in general graphs has recently been studied in the centralized setting as well \cite{AbboudSW26}.

For comparison, we describe why the assumption of induced-$C_4$-freeness does not seem to speed up the $O(n\log^2 n)$-round algorithm of \cite{miyamoto2025pk}. At a high level, the vertices first compute the diameter of the graph in $O(n)$ rounds. If the depth is at least 4, the graph necessarily contains an induced $P_5$. The algorithm then treats the cases of depth 2 or 3 separately. Already in case of depth 2, induced-$C_4$-freeness does not seem to help: every vertex randomly joins one of $\Omega(\sqrt{m/n})$ subsets, where $m$ is the number of edges. This guarantees that with high probability each pair of subsets is connected by roughly $O(n)$ edges. For every such pair, these edges are sent to a unique representative vertex among the neighbors of the \emph{root} (the vertex with the maximal degree), which is possible due to the diameter guarantee. The representative vertices in turn forward the $O(n)$ edges to the root in $O(n)$ rounds, which thus learns the entire graph and can detect an induced $P_5$. In a sufficiently dense graph, such as an almost complete graph, one can plant an induced $P_5$ without creating an induced $C_4$. This would force some representative to send $\Theta(n)$ edges to the root, which implies no gain in the complexity. Since each representative holds only a partial set of edges, detecting an induced $P_5$ by that vertex may require it to receive too many edges from other vertices, concluding no gain in the complexity.

~\\For induced $P_6$, we use our framework to show a fast detection algorithm, as we prove in Section \ref{sec:induced-p6-detection}. Our algorithm obtains a similar round complexity of $O(\Delta)$.

\begin{restatable}{theorem}{inducedpsixdetection}\label{thm:induced-p6-signal}
\ifinformalstatement
\textup{(informal)} There exists a randomized $O(\Delta)$-round distributed algorithm which detects induced $P_6$ with high probability.
\else
There exists a randomized $O(\Delta)$-round algorithm that solves induced-$P_6$ detection with high probability in broadcast \congest.
\fi
\end{restatable}
\informalstatementfalse

To our knowledge, the previously best complexity is due to a more general algorithm for detecting any (induced or non-induced) $k$-vertex subgraph in $\tilde{O}(n^{2-\frac{2}{3k+1}})$ rounds \cite{EFFKO19}. We mention that in this algorithm, our detection variant is the traditional one: all vertices accept if the graph is induced-$P_6$-free, whereas at least one vertex rejects if the graph contains an induced $P_6$.

\subsection{Technical Overview}

In \cite{EvenEtAl17}, it is shown how to detect a (non-induced) $P_k$ for constant $k$ in a constant number of rounds using the color-coding technique~\cite{AYZ95}. Every vertex randomly and independently picks a color in $\{1,\ldots, k\}$. A signal of ordered colors is then propagated by the vertices, by having a vertex of color $i$ forward the signal iff it receives a signal from a neighbor with color $i-1$. If a $P_k$ exists and happens to be colored correctly by $1,\dots,k$, then the vertex colored $k$ detects this. Repeating this $k$-round trial for $O(\log n)$ times gives a high probability of a good coloring for a copy of $P_k$, and since a single trial requires only a single bit per message, these trials can be packed in parallel into a constant number of rounds. 

The challenge in extending the above to detect \emph{induced} $P_k$ lies in the fact that vertex identifiers are not sent with the signal, and hence it does not distinguish an induced path from a cycle, or from any other subgraph that is a path with some chords. However, forwarding identifiers of each well-colored prefix of a path is too costly in \congest.

\paragraph{Our local-counting approach.} Our key observation is that the existence of chords in a $P_k$ may imply additional non-induced paths length $k$ and that while identifying specific vertices of a path is costly, adding a counter to the signal is not. This lies at the heart of our local-counting approach, which leverages these additional non induced copies of $P_k$ to distinguish such a subgraph from an induced $P_k$. Counting various subgraphs for the goal of deciding $H$-freeness has been previously proven successful in the centralized setting \cite{KKM00,BCLSV06,AFS12,FLRSR12,KLL13,FLLL15,FKLL15,WWWY15,CurticapeanDM17,BKS18,Bressan21,KKMS23}. However, implementing these in the \congest model is too costly, requiring us to design a counting approach that is \emph{local}.

\begin{figure}[t]
\centering
\begin{tikzpicture}[x=1.35cm,y=1.35cm]
\begin{scope}[shift={(0,0)}]
  \coordinate (a) at (0,0); \coordinate (b) at (0.6,0.8); \coordinate (c) at (1.2,0);
  \draw[paper highlighted edge] (a)--(b) node[midway,auto,paper highlighted label] {$1$};
  \draw[paper highlighted edge] (b)--(c) node[midway,auto,paper highlighted label] {$1$};
  \foreach \p in {a,b,c}{\node[paper vertex] at (\p) {};}
  \node[paper panel] at (0.6,-0.4) {induced $P_3$};
\end{scope}
\begin{scope}[shift={(2.6,0)}]
  \coordinate (a) at (0,0); \coordinate (b) at (0.6,0.8); \coordinate (c) at (1.2,0);
  \draw[paper edge] (a)--(b) node[midway,auto,paper edge label] {$2$};
  \draw[paper edge] (b)--(c) node[midway,auto,paper edge label] {$2$};
  \draw[paper edge] (c)--(a) node[midway,auto,paper edge label] {$2$};
  \foreach \p in {a,b,c}{\node[paper vertex] at (\p) {};}
  \node[paper panel] at (0.6,-0.4) {triangle};
\end{scope}
\end{tikzpicture}
\caption{The number of non-induced $P_3$ that end with each edge, for each subgraph on $3$ vertices that includes a non-induced $P_3$. A red edge is an edge with an odd count.}
\label{fig:p3-contributions}
\end{figure}
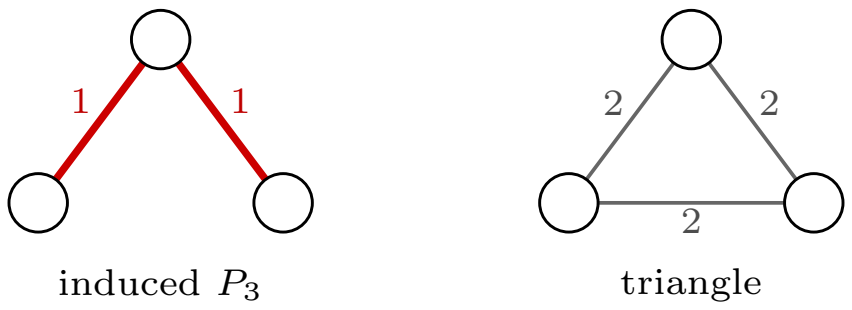

\paragraph{Warm-up: the case of $k=3$.} An induced $P_3$ implies two copies of a non-induced $P_3$, one in each direction. In contrast, a triangle implies four such copies. Suppose each vertex $v$ counts the number of non-induced $P_3$ that end at each edge $\{u,v\}$ in either of its directions. An induced $P_3$ contributes 1 to that count, while a triangle contributes 2 (see Figure~\ref{fig:p3-contributions}). Thus, keeping track of the parity of this sum is sufficient for detecting an induced $P_3$, given a good way to count the above. To implement the counting, each vertex chooses a random color $c_v$ from the set $\{1,2,3, \phi\}$, where $\phi$ stands for an uncolored vertex in that trial. Then, the vertices iterate over all possible permutations of $\{1,2,3\}$, and for every permutation $\pi$, vertex $v$ sets its color to $\pi(c_v)$. The vertices propagate a \emph{path signal} according to the order of colors as in \cite{EvenEtAl17}, but now the signal also contains the parity of the well-colored prefixes. At the end, vertex $v$ colored $\pi(c_v)=3$ holds the parity for an edge $\{u,v\}$ and sums it up with the parity $u$ has for that edge, to get the parity in either direction. It can be shown that for a set $H$ of three vertices and an edge $e$, $H$ adds an odd value to the sum if and only if it is a colorful induced $P_3$. With some care, it can also be shown that if there exists an induced $P_3$ in the graph then the probability that the counts with respect to the randomly chosen coloring is odd is constant. Hence, one can pack a logarithmic number of such trials within a constant number of rounds to get a high probability for detecting an induced $P_3$. Notice that this even provides an \emph{edge membership detection} guarantee by using a simple union bound.

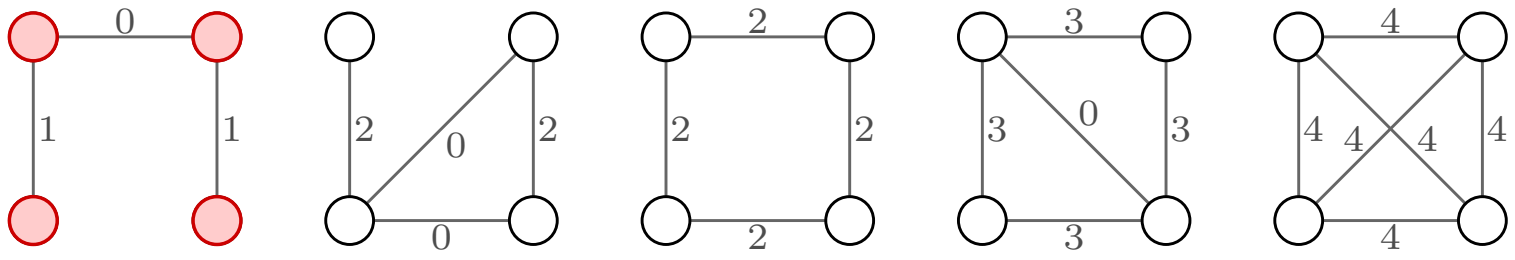
\begin{figure}[t]
\centering
\begingroup
\newcommand{\piv}[2]{\coordinate (pi#1) at #2;}
\newcommand{\pie}[3]{%
  \draw[paper edge] (pi#1)--(pi#2)
    node[midway,auto,paper edge label] {$#3$};
}
\newcommand{\pieoff}[4]{%
  \draw[paper edge] (pi#1)--(pi#2);
  \node[paper edge label]
    at ($(pi#1)!0.5!(pi#2)+#4$) {$#3$};
}
\newcommand{\piwhite}[1]{\node[paper vertex] at (pi#1) {};}
\newcommand{\pired}[1]{\node[paper representative] at (pi#1) {};}
\newcommand{\pivertices}{%
  \piv{0}{(-0.45,0.45)}\piv{1}{(0.45,0.45)}
  \piv{2}{(0.45,-0.45)}\piv{3}{(-0.45,-0.45)}
}
\newcommand{\piwhites}{\foreach \i in {0,...,3}{\piwhite{\i}}}
\begin{tikzpicture}[
  x=1.35cm,
  y=1.35cm,
  paper edge/.append style={line width=0.55pt},
  paper edge label/.append style={inner sep=0.5pt},
  paper vertex/.append style={minimum size=9pt,inner sep=0pt,line width=0.55pt},
  paper representative/.append style={minimum size=9pt,inner sep=0pt,line width=0.65pt}
]
\begin{scope}[shift={(0,0)}]
  \pivertices
  \pie{0}{1}{0}\pie{0}{3}{1}\pie{1}{2}{1}
  \piwhites\foreach \i in {0,...,3}{\pired{\i}}
\end{scope}
\begin{scope}[shift={(1.55,0)}]
  \pivertices
  \pie{0}{3}{2}\pie{1}{2}{2}\pie{1}{3}{0}\pie{2}{3}{0}
  \piwhites
\end{scope}
\begin{scope}[shift={(3.10,0)}]
  \pivertices
  \pie{0}{1}{2}\pie{0}{3}{2}\pie{1}{2}{2}\pie{2}{3}{2}
  \piwhites
\end{scope}
\begin{scope}[shift={(4.65,0)}]
  \pivertices
  \pie{0}{1}{3}\pie{0}{2}{0}\pie{0}{3}{3}\pie{1}{2}{3}\pie{2}{3}{3}
  \piwhites
\end{scope}
\begin{scope}[shift={(6.20,0)}]
  \pivertices
  \pie{0}{1}{4}\pieoff{0}{2}{4}{(-0.18,-0.05)}
  \pie{0}{3}{4}\pie{1}{2}{4}\pieoff{1}{3}{4}{(0.18,-0.05)}
  \pie{2}{3}{4}\piwhites
\end{scope}
\end{tikzpicture}%
\endgroup
\caption{The number of non-induced $P_4$ that end with each edge, for each subgraph on $4$ vertices that includes a non-induced $P_4$. A red vertex has an odd total count over its incident edges.}
\label{fig:p4-contributions}
\end{figure}

\paragraph{Parity over edges is insufficient for $k=4$.} At a first glance, it seems that the above approach should work for detecting induced $P_k$ for any constant $k$. However, already $k=4$ introduces additional challenges, since there are additional $4$-vertex graphs beyond a path and a cycle. To see why this raises an issue, consider the diamond, which is a $4$-clique minus one edge. In a diamond $\{x,y,z,w\}$ with the non-edge being $\{x,z\}$, there are no non-induced copies of $P_4$ that end in $\{y,w\}$, and there are 3 such copies for any other edge. This means that keeping track of the parity of such paths per edge $e$ may lead to false positives. To solve this issue, each vertex $v$ considers the parity of the number of all such paths ending in all of its incident edges. Thus, in such a diamond, each vertex counts 6 paths in total for all of its edges, as opposed to an induced $P_4$ which gives a count of 1 for the end edges and 0 for the middle edge, which sums to 1 at all vertices (see Figure~\ref{fig:p4-contributions}). We prove that this separates induced $P_4$ from \emph{all} other $4$-vertex subgraphs, and thus apply our local-counting approach to obtain \emph{membership detection}.

\paragraph{Parity over vertices is insufficient for $k=5$.} The case of $k=5$ is more involved. Consider the \emph{bull subgraph}, which is an induced $P_5$ on vertices $x,y,z,w,v$ with a chord $\{y,w\}$. It has the same two non-induced copies of $P_5$ as an induced $P_5$, and thus merely counting non-induced copies of $P_5$ is insufficient.
For this reason, we require further counts. That is, we count additional subgraphs rather than only non-induced copies of $P_5$, in order for the parity of the total count to reflect the existence of an induced $P_5$. We refer to these as \emph{local corrections}. 

We compute the local corrections as follows. In $O(\Delta)$ rounds, every vertex sends its list of neighbors to all of its neighbors. Vertices then propagate path signals as before, to count the number of paths that pass through every edge. We show that for every $5$-vertex set that is not an induced $P_5$ and contributes an odd count of $P_5$ to an edge $e$, at least one endpoint of $e$ has degree at least $3$ in the induced graph on this set. That endpoint thus learns the entire induced $5$-vertex subgraph and can therefore count the corresponding local correction. This promises that each edge $e$ gets an odd contribution only from colorful induced $P_5$ that include it, and we prove that there is a constant probability that it is odd if there exists an induced $P_5$ that includes it, thus we obtain \emph{edge membership detection}. We stress that while an $O(\Delta)$ round complexity is trivial for induced $P_4$, this is not the case for induced $P_5$, as no vertex sees the induced subgraph just by obtaining the lists of neighbors of its neighbors.

\begin{figure}[htbp]
    \centering
    \begin{tikzpicture}[
        scale=1.35,
        base/.style={paper path},
        extra/.style={paper extra edge},
        last/.style={paper highlighted edge},
        vertex/.style={paper vertex},
        panel/.style={paper panel}
    ]

    \begin{scope}[shift={(0,0)}]
    \coordinate (p1) at (-0.90,0.00);
    \coordinate (p2) at (-0.45,0.42);
    \coordinate (p3) at (0.00,0.00);
    \coordinate (p4) at (0.45,0.42);
    \coordinate (p5) at (0.90,0.00);
    \draw[base] (p1)--(p2)--(p3)--(p4)--(p5);
    \draw[last] (p4)--(p5);
    \node[vertex] at (p1) {$1$};
    \node[vertex] at (p2) {$2$};
    \node[vertex] at (p3) {$3$};
    \node[vertex] at (p4) {$4$};
    \node[vertex] at (p5) {$5$};
    \node[panel] at (0,-1.05) {path $\mathcal{P}$};
    \end{scope}

    \begin{scope}[shift={(2.85,0)}]
    \coordinate (b1) at (-1.05,-0.28);
    \coordinate (b2) at (-0.50,0.00);
    \coordinate (b3) at (0.00,0.58);
    \coordinate (b4) at (0.50,0.00);
    \coordinate (b5) at (1.05,-0.28);
    \draw[base] (b1)--(b2)--(b3)--(b4)--(b5);
    \draw[extra] (b2)--(b4);
    \draw[last] (b4)--(b5);
    \node[vertex] at (b1) {$1$};
    \node[vertex] at (b2) {$2$};
    \node[vertex] at (b3) {$3$};
    \node[vertex] at (b4) {$4$};
    \node[vertex] at (b5) {$5$};
    \node[panel] at (0,-1.05) {bull $\mathcal{B}$};
    \end{scope}

    \begin{scope}[shift={(5.70,0)}]
    \coordinate (d1) at (-1.12,0.00);
    \coordinate (d2) at (-0.58,0.00);
    \coordinate (d3) at (0.00,0.58);
    \coordinate (d4) at (0.00,-0.58);
    \coordinate (d5) at (0.72,0.00);
    \draw[base] (d1)--(d2)--(d3)--(d4)--(d5);
    \draw[extra] (d2)--(d4);
    \draw[extra] (d3)--(d5);
    \draw[last] (d4)--(d5);
    \node[vertex] at (d1) {$1$};
    \node[vertex] at (d2) {$2$};
    \node[vertex] at (d3) {$3$};
    \node[vertex] at (d4) {$4$};
    \node[vertex] at (d5) {$5$};
    \node[panel] at (0,-1.05) {diamond-pendant $\mathcal{D}$};
    \end{scope}

    \begin{scope}[shift={(9.05,0)}]
    \coordinate (g1) at (0.00,-0.10);
    \coordinate (g2) at (-0.82,-0.52);
    \coordinate (g3) at (-0.44,0.54);
    \coordinate (g4) at (0.44,0.54);
    \coordinate (g5) at (0.82,-0.52);
    \draw[base] (g1)--(g2)--(g3)--(g4)--(g5);
    \draw[extra] (g1)--(g3);
    \draw[extra] (g1)--(g4);
    \draw[extra] (g1)--(g5);
    \draw[last] (g4)--(g5);
    \node[vertex] at (g1) {$1$};
    \node[vertex] at (g2) {$2$};
    \node[vertex] at (g3) {$3$};
    \node[vertex] at (g4) {$4$};
    \node[vertex] at (g5) {$5$};
    \node[panel] at (0,-1.05) {gem $\mathcal{G}$};
    \end{scope}

    \end{tikzpicture}
    \caption{The four subgraphs corresponding to local corrections.  All
    subgraphs share the ordered path $1-2-3-4-5$, drawn with \emph{bold} edges; the additional edges of each
    subgraph are drawn \emph{dashed}.  The
    edge between colors $4$ and $5$ is highlighted in red.}
    \label{fig:p5-subgraphs}
\end{figure}
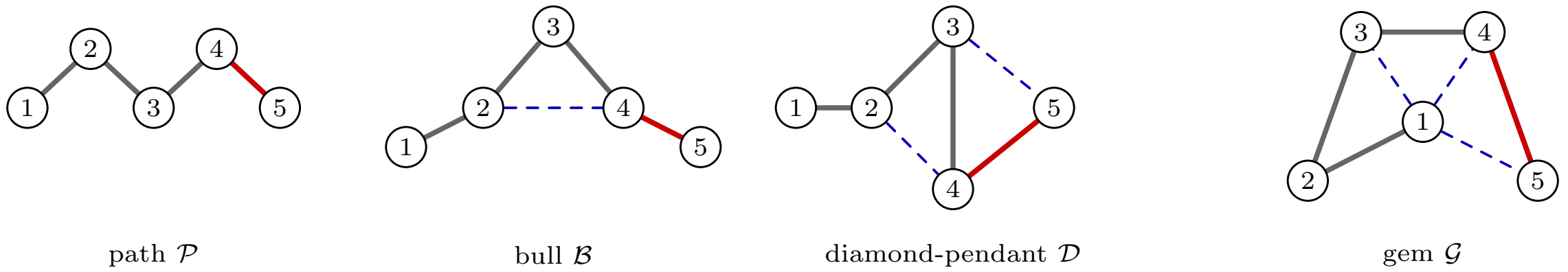

We notice that we can do better than $O(\Delta)$ rounds for large $\Delta$, given a promise that the input graph is induced-$C_4$ free. In more detail, consider the aforementioned bull graph. We show how using a slight modification to the triangle-listing algorithm of \cite{CPSZ21} allows us to count such bull graphs. Our goal would then be to decrement their number from the number of non-induced $P_5$, to get the number of induced $P_5$. However, bull graphs are subgraphs of other $5$-vertex graphs and these get counted when counting bull graphs, thus requiring further local corrections. The promise that the graph contains no induced $C_4$ allows this sequence of local corrections to require only listing triangles rather than the entire neighborhoods (see Figure \ref{fig:p5-subgraphs}), which in turn yields a round complexity of $\tilde{O}(n^{1/3})$ rather than $O(\Delta)$. Notice that the sequence of subgraphs for these local corrections enjoys the property that each one is inclusive of all of its predecessors, thus counting a subgraph in the sequence does not alter the count of prior subgraphs. Notice that if we were to count subgraphs that contain copies of induced $C_4$, we would necessarily incur a number of rounds that is near-linear in $n$, due to the lower bound of \cite{GM21}.

\paragraph{Neighbors of neighbors are insufficient for $k=6$.} For detecting induced $P_6$, even allowing a budget of $O(\Delta)$ rounds is insufficient for directly extending the induced $P_5$ detection algorithm. The reason is that with $6$ vertices there may be subgraphs for which no vertex can learn the entire induced subgraph within $O(\Delta)$ rounds. For example, consider the $6$-vertex graph which is an induced $C_5$ on vertices $\{x,y,z,w,u\}$ with a pendant edge $\{u,v\}$. The vertices $x$ and $w$ that are neighbors of $u$ both count one path signal for this subgraph, which would cause false positives. However, no vertex on a copy of this subgraph can see the entire subgraph within $O(\Delta)$ rounds and thus local corrections are insufficient. Notice that no vertex can count an induced $C_6$ as well, but for an induced $C_6$ the counts for the path signal are 2 for each edge, and thus it does not cause false positives.

Our solution is as follows. In the subgraph above of an induced $C_5$ with a pendant, there are two different vertices that each count a single path of length $6$. To overcome this, the vertices sum all of their counts at a single vertex, which can be done in a constant number of rounds over a BFS tree (since if the depth of the BFS tree in the graph is greater than $5$ then there must be an induced $P_6$). This way, the two paths of the $C_5$ with a pendant subgraph that are counted by different vertices get canceled out in the total count. 

This alone is insufficient and still requires the vertices to count local corrections, as otherwise we may have a single vertex that counts a certain subgraph as adding 1 to the path signal counts. Since there are many local corrections in this case, rather than counting certain subgraphs, we count \emph{patterns}. A pattern is a set of required edges on $6$ vertices along with a set of required non-edges. These may not cover all pairs of vertices, and rather there may be pairs of vertices for which the above is not specified. Each pattern is associated with a single vertex that counts it, and it is associated with a coefficient that reflects what it should add to the total count, in order for the latter to eventually cancel out subgraphs that are not induced $P_6$.

\subsection{Discussion}
Since only checking whether a colorful path signal reaches an edge is insufficient for finding \emph{induced} paths, we introduce the \emph{local-counting} approach: each edge counts these signals over all permutations of the random coloring. In some cases, this is insufficient as well, and \emph{local corrections} need to be counted, to cancel the contributions of other subgraphs to the count. This approach yields our improved algorithms for induced $P_4$, $P_5$, and $P_6$. We hope that, combined with additional techniques, the local-counting approach can be extended to find additional induced subgraphs, such as induced $P_k$ for larger $k$, or induced cycles. The complexities of these problems (also for deterministic algorithms) are important remaining open questions.

\subsection{Additional Related Work}
There is extensive literature on distributed subgraph finding. In what follows, we focus on the detection variant in the \congest model.

First, we stress that the aforementioned color-coding based algorithm of \cite{EvenEtAl17} extends from paths to any non-induced tree on $k$ vertices. Deterministic algorithms for detecting paths are given in \cite{KorhonenR17}.
For detecting induced paths of larger $k$, a lower bound of $\tilde\Omega(n^{2-1/d})$ rounds was proven in \cite{miyamoto2025pk} for every $k\ge 11d$ when $d\le 2$, and for every $k\ge 8d$ for $d\ge 3$.

For cycles, the fastest triangle detection algorithm is the $\tilde O(n^{1/3})$-round algorithm of \cite{CPSZ21}, and the later deterministic algorithm of \cite{CensorHillelLV24}, which both provide the stronger guarantee of listing (each vertex outputs a list of triangles such that the union of the lists is the set of all triangles). For listing, the above is tight \cite{IzumiG17,Pandurangan0S18}, but it is not known to be tight for detection and, in fact, in the related \clique model where vertices can send $O(\log n)$ bit messages to \emph{all} vertices in the graph in each round, triangle detection can be solved faster than listing by using fast matrix multiplication \cite{Censor-HillelKK19}, which is not known in the \congest model. Lower bounds of two rounds were proven in \cite{AbboudCKL20,FischerGKO18} and recently an $\Omega(\log\log n)$ lower bound was given in \cite{AssadiS25}. 
Larger cycles were addressed in several papers \cite{DruckerKO13, KorhonenR17, Censor-HillelKK19, Censor-HillelFG20, FraigniaudLMT24, FraigniaudLT24}, with the state of the art being the deterministic $C_{2k}$-detection algorithm in $\tilde{O}(n^{1-1/k})$ rounds of \cite{FraigniaudLMT25}, which is tight due to the lower bound of \cite{DruckerKO13}. It is known that the complexity of $C_{2k+1}$ detection is near linear \cite{DruckerKO13,KorhonenR17}.

Beyond paths and cycles, \cite{EFFKO19} give an $\tilde{O}(n^{2-\frac{2}{3k+1}})$-round algorithm for detecting any fixed $k$-vertex pattern. This nearly matches the $\tilde{\Omega}(n^{2-\Theta(1/k)})$ lower bound for some $k$-vertex patterns~\cite{FischerGKO18}.

\section{Induced $P_4$ Detection}
\label{sec:induced-p4-detection}
In this section, we present our algorithm for induced $P_4$ membership detection.

\inducedpfourdetection*

We prove Theorem~\ref{thm:induced-p4-detection} by giving the local-counting algorithm below. We first present a single trial of Algorithm~\ref{alg:induced-p4-detection}, in which every vertex $v$ independently samples a color $c_v\in\{1,2,3,4,\phi\}$; vertices colored $\phi$ do not participate, and the subsequent signal-propagation steps are executed on the remaining vertices. In each communication round, every vertex sends one bit that records the \emph{parity} of the number of signals traveling along its edges that follow the color order in that round. Vertex $v$ stores this parity in $\omega_v$, and maintains, for each incident edge $e$, a counter $s_e^v$ that holds the parity of the number of \emph{colorful} non-induced copies of $P_4$ that end with $e$. The same propagation is repeated for each of the $24$ permutations of the four colors, ensuring that all paths on a colorful subset of 4 vertices are counted, and only the induced paths affect the parity at a vertex. The vertices run $\alpha\log n$ independent trials of Algorithm~\ref{alg:induced-p4-detection} in parallel, where $\alpha$ is a sufficiently large constant yet to be determined; the pseudocode and lemmas follow.

\begin{algorithm}[H]
\DontPrintSemicolon
\SetAlgorithmName{Algorithm}{alg}{List}
$s_e^v$ $\gets 0$ for all $e$ incident to $v$ \;

$c_v \gets$ uniform random color in $\{1,2,3,4,\phi\}$ \;
send $c_v$ to every $u\in N(v)$\;

\If{$c_v = \phi$}{
    \textbf{accept} \;
}

\ForEach{permutation $\pi$ of $\{1,2,3,4\}$}{ \label{step: perm}
    $k\gets\pi(c_v)$ \tcp*[h]{$k=\pi(c_v)$ is the permuted color of $v$} \;
    $N_j(v)\gets\{u\in N(v):\pi(c_u)=j\}$ for each $j\in\{1,\dots,4\}$\;
    $\mathrm{in}(u) \gets 0$ for every $u\in N(v)$\;
    \For{$i \gets 1$ \KwTo $4$}{ \label{step: colors}
        \If(\tcp*[h]{$k=\pi(c_v)$ is the current color of $v$}){$k=i$}{
            \eIf{$i=1$}{send the value $1$ to every $u\in N_2(v)$\;}{
                $\omega_v\gets\sum_{u\in N_{i-1}(v)}\mathrm{in}(u) \pmod 2$\;
                \lIf{$i\le 3$}{send $\omega_v$ to every $u\in N_{i+1}(v)$}
            }
            \lIf{$i=3$}{$s_{\{v,u\}}^v$ $\gets$ $s_{\{v,u\}}^v$ $+\,\omega_v \pmod 2$ for every $u\in N_4(v)$} \label{step: count edges 3}
            \lIf{$i=4$}{$s_{\{v,u\}}^v$ $\gets$ $s_{\{v,u\}}^v$ $+\,\mathrm{in}(u) \pmod 2$ for every $u\in N_3(v)$} \label{step: count edges}
        }
        \If{$k=i+1$}{$\mathrm{in}(u)\gets$ value received from $u$, for every $u\in N_i(v)$\;}
    }
}

$s_v \gets \sum_{e \text{ incident to } v}$ $s_e^v$ $\pmod{2}$ \;
\If{$s_v \neq 0$}{
    \textbf{reject} \;
}

\textbf{accept} \;
\caption{Single Trial for Induced $P_4$ Membership Detection for a Vertex $v$}
\label{alg:induced-p4-detection}
\end{algorithm}

Each vertex $v$ holds its own copy $s_e^v$ of the counter of every incident edge $e$. We show that both endpoints of an edge hold the same value, which lets us write $s_e$ without a superscript.
\begin{lemma}\label{lem:p4-se-symmetric}
For every edge $e=\{u,v\}$, it holds that $s_e^u=s_e^v$ at the end of every trial of Algorithm~\ref{alg:induced-p4-detection}.
\end{lemma}
\begin{proof}
The counters of $e$ change only in Lines~\ref{step: count edges 3} and~\ref{step: count edges}, in an iteration $\pi$ in which one endpoint of $e$, say $u$, satisfies $\pi(c_u)=3$ and the other satisfies $\pi(c_v)=4$. In this iteration, $u$ adds $\omega_u$ to $s_e^u$ and sends $\omega_u$ to $v$, which receives it as $\mathrm{in}(u)$ and adds it to $s_e^v$. Hence, both endpoints add the same value in every iteration.
\end{proof}

A coloring $c: V\to \{1,2,3,4,\phi\}$ assigns to each vertex either a color in $\{1,2,3,4\}$ or $\phi$. We say a subset of 4 vertices is \emph{active} if none of them is colored $\phi$, and \emph{colorful} with respect to $c$ if it is active and the four vertices are colored with distinct colors. When $c$ is clear from context we simply say the subset is colorful.
Define the \emph{contribution} of a subset of 4 vertices $H\subseteq V$ to an edge $e$, denoted $C(H,e)$, as the number of non-induced copies of $P_4$ in $H$ that end at $e$, provided $H$ is colorful; if $H$ is not colorful, then it holds that $C(H,e)=0$. Furthermore, define
\[
    C(H,v) = \sum_{\substack{e \text{ incident to } v \\ e\in E(H)}}C(H, e).
\]
In the following lemma, we connect this definition to the quantity $s_e$ accumulated on each edge $e$ in Algorithm~\ref{alg:induced-p4-detection}.
\begin{restatable}{lemma}{pfourcontribution}
\label{lem: contribution} 
For every $e\in E$, it holds that
\[
    s_e \equiv \sum_{H\in \binom{V}{4}}{C(H,e)} \pmod 2,
\]
and hence the value $s_v$ satisfies $s_v \equiv \sum_{H\in \binom{V}{4}}{C(H,v)} \pmod 2$ for each $v\in V$.
\end{restatable}

Then, we prove the correctness guarantee that ensures no subgraph can cause false positives.

\begin{restatable}{lemma}{pfourparitycounting}
\label{lem:parity_counting}
For every subset of 4 vertices $H\subseteq V$ and every vertex $v\in V$, it holds that $C(H,v) \equiv 1 \pmod 2$ if and only if $v\in H$, $H$ is colorful, and $G[H]$ is an induced $P_4$.
\end{restatable}

By Lemma~\ref{lem: contribution}, $s_v$ equals modulo $2$ the sum of the contributions $C(H,v)$ over all four-vertex sets $H$. Lemma~\ref{lem:parity_counting} shows that precisely the colorful induced copies of $P_4$ containing $v$ contribute $1$. Thus, $s_v$ is odd exactly when their number is odd. The following lemma shows that, whenever $v$ belongs to an induced $P_4$, this occurs for sufficiently many colorings.

\begin{lemma} \label{lem:p4_colorings}
    If $v$ is a vertex in an induced $P_4$, then there are at least $5^{n-4}$ colorings for which $s_v$ is odd. 
\end{lemma}

The above is a special case of the following lemma, which we will use in other sections as well. After its statement, we explain how to instantiate it to get Lemma~\ref{lem:p4_colorings}. 

\begin{restatable}{lemma}{pkcolorings}
\label{lem:pk-colorings}
Let $k\ge 1$ be a constant, and let $M \subseteq \binom{V}{k}$ be a nonempty set of induced copies of $P_k$. For a coloring $c: V \to \{1,2,\dots,k,\phi\}$ of $V$, let $M_c := \{H\in M : H\text{ is colorful under } c\}$. Then there are at least $(k+1)^{n-k}$ colorings $c$ for which $|M_c| \not\equiv 0 \pmod 2$.
\end{restatable}

To obtain Lemma~\ref{lem:p4_colorings}, fix a vertex $v$ contained in an induced $P_4$ and apply Lemma~\ref{lem:pk-colorings} with $k=4$ and $M$ equal to the nonempty set of induced copies of $P_4$ containing $v$. Then $|M_c|\equiv s_v\pmod 2$, so the lemma gives at least $(4+1)^{n-4}=5^{n-4}$ colorings for which $s_v$ is odd.

We are now ready to prove the main theorem of this section, given the above lemmas. 

\begin{proof}[Proof of Theorem~\ref{thm:induced-p4-detection}]

    Vertices run $\alpha\log n$ independent trials of Algorithm~\ref{alg:induced-p4-detection} in parallel, for a sufficiently large constant $\alpha$. Since each trial of Algorithm~\ref{alg:induced-p4-detection} uses 1-bit bandwidth, the $\alpha\log n$ trials of this algorithm are packed into $O(1)$ \congest chunks, and each chunk is sent in a single round. A vertex rejects if it rejects in at least one trial of this algorithm, and accepts otherwise. Each trial of Algorithm~\ref{alg:induced-p4-detection} takes a constant number of rounds to iterate over 24 permutations and 4 iterations of the loop in Line~\ref{step: colors}. Since the cost of running the $\alpha \log n$ trials of this algorithm in parallel is only another constant factor, the total number of rounds is $O(1)$.

\medskip\noindent\textbf{Correctness.} By Lemma~\ref{lem: contribution}, for every vertex $v$, the value $s_v$ in a single trial of Algorithm~\ref{alg:induced-p4-detection} satisfies
\[
s_v \equiv \sum_{H \in \binom{V}{4}} C(H, v) \pmod 2.
\]
By Lemma~\ref{lem:parity_counting}, it holds that $C(H,v) \equiv 0\text{ modulo }2$ for any $H$ that is not a colorful induced $P_4$. Therefore, the value $s_v$ equals the number of colorful induced copies of $P_4$ containing $v$ modulo $2$. If no induced $P_4$ exists, then it holds that $s_v = 0$ for every $v \in V$ in every trial of Algorithm~\ref{alg:induced-p4-detection}, and every vertex correctly accepts.

Otherwise, let $v$ be a vertex on an induced $P_4$. By Lemma~\ref{lem:p4_colorings}, there are at least $5^{n-4}$ colorings for which $s_v \equiv 1 \pmod 2$, hence the probability that $v$ rejects in a single trial of Algorithm~\ref{alg:induced-p4-detection} is at least $1/5^4$. 
The $\alpha\log n$ trials of Algorithm~\ref{alg:induced-p4-detection} are independent. For a fixed vertex $v$ belonging to an induced $P_4$, the probability that all $\alpha\log n$ trials of this algorithm accept is at most
\[
(1-5^{-4})^{\alpha\log n}=n^{-\kappa\alpha},
\]
where $\kappa=-\log(1-5^{-4})>0$. By a union bound over all vertices, the probability that any vertex belonging to an induced $P_4$ accepts is at most $n^{1-\kappa\alpha}$. Thus, for sufficiently large $\alpha$, the algorithm succeeds with high probability.
\end{proof}

We now return to proving Lemmas~\ref{lem: contribution}, \ref{lem:parity_counting}, and \ref{lem:pk-colorings}.

\pfourcontribution*

\begin{proof}
Let a single trial of Algorithm~\ref{alg:induced-p4-detection} be fixed, and let $c$ be its corresponding coloring.

Let $e\in E$ be an edge, let $\pi$ be a permutation of $\{1,2,3,4\}$, and let $s_e^\pi$ be the value that the endpoints of $e$ add to $s_e$ during the iteration $\pi$ of Algorithm~\ref{alg:induced-p4-detection}, in Lines~\ref{step: count edges 3} and~\ref{step: count edges}; thus, $s_e\equiv\sum_{\pi} s_e^\pi \pmod 2$, summed over all permutations $\pi$ of $\{1,2,3,4\}$.

We show by induction on $i$ that, for every vertex $v$ with $\pi(c_v)=i$, the value sent (if $i=1$) or computed as $\omega_v$ (if $i\ge 2$) is equal modulo $2$ to the number of colorful ordered paths $v_1,\ldots,v_i$ such that $v_i=v$ and $\pi(c_{v_j})=j$ for every $j\le i$. For $i=1$ this is the value $1$ sent by $v$, since the only path of length 1 is simply the vertex $v$ itself. For $i\ge 2$, $\omega_v=\sum_{u\in N_{i-1}(v)}\mathrm{in}(u) \pmod 2$, where $\mathrm{in}(u)$ is the value $u$ sent in the previous round. By the induction hypothesis this value is equal modulo $2$ to the number of colorful ordered paths $v_1,\ldots,v_{i-1}$ such that $v_{i-1}=u$. Summing over $u\in N_{i-1}(v)$ counts, modulo $2$, the colorful ordered paths $v_1,\ldots,v_{i-1},v$.

Consequently, at $i=3$ the quantity $\omega_{v_3}$ computed for a vertex $v_3$ with $\pi(c_{v_3})=3$ is equal modulo $2$ to the number of colorful ordered paths $v_1v_2v_3$ with $\pi(c_{v_j})=j$, and in Line~\ref{step: count edges 3}, $v_3$ adds this value to $s_{\{v_3,u\}}$ for every neighbor $u$ with $\pi(c_u)=4$; symmetrically, in Line~\ref{step: count edges}, a vertex $v_4$ with $\pi(c_{v_4})=4$ adds the value $\mathrm{in}(v_3)$ received from each neighbor $v_3$ with $\pi(c_{v_3})=3$, which by the same induction is equal modulo $2$ to the number of colorful ordered paths $v_1v_2v_3$ ending at $v_3$. Hence, for every edge $e$, both endpoints add, modulo $2$, one unit for every non-induced copy of $P_4$ whose colors under $\pi$ are 1,2,3,4 in order and whose last edge is $e$.

Summing over all $24$ permutations $\pi$ and all subsets of 4 vertices $H\in\binom{V}{4}$ gives
\[
    s_e \equiv \sum_{H\in \binom{V}{4}} C(H,e) \pmod 2.
\]
Consequently,
\begin{align*}
s_v
&= \sum_{e \text{ incident to } v} s_e \equiv \sum_{e \text{ incident to } v} \sum_{H\in\binom{V}{4}} C(H,e) \pmod 2 \\
&\equiv \sum_{H\in\binom{V}{4}}
    \sum_{\substack{e \text{ incident to } v \\ e\in E(H)}} C(H,e) \pmod 2
\equiv \sum_{H\in\binom{V}{4}} C(H,v) \pmod 2.
\end{align*}
The first transition applies the preceding identity to every edge incident to $v$, the second exchanges the order of summation and uses $C(H,e)=0$ for $e\notin E(H)$, and the final transition follows from the definition of $C(H,v)$.
\end{proof}

Next, we prove Lemma~\ref{lem:parity_counting}, to ensure no false positives occur.

\pfourparitycounting*

\begin{proof}
If the subset $H$ is not colorful or the vertex $v$ does not belong to $H$, then it holds that $C(H,v)=0$ by definition. Otherwise, we prove the claim by an exhaustive computation over all isomorphism classes on 4 vertices, using the SageMath script shown in Figure~\ref{lst:p4-sage-check}. The values computed by this script are shown in Figure~\ref{fig:p4-contributions}.
\end{proof}

\begin{figure}[htbp]
\begin{lstlisting}[style=sagepy]
from itertools import permutations

V = range(4)
PATH = [(0, 1), (1, 2), (2, 3)]          
k = lambda u, v: tuple(sorted((u, v)))

def C_edge(E, e):
    return sum(1 for t in permutations(V)
               if all(k(t[i], t[j]) in E for i, j in PATH) and k(t[2], t[3]) == e)

def C_vertex(E, v):
    # C(H,v): sum of C(H,e) over the edges e of H incident to v
    return sum(C_edge(E, e) for e in E if v in e)

for G in graphs.nauty_geng("4"):
    E = set(G.edges(labels=False))
    is_p4 = G.is_isomorphic(graphs.PathGraph(4))
    for v in V:
        assert C_vertex(E, v) % 2 == (1 if is_p4 else 0)

print("Verified: C(H,v) is odd iff H is a colorful induced P4, for every v and every H.")
\end{lstlisting}
\caption{The SageMath script verifying Lemma~\ref{lem:parity_counting} by exhaustive search over all four-vertex graphs.}
\label{lst:p4-sage-check}
\end{figure}

Finally, we prove Lemma~\ref{lem:pk-colorings} to claim there are sufficiently many ``good'' colorings.

\pkcolorings*

\begin{proof}
Let $M$ be the set of induced $P_k$ in $G$. We show that there are at least ${(k+1)}^{n-k}$ colorings $c$ such that $|M_c| \not\equiv 0 \pmod 2$, as follows. Since $M$ is not empty, there exists an $H\in M$, i.e., $H$ is an induced $P_k$. Consider a coloring of $V\setminus H$. We prove that one can color the vertices of $H$ such that $|M_c|\not\equiv 0\pmod 2$, which proves the statement. 

Consider a fixed injective coloring of $H$. Let $S$ be a subset of $H$, and let $c_S$ be the coloring in which the vertices of $S$ inherit the injective coloring, the vertices of $H\setminus S$ are colored $\phi$, and the vertices of $V\setminus H$ retain their fixed colors. Define $\sigma(S):=|M_{c_S}|$. We prove that there exists at least one subset $S$ such that $\sigma(S)\not\equiv 0\pmod 2$.

If for some $S\subsetneq H$ it holds that $\sigma(S)\not\equiv 0\pmod 2$, the claim is correct. Otherwise, let $\sigma(S)\equiv 0\pmod 2$ for every such set $S\subsetneq H$. We prove that $\sigma(H)\not\equiv 0\pmod 2$. Define
\[
r_S \;:=\; \bigl|\{A\in M_{c_S}:\ A\cap H = S\}\bigr|,
\]
Thus, $r_S$ counts the colorful sets $A\in M$ satisfying $A\cap H=S$. The value $r_S$ depends only on the colors of $S$ and on the fixed coloring of $V\setminus H$. We first claim that
\begin{equation}\label{eq:pk-zeta}
\sigma(S) \;=\; \sum_{S'\subseteq S} r_{S'}.
\end{equation}

Indeed, the coloring defining $\sigma(S)$ colors every vertex of $H\setminus S$ by $\phi$. A colorful set contains no vertex colored $\phi$; therefore, every colorful $A\in M_{c_S}$ satisfies $A\cap H\subseteq S$. Grouping these sets $A$ by the value $S'=A\cap H$, which ranges over all $S'\subseteq S$, gives Equation~\ref{eq:pk-zeta}.

We claim that, under the assumption, $r_S\equiv 0\pmod 2$ for every proper subset $S\subsetneq H$ by induction on $|S|$. The base case is $S=\emptyset$; hence, Equation~\ref{eq:pk-zeta} becomes $\sigma(\emptyset)=r_\emptyset$. As the left-hand side is equal to $0$ modulo $2$ by the assumption, it holds that $r_\emptyset\equiv 0\pmod 2$. We now assume the claim holds for all subsets $S'\subsetneq S$ and prove it for a proper subset $S$ of $H$.

Let $S\subsetneq H$ and assume that $r_{S'}\equiv 0\pmod 2$ for every $S'\subsetneq S$. Rewrite Equation~\ref{eq:pk-zeta} as
\[
\sigma(S) \;\equiv\; r_S + \sum_{S'\subsetneq S} r_{S'} \;\equiv\; r_S \pmod 2,
\]
since each $r_{S'}$ for $S'\subsetneq S$ is equal to $0$ modulo $2$ by the induction hypothesis. As $\sigma(S)\equiv 0\pmod 2$, it follows that $r_S\equiv 0\pmod 2$.

Hence, applying Equation~\ref{eq:pk-zeta} for $S=H$, we get
\[
\sigma(H)\equiv r_H+\sum_{S'\subsetneq H}r_{S'}\equiv r_H\pmod 2.
\]
Finally, a set $A\in M$ satisfying $A\cap H=H$ must satisfy $A=H$, since $|A|=|H|=k$. Thus the set defining $r_H$ contains only $H$. Since $H\in M$ and $H$ is colorful under the fixed injective coloring, it holds that $r_H=1$, and hence $\sigma(H)\equiv 1\pmod 2$.

Thus, for every fixed coloring of $V\setminus H$, there exists a coloring of the vertices in $H$ for which $|M_c|\not\equiv 0\pmod 2$. Since there are $(k+1)^{n-k}$ possible colorings of $V\setminus H$, the lemma follows.
\end{proof}

\section{Induced $P_5$ Detection}
\label{sec:induced-p5-detection}

In this section, we present two algorithms for detecting induced $P_5$ as stated in Theorem \ref{thm:induced-p5-delta} and Theorem \ref{thm:induced-p5-signal}.

\subsection{Induced $P_5$ Detection in General Graphs}

\label{sec:induced-p5-delta}
We now present our \congest algorithm for edge-membership detection of induced $P_5$ in general graphs. At the end of this algorithm, every vertex $v$ outputs a set $L_v$ of all incident edges that are in some induced $P_5$, with high probability.
\inducedpfivedeltadetection*

The algorithm first sends its list of neighbors to all of its neighbors, which takes $O(\Delta)$ rounds. It then runs $O(\log n)$ trials of Algorithm~\ref{alg:p5-delta-detection}, in parallel, since every such trial requires a bandwidth of 3 bits. Algorithm~\ref{alg:p5-delta-detection} presents the pseudocode of a single trial of this algorithm. The output set $L_v$ of a vertex $v$ is the union of the edges detected in all trials of this algorithm.

Fix a single trial of Algorithm~\ref{alg:p5-delta-detection}. Each vertex $v$ maintains two counters for every incident edge $e=\{u,v\}$, denoted as $\mathcal P_e^v $ and $ \mathcal C_e^v$.  
Each vertex $v$ sends its counter $\mathcal C_e^v$ to its neighbors and computes the value 
\[
s_e^v = \mathcal P_e^v - \mathcal C_e^v - \mathcal C_e^u
\]
for every incident edge $e=\{u,v\}$, and adds every edge $e$ with an odd $s_e^v$ to $L_v$. Lemma~\ref{lem:p5-delta-se-symmetric} proves that both endpoints obtain the same value, which we denote by $s_e$. 
All counters are recorded modulo 2 since we only care about the parity of $s_e$. 

\textbf{The counter $\mathcal P_e^v$.} For a $5$-vertex set $H$ and an edge $e$, let $R(H,e)$ be the number of non-induced copies of $P_5$ on $H$ containing $e$. 
The counter $\mathcal P_e^v$ records, modulo $2$, the sum of $R(H,e)$ over the colorful sets of five vertices $H$. Algorithm~\ref{alg:p5-delta-count-path} computes this counter by propagating path signals. The signals count every non-induced copy of $P_5$ containing $e$ twice, once in each direction. We thus divide their total by $2$ to obtain the desired count, whose parity is recorded in $\mathcal P_e^v$. Throughout the propagation, signal values are sent modulo $4$. The final value of $\mathcal{P}_e^v$ modulo 4 is twice the number of (non-induced) paths including $e$, and thus we divide it by 2 and get the \emph{parity} of the number of non-induced paths including $e$. 

Formally, for a permutation $\pi$, vertex $v$ has $k=\pi(c_v)$ for its current color and $N_j(v)=\{u\in N(v):\pi(c_u)=j\}$ for its neighbors of current color $j$. In Algorithm~\ref{alg:p5-delta-count-path}, $F_v$ counts modulo $4$ the partial paths ending at $v$ whose colors under $\pi$ are $1,\ldots,k$ in order, while $B_v$ counts modulo $4$ those whose colors are $5,\ldots,k$ in order. The final value of $\mathcal P_e^v$ sums the counts of the partial paths ending at $v$ and $u$ for each edge $e=\{u,v\}$, modulo $4$.

\textbf{The counter $\mathcal C_e^v$.} The counter $\mathcal C_e^v$ records the \emph{local corrections}. Local corrections are counts of subgraphs which are not copies of induced $P_5$, added by $v$ to the previous counter to avoid false positives: without these, some subgraphs that contain a (non-induced) $P_5$ appear to such counting algorithm the same as a subgraph which is an induced $P_5$. Algorithm \ref{alg:p5-delta-count-corrections} computes $\mathcal C_e^v$ using the information obtained by $v$ when receiving the list of neighbors from all neighbors by processing all subgraphs that might cause false positives. This is possible since subgraphs that need to be processed by a vertex are contained in the two-hop neighborhood of this vertex.

In more detail, in Algorithm~\ref{alg:p5-delta-count-corrections}, vertex $v$ considers every connected colorful five-vertex set $H$ containing $v$ for which $\deg_{G[H]}(v)\geq 3$. From the lists of neighbors received earlier, $v$ reconstructs $G[H]$ and computes $R(H,e)$ for every incident edge $e\in E(G[H])$. If $G[H]$ is not an induced $P_5$, the eligible endpoint of $e$ with minimum identifier adds $R(H,e)$ modulo $2$ to its counter $\mathcal C_e^v$; the uniqueness of the endpoint prevents double counting.

\begin{algorithm}[H]
\DontPrintSemicolon
\SetAlgorithmName{Algorithm}{alg}{List}
$c_v\gets$ uniform random color in $\{1,2,3,4,5,\phi\}$\;
send $c_v$ to every $u\in N(v)$\;
receive $c_u$ from every $u\in N(v)$, and send the colors $\{c_u:u\in N(v)\}$ to every neighbor\;
\lIf{$c_v=\phi$}{\textbf{halt}}
$\mathcal P_e^v\gets 0$ and $\mathcal C_e^v\gets 0$ for every edge $e$ incident to $v$\;
\ForEach{permutation $\pi$ of $\{1,2,3,4,5\}$}{
  $k\gets\pi(c_v)$\;
  $N_j(v)\gets\{u\in N(v):\pi(c_u)=j\}$ for each $j\in\{1,\ldots,5\}$\;
  \FuncSty{CountPathThroughEdges}() \tcp*[l]{Compute $\mathcal{P}_e^v$}
  \FuncSty{CountCorrections}() \tcp*[l]{Compute $\mathcal{C}_e^v$}
}

$\mathcal P_e^v\gets\mathcal P_e^v/2\pmod 2$ for every edge $e$ incident to $v$\;
\lForEach{incident edge $e=\{u,v\}$}{send $\mathcal C_e^v$ to the neighbor $u$}
\lForEach{edge $e=\{u,v\}$ incident to $v$}{set $s_e^v\gets\mathcal P_e^v-\mathcal C_e^v-\mathcal C_e^u$} \label{step:p5-delta-se}
\lForEach{incident edge $e$ with an odd value of $s_e^v$}{set $L_v\gets L_v\cup\{e\}$}
\caption{Single Trial for Induced $P_5$ Edge-Membership Detection for a Vertex $v$}
\label{alg:p5-delta-detection}
\end{algorithm}

\begin{algorithm}[H]
\DontPrintSemicolon
\SetAlgorithmName{Algorithm}{alg}{List}
$F_v\gets 1$ if $k=1$, and $F_v\gets 0$ otherwise\;
$B_v\gets 1$ if $k=5$, and $B_v\gets 0$ otherwise\;
\For{$i\gets 2$ \KwTo $5$}{
  send $F_v$ to every vertex in $N_i(v)$ if $k=i-1$\;
  \lIf{$k=i$}{$F_v\gets\sum_{u\in N_{i-1}(v)}F_u \pmod 4$}
}
\For{$i\gets 4$ \KwTo $1$}{
  send $B_v$ to every vertex in $N_i(v)$ if $k=i+1$\;
  \lIf{$k=i$}{$B_v\gets\sum_{u\in N_{i+1}(v)}B_u \pmod 4$}
}
\ForEach{incident edge $e=\{v,u\}$ colored $\{j,j+1\}$, for some $j\in\{1,\ldots,4\}$}{
  \lIf{$k=j$}{$\mathcal P_e^v\gets\mathcal P_e^v+F_vB_u \pmod 4$} \label{step:p5-delta-path-lower}
  \lIf{$k=j+1$}{$\mathcal P_e^v\gets\mathcal P_e^v+F_uB_v \pmod 4$} \label{step:p5-delta-path-upper}
}

\caption{\FuncSty{CountPathThroughEdges}() for a Vertex $v$}
\label{alg:p5-delta-count-path}
\end{algorithm}

\begin{algorithm}[H]
\DontPrintSemicolon
\SetAlgorithmName{Algorithm}{alg}{List}
  \ForEach{connected colorful five-vertex set $H$ containing $v$ with $\deg_{G[H]}(v)\geq3$ }{
    \ForEach{edge $e\in E(G[H])$ incident to $v$}{
      \If{$G[H]\not\equiv P_5$ and $v$ has minimum identifier among the endpoints of $e$ with degree at least $3$ in $G[H]$}{\label{step:p5-delta-correction-owner}
        calculate $R(H,e)$ \tcp*[r]{$v$ knows the induced graph $G[H]$}
        $\mathcal C_e^v\gets\mathcal C_e^v+R(H,e) \pmod 2$\;
      }
    }
  }
\caption{\FuncSty{CountCorrections}() for a Vertex $v$}
\label{alg:p5-delta-count-corrections}
\end{algorithm}

We first claim that $s_e^v$, as computed by Algorithm~\ref{alg:p5-delta-detection}, does not depend on which endpoint $v$ of $e$ computes it, which lets us use $s_e$ without a superscript in our analysis.

\begin{lemma}\label{lem:p5-delta-se-symmetric}
For every edge $e=\{u,v\}$, it holds that $s_e^v=s_e^u$ at the end of every trial of Algorithm~\ref{alg:p5-delta-detection}.
\end{lemma}

\begin{proof}
The values $\mathcal P_e^v$ and $\mathcal P_e^u$  are increased equally by both endpoints of $e=\{u,v\}$ in Lines~\ref{step:p5-delta-path-lower} and~\ref{step:p5-delta-path-upper} of Algorithm~\ref{alg:p5-delta-count-path}, and hence we will refer to this value as $\mathcal{P}_e$ without the superscript. Moreover, the endpoints exchange $\mathcal C_e^u$ and $\mathcal C_e^v$ before computing $s_e$. Hence, the endpoints compute $s_e^v = s_e^u = \mathcal P_e-\mathcal C_e^u-\mathcal C_e^v$, and therefore obtain the same value.
\end{proof}

We thus denote by $s_e$ and $\mathcal{P}_e$ the (equal) pairs of values $s_e^v,s_e^u$ and $P_e^v,P_e^u$, respectively, accumulated on both endpoints of $e=\{u,v\}$. 

Next, we claim the following structural guarantee that enables the vertices deduce the value of the counter for local corrections.

\begin{restatable}{lemma}{pfivedeltahighdegreeendpoint}\label{lem:p5-delta-high-degree-endpoint}
Let $H$ be a graph on five vertices, let $e=\{u,v\}\in E(H)$, and suppose that $H\not\equiv P_5$. If $R(H,e) \equiv 1 \pmod 2$, then
\[
\max\{\deg_H(u),\deg_H(v)\}\geq 3.
\]
Consequently, exactly one endpoint of $e$ adds $R(H,e)$ to its local counter $\mathcal{C}_e^v$.
\end{restatable}

Let $H$ be a subset of 5 vertices and let $e$ be some edge. The contribution of $H$ to $e$, denoted by $C(H,e)$, is 1 if $H$ is a colorful induced $P_5$ with $e$ in it, and 0 otherwise. We now formalize the relation between the value $s_e$ and the contribution $C(H,e)$ defined above.
\begin{restatable}{lemma}{pfivedeltaedgesignal}\label{lem:p5-delta-edge-signal}
For each edge $e$, it holds that
\[
    s_e\equiv\sum_{H\in\binom{V}{5}}C(H,e)\pmod 2.
\]
\end{restatable}

Then, we state the following lemma that proves there are sufficiently many ``good'' colorings, and thus the vertices correctly detect an induced $P_5$ with high probability.

\begin{lemma}\label{lem:p5-delta-colorings}
If $e$ is an edge of an induced $P_5$, then there are at least $6^{n-5}$ colorings for which $s_e \equiv 1 \pmod 2$.
\end{lemma}

Let $M$ be the nonempty set of induced copies of $P_5$ containing $e$, and let $M_c$ be its subset of copies that are colorful under a coloring $c$. By Lemma~\ref{lem:p5-delta-edge-signal}, $s_e\equiv |M_c|\pmod 2$. Applying Lemma~\ref{lem:pk-colorings} with $k=5$ gives at least $6^{n-5}$ colorings for which $|M_c|$, hence $s_e$ is odd.

Finally, we prove Theorem~\ref{thm:induced-p5-delta}, given Lemmas~\ref{lem:p5-delta-high-degree-endpoint}, \ref{lem:p5-delta-edge-signal}, and~\ref{lem:p5-delta-colorings}. 

\begin{proof}[Proof of Theorem~\ref{thm:induced-p5-delta}]

Fix an edge $e$ and a single trial of Algorithm~\ref{alg:p5-delta-detection}. By Lemma~\ref{lem:p5-delta-edge-signal}, if no induced $P_5$ contains $e$, then every set of five vertices contributes 0 modulo 2 to $s_e$, in which case neither endpoint includes $e$ in its output set. Otherwise, by Lemma~\ref{lem:p5-delta-colorings}, there are at least $6^{n-5}$ colorings for which $s_e \equiv 1 \pmod 2$, and therefore both endpoints include $e$ in their output sets in a single trial of this algorithm with probability at least $6^{-5}$.

Repeating this for $O(\log n)$ independent trials of Algorithm~\ref{alg:p5-delta-detection}, the probability that a fixed edge $e$ belonging to an induced $P_5$ does not have an odd $s_e$ in any trial of this algorithm is $(1-6^{-5})^{O(\log(n))}\leq 1/n^\gamma$, for any desired constant $\gamma>0$, by choosing the constant in the number of trials of this algorithm sufficiently large. By a union bound over the edges that belong to an induced $P_5$, the probability that some such edge is omitted from the output of an  endpoint is at most $1/n^{\gamma-2}$. Setting $\gamma=\kappa+2$ shows that, with probability at least $1-1/n^\kappa$, every vertex correctly detects every incident edge in an induced $P_5$.

For the round complexity, note that sending the neighbor lists takes $O(\Delta)$ rounds. The path signal requires $O(1)$ rounds per permutation, and exchanging the counter for the local corrections requires one additional round. Therefore, each trial of Algorithm~\ref{alg:p5-delta-detection} requires $O(\Delta)$ rounds.  Each such trial requires a bandwidth of 3 bits since all values are sent modulo 2 and colors require at most 3 bits. Therefore, the vertices can simulate each round of a trial for $O(\log n)$ trials in $O(1)$ rounds by compressing the values sent over each edge in a round into $O(1)$ \congest messages. Hence the total complexity of the algorithm is $O(\Delta)$ rounds. 
Therefore, the total round complexity is $O(\Delta)$, concluding the proof.
\end{proof}

We now prove Lemmas~\ref{lem:p5-delta-high-degree-endpoint} and~\ref{lem:p5-delta-edge-signal}.

\pfivedeltahighdegreeendpoint*

\begin{figure}[p]
\centering
\begingroup
\newcommand{\pfv}[2]{\coordinate (pf#1) at #2;}
\newcommand{\pfe}[3]{%
  \draw[paper edge] (pf#1)--(pf#2)
    node[midway,auto,paper edge label] {$#3$};
}
\newcommand{\pfedges}[1]{\foreach \x/\y/\z in {#1}{\pfe{\x}{\y}{\z}}}
\newcommand{\pfwhite}[1]{\node[paper vertex] at (pf#1) {};}
\newcommand{\pfred}[1]{\node[paper representative] at (pf#1) {};}
\newcommand{\pfvertices}{%
  \pfv{0}{(0,0.90)}\pfv{1}{(0.86,0.28)}
  \pfv{2}{(0.53,-0.74)}\pfv{3}{(-0.53,-0.74)}
  \pfv{4}{(-0.86,0.28)}
}
\newcommand{\pfwhites}{\foreach \i in {0,...,4}{\pfwhite{\i}}}
\begin{tikzpicture}[x=1.38cm,y=1.38cm]
\begin{scope}[shift={(0,0)}]
  \pfvertices
  \pfedges{0/1/1,0/4/1,1/2/1,2/3/1}
  \pfwhites
\end{scope}
\begin{scope}[shift={(2.85,0)}]
  \pfvertices
  \pfedges{0/4/2,1/2/2,1/3/1,2/3/1,3/4/2}
  \pfwhites\pfred{3}
\end{scope}
\begin{scope}[shift={(5.70,0)}]
  \pfvertices
  \pfedges{0/1/2,1/3/1,1/4/1,2/3/2,2/4/2}
  \pfwhites\pfred{1}
\end{scope}
\begin{scope}[shift={(8.55,0)}]
  \pfvertices
  \pfedges{0/1/4,0/4/4,1/2/4,2/3/4,3/4/4}
  \pfwhites
\end{scope}
\begin{scope}[shift={(0,-2.35)}]
  \pfvertices
  \pfedges{0/1/1,0/2/0,0/4/1,1/2/1,2/3/1}
  \pfwhites\pfred{0}\pfred{2}
\end{scope}
\begin{scope}[shift={(2.85,-2.35)}]
  \pfvertices
  \pfedges{0/2/4,0/3/4,0/4/4,1/2/4,1/3/4,1/4/4}
  \pfwhites
\end{scope}
\begin{scope}[shift={(5.70,-2.35)}]
  \pfvertices
  \pfedges{0/1/4,0/4/2,1/4/2,2/3/4,2/4/2,3/4/2}
  \pfwhites
\end{scope}
\begin{scope}[shift={(8.55,-2.35)}]
  \pfvertices
  \pfedges{0/1/4,1/3/2,1/4/2,2/3/3,2/4/3,3/4/2}
  \pfwhites\pfred{3}\pfred{4}
\end{scope}
\begin{scope}[shift={(0,-4.70)}]
  \pfvertices
  \pfedges{0/1/2,1/2/0,1/3/1,1/4/1,2/3/2,2/4/2}
  \pfwhites\pfred{1}
\end{scope}
\begin{scope}[shift={(2.85,-4.70)}]
  \pfvertices
  \pfedges{0/1/5,0/3/2,0/4/5,1/2/6,2/3/5,3/4/5}
  \pfwhites\pfred{0}\pfred{3}
\end{scope}
\begin{scope}[shift={(5.70,-4.70)}]
  \pfvertices
  \pfedges{0/3/4,0/4/4,1/3/4,1/4/4,2/3/4,2/4/4,3/4/0}
  \pfwhites
\end{scope}
\begin{scope}[shift={(8.55,-4.70)}]
  \pfvertices
  \pfedges{0/4/6,1/2/4,1/3/4,1/4/2,2/3/4,2/4/2,3/4/2}
  \pfwhites
\end{scope}
\begin{scope}[shift={(0,-7.05)}]
  \pfvertices
  \pfedges{0/2/7,0/3/10,0/4/7,1/2/7,1/3/10,1/4/7,2/4/8}
  \pfwhites\pfred{0}\pfred{1}\pfred{2}\pfred{4}
\end{scope}
\begin{scope}[shift={(2.85,-7.05)}]
  \pfvertices
  \pfedges{0/1/6,0/4/8,1/2/6,1/3/3,1/4/3,2/3/8,3/4/6}
  \pfwhites\pfred{1}\pfred{3}\pfred{4}
\end{scope}
\begin{scope}[shift={(5.70,-7.05)}]
  \pfvertices
  \pfedges{0/1/12,0/3/8,0/4/8,1/3/8,1/4/8,2/3/12,2/4/12,3/4/4}
  \pfwhites
\end{scope}
\begin{scope}[shift={(8.55,-7.05)}]
  \pfvertices
  \pfedges{0/1/14,0/3/14,0/4/10,1/2/14,1/4/10,2/3/14,2/4/10,3/4/10}
  \pfwhites
\end{scope}
\begin{scope}[shift={(2.85,-9.40)}]
  \pfvertices
  \pfedges{0/1/18,0/3/18,0/4/18,1/2/18,1/3/12,1/4/12,2/3/18,2/4/18,3/4/12}
  \pfwhites
\end{scope}
\begin{scope}[shift={(5.70,-9.40)}]
  \pfvertices
  \foreach \x/\y in {0/1,0/2,0/3,0/4,1/2,1/3,1/4,2/3,2/4,3/4}
    {\draw[paper edge] (pf\x)--(pf\y);}
  \node[paper edge label] at (0,-1.12) {$\forall e,\ R(H,e)=24$};
  \pfwhites
\end{scope}
\end{tikzpicture}%
\endgroup
\caption{The value $R(H,e)$ for every edge $e$ in each of the $18$ five-vertex isomorphism classes $H$ that contain a non-induced copy of $P_5$. Each edge is annotated with $R(H,e)$, the number of non-induced copies of $P_5$ on $H$ that contain $e$. An endpoint of an edge with odd $R(H,e)$ is colored red if it has degree at least $3$. Only the endpoint with the minimal ID adds $R(H,e)$ to its local counter.}
\label{fig:p5-delta-r-counts}
\end{figure}
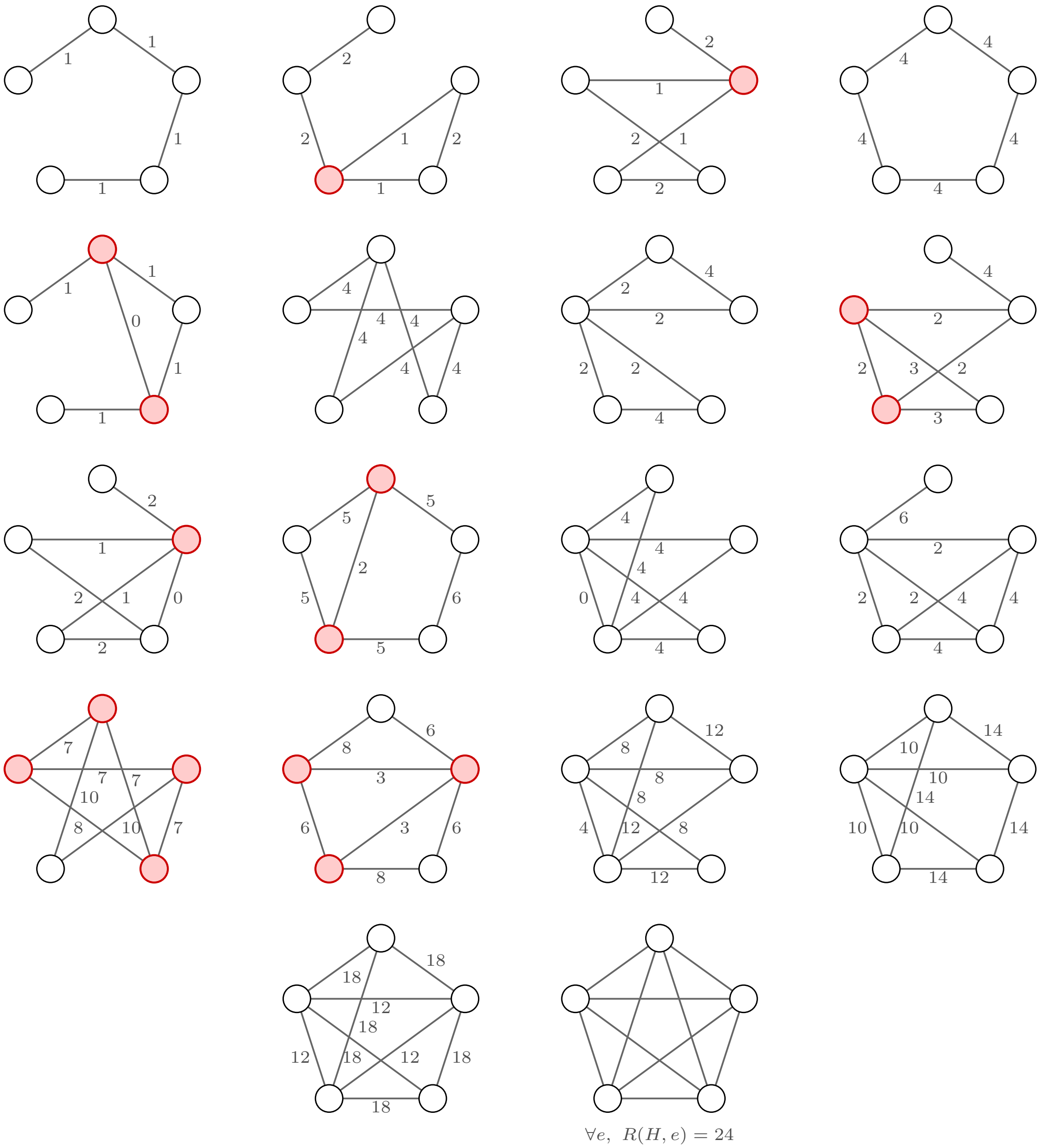

\begin{proof}
We verify the claim by exhaustive computation over all isomorphism classes of graphs on five vertices using the SageMath script in Figure~\ref{lst:p5-delta-high-degree-check}.

Figure~\ref{fig:p5-delta-r-counts} displays the values $R(H,e)$ used in this verification for every five-vertex isomorphism class that contains a non-induced copy of $P_5$.

It remains to prove that the claim allows exactly one endpoint of the edge to compute $R(H,e)$. Let $v$ be an endpoint of $e$ with $\deg_H(v)\geq 3$. This means that at most one vertex $z\in H$ is not in $N[v]$. Notice that $N(z) \cap H\subseteq N[v]$, and hence $v$ has learned the induced graph on $H$, and can therefore compute $R(H,e)$ for every incident edge $e$. Notice that by Line~\ref{step:p5-delta-correction-owner} only one such endpoint adds $R(H,e)$ to its local corrections, which is the endpoint with a degree at least 3 in $G[H]$ with the minimal ID.
\end{proof}

\begin{figure}[H]
\begin{lstlisting}[style=sagepy]
from itertools import permutations

V = range(5)
key = lambda u, v: tuple(sorted((u, v)))

def path_edges(t):
    return {key(t[i], t[i + 1]) for i in range(4)}

def R(E, e):
    paths = set()
    for t in permutations(V):
        if path_edges(t).issubset(E):
            paths.add(min(t, t[::-1]))
    return sum(1 for t in paths if e in path_edges(t))

for H in graphs.nauty_geng("5"):
    E = {key(u, v) for u, v in H.edges(labels=False)}
    if H.is_isomorphic(graphs.PathGraph(5)):
        continue
    for e in E:
        if R(E, e) % 2 == 1:
            assert max(H.degree(e[0]), H.degree(e[1])) >= 3

print("Verified the high-degree endpoint claim.")
\end{lstlisting}
\caption{The SageMath script verifying Lemma~\ref{lem:p5-delta-high-degree-endpoint} over all five-vertex graphs.}
\label{lst:p5-delta-high-degree-check}
\end{figure}

We now show that our algorithm correctly counts the number of (non-induced) copies of $P_5$. 
\pfivedeltaedgesignal*

\begin{proof}
Let $e\in E$ be an edge, let $\pi$ be a permutation of $\{1,\ldots,5\}$, and let $\mathcal P_e^\pi$ be the value that an endpoint of $e$ adds to $\mathcal P_e$ during the iteration of Algorithm~\ref{alg:p5-delta-detection} that corresponds to $\pi$. We first prove by induction on $i$ that, for every vertex $v$ whose current color $\pi(c_v)$ is $i$, $F_v$ is equal modulo $4$ to the number of colorful ordered paths following colors $1,\ldots,i$ and ending at $v$. For $i=1$, the only such path is the single vertex $v$, and the algorithm sets $F_v=1$. Suppose that the claim holds for $i-1$. Every ordered path following colors $1,\ldots,i$ and ending at $v$ consists of an ordered path following colors $1,\ldots,i-1$ and ending at some $u\in N_{i-1}(v)$, followed by the edge $\{u,v\}$. Conversely, every such path ending at $u$ extends uniquely to a path ending at $v$. Therefore, the assignment
\[
F_v\gets\sum_{u\in N_{i-1}(v)}F_u\pmod 4
\]
counts all such paths modulo $4$, proving the claim.

Similarly, for every vertex $u$ whose current color $\pi(c_u)$ is $i$, $B_u$ is equal, modulo 4, to the number of colorful ordered paths following colors $5,4,\ldots,i$ and ending at $u$.

Consequently, for an edge $e=\{u,v\}$ such that $u$ has color $i$ and $v$ has color $i+1$, the product $F_uB_v$ is equal, modulo $4$, to the number of colorful paths ordered according to $\pi$ that pass through $e$. Summing over all $5!$ permutations counts both directions of every non-induced copy of $P_5$. After dividing $\mathcal P_e$ by $2$, it therefore holds that
\[
\mathcal P_e\equiv\sum_{\substack{H\in\binom{V}{5}\\H\text{ colorful}}}R(H,e)\pmod 2.
\]
Therefore,
\begin{align*}
    s_e
    &\equiv \mathcal P_e-\mathcal C_e^u-\mathcal C_e^v
    \equiv \sum_{\substack{H\in\binom{V}{5}\\H\text{ colorful}}}R(H,e)
    -\sum_{\substack{H\in\binom{V}{5}\\H\text{ colorful}\\G[H]\not\equiv P_5}}R(H,e) \\
    &\equiv \sum_{\substack{H\in\binom{V}{5}\\H\text{ colorful}\\G[H]\equiv P_5}}R(H,e)
    \equiv \sum_{H\in\binom{V}{5}}C(H,e)\pmod 2.
\end{align*}

The second equality holds by Lemma \ref{lem:p5-delta-high-degree-endpoint}. The last equality holds because when $G[H]$ is an induced $P_5$ containing $e$, there exists one non-induced copy of $P_5$ on $H$, and it contains $e$. Thus, for such $H$ it holds that $R(H,e)=C(H,e)=1$.
\end{proof}

\subsection{Induced $P_5$ Detection in Induced-$C_4$-Free Graphs}
\label{sec:induced-p5-c4-free}
We now turn to presenting our \congest algorithm for end-edge membership detection for induced $P_5$ in induced-$C_4$-free graphs. We prove the following result.
\inducedpfivedetection*

\begin{figure}[htbp]
    \centering
    \begin{tikzpicture}[
        scale=1.35,
        base/.style={paper path},
        extra/.style={paper extra edge},
        last/.style={paper highlighted edge},
        vertex/.style={paper vertex},
        panel/.style={paper panel}
    ]

    \begin{scope}[shift={(0,0)}]
    \coordinate (p1) at (-0.90,0.00);
    \coordinate (p2) at (-0.45,0.42);
    \coordinate (p3) at (0.00,0.00);
    \coordinate (p4) at (0.45,0.42);
    \coordinate (p5) at (0.90,0.00);
    \draw[base] (p1)--(p2)--(p3)--(p4)--(p5);
    \draw[last] (p4)--(p5);
    \node[vertex] at (p1) {$1$};
    \node[vertex] at (p2) {$2$};
    \node[vertex] at (p3) {$3$};
    \node[vertex] at (p4) {$4$};
    \node[vertex] at (p5) {$5$};
    \node[panel] at (0,-1.05) {path $\mathcal{P}$};
    \end{scope}

    \begin{scope}[shift={(2.85,0)}]
    \coordinate (b1) at (-1.05,-0.28);
    \coordinate (b2) at (-0.50,0.00);
    \coordinate (b3) at (0.00,0.58);
    \coordinate (b4) at (0.50,0.00);
    \coordinate (b5) at (1.05,-0.28);
    \draw[base] (b1)--(b2)--(b3)--(b4)--(b5);
    \draw[extra] (b2)--(b4);
    \draw[last] (b4)--(b5);
    \node[vertex] at (b1) {$1$};
    \node[vertex] at (b2) {$2$};
    \node[vertex] at (b3) {$3$};
    \node[vertex] at (b4) {$4$};
    \node[vertex] at (b5) {$5$};
    \node[panel] at (0,-1.05) {bull $\mathcal{B}$};
    \end{scope}

    \begin{scope}[shift={(5.70,0)}]
    \coordinate (d1) at (-1.12,0.00);
    \coordinate (d2) at (-0.58,0.00);
    \coordinate (d3) at (0.00,0.58);
    \coordinate (d4) at (0.00,-0.58);
    \coordinate (d5) at (0.72,0.00);
    \draw[base] (d1)--(d2)--(d3)--(d4)--(d5);
    \draw[extra] (d2)--(d4);
    \draw[extra] (d3)--(d5);
    \draw[last] (d4)--(d5);
    \node[vertex] at (d1) {$1$};
    \node[vertex] at (d2) {$2$};
    \node[vertex] at (d3) {$3$};
    \node[vertex] at (d4) {$4$};
    \node[vertex] at (d5) {$5$};
    \node[panel] at (0,-1.05) {diamond-pendant $\mathcal{D}$};
    \end{scope}

    \begin{scope}[shift={(9.05,0)}]
    \coordinate (g1) at (0.00,-0.10);
    \coordinate (g2) at (-0.82,-0.52);
    \coordinate (g3) at (-0.44,0.54);
    \coordinate (g4) at (0.44,0.54);
    \coordinate (g5) at (0.82,-0.52);
    \draw[base] (g1)--(g2)--(g3)--(g4)--(g5);
    \draw[extra] (g1)--(g3);
    \draw[extra] (g1)--(g4);
    \draw[extra] (g1)--(g5);
    \draw[last] (g4)--(g5);
    \node[vertex] at (g1) {$1$};
    \node[vertex] at (g2) {$2$};
    \node[vertex] at (g3) {$3$};
    \node[vertex] at (g4) {$4$};
    \node[vertex] at (g5) {$5$};
    \node[panel] at (0,-1.05) {gem $\mathcal{G}$};
    \end{scope}

    \end{tikzpicture}
    \caption{The four subgraphs corresponding to local corrections.  All
    subgraphs share the ordered path $1-2-3-4-5$, drawn with \emph{bold} edges; the additional edges of each
    subgraph are drawn \emph{dashed}.  The
    edge between colors $4$ and $5$ is highlighted in red.}
    \label{fig:p5-subgraphs-copy}
\end{figure}
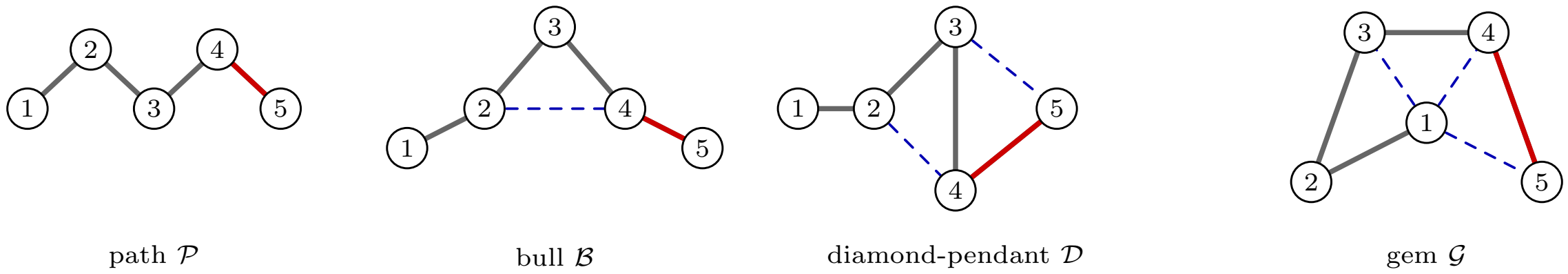

We use our counting-based approach to detect the existence of an induced $P_5$ as follows. The vertices run $O(\log n)$ \emph{trials} of Algorithm~\ref{alg:restricted-p5-detection}. In each trial of this algorithm, every vertex $v$ draws a random color $c_v$, and iterates over all $5!$ permutations of $\{1,\ldots,5\}$. For each permutation $\pi$, each vertex sets its current color as $\pi(c_v)$. Algorithm~\ref{alg:p5-count-path} counts the number of colorful copies of the (non-induced) $P_5$ ending at each incident edge $e$ using signal propagation, and adds it to a local counter $\mathcal{P}_e^v$. In Algorithms~\ref{alg:p5-count-bull}--\ref{alg:p5-count-gem}, each vertex counts and adds the local corrections $\mathcal{B}_e^v$, $\mathcal{D}_e^v$, and $\mathcal{G}_e^v$, which count instances of the bull, diamond-pendant, and gem subgraphs, respectively (see Figure~\ref{fig:p5-subgraphs-copy}). Finally, vertices compute the value

\[
   s_e^v = \mathcal{P}_e^v - \mathcal{B}_e^v - \mathcal{D}_e^v - \mathcal{G}_e^v.
\]

Each vertex $v$ initializes an output set $L_v\gets\emptyset$, adds every incident edge $e$ with $s_e^v\not\equiv 0\pmod 2$ in some trial of Algorithm~\ref{alg:restricted-p5-detection} to $L_v$, and halts after all trials of this algorithm.

Computing the local corrections is done by counting, for each edge, the number of triangles including it with edge multiplicities, using a slightly modified version of the triangle counting algorithm of \cite{CPSZ21}, which we now explain. The adjustment is required for two reasons: 
\begin{enumerate}
  \item We need the endpoints of each edge to learn the exact number of triangles on this edge.
  \item We need to count with multiplicities that represent the number of triangles with specific colors that include each edge. 
\end{enumerate}

\paragraph{The triangle-counting algorithm.}
The local corrections are computed by counting colored triangles.
We use the following form of triangle counting.

Assume each edge $e$ is associated with a multiplicity $\mu(e)$, which is known to its endpoints, and each vertex is colored by a coloring function $c: V \to \{1 ,\ldots, 5\} $, such that every vertex is only aware of its own color locally. For a triangle $T=\{u,v,w\}$, define its \emph{triangle weight} as
\[
    W_T \;=\; \prod_{f\in E(T)} \mu(f).
\]

For a triplet of distinct colors $C=\{\alpha,\beta,\gamma\}$ we say that a triangle $T$ is \emph{$C$-colored} if its vertices are distinctly colored by $c$ with colors of $C$.
For an edge $e\in E$, and a triplet of distinct colors $C$, define the \emph{weighted triangle count} as
\[
    \WT(e,C)
    \;=\;
    \sum_{\substack{T:\ e\in E(T)\\ T\text{ is }C\text{-colored}}} W_T.
\]

If $\mu(e)=1$ for every edge $e\in E$, then $W_T=1$ for every triangle $T$, and the weighted triangle count $\WT(e,C)$ is simply the number of triangles containing $e$ that are $C$-colored.
We prove the following lemma which will be used to count triangles with multiplicities. The proof relies on an adaptation of the triangle listing algorithm of \cite{CPSZ21}. 
\begin{restatable}{lemma}{triangleedgecounting}\label{lem:triangle-to-edge-counting}
For a given color triplet $C$ and an integer multiplicity $\mu(e)\in\poly(n)$, there exists an $\tilde O(n^{1/3})$-round \congest algorithm such that upon its termination, with high probability, the endpoints of every edge $e$ learn $\WT(e,C)$.
\end{restatable}

We use the triangle counting algorithm of Lemma~\ref{lem:triangle-to-edge-counting} as a black box: after invoking it with a color triplet $C$ and multiplicities $\mu$, each vertex knows $\WT(e,C)$ for every incident edge $e$.  We write $\TC(C,\mu)$ for a single call to this algorithm with the color triplet $C$ and the current edge multiplicities $\mu$. To compute $\mathcal{B}_e^v$, $\mathcal{D}_e^v$, and $\mathcal{G}_e^v$, vertices compute the number of triangles on each edge by invoking $\TC(C,\mu)$ multiple times. To illustrate this process, consider the diamond-pendant pattern $\mathcal D$. For each $\{2,3\}$-colored edge, its multiplicity is set to the number of possible color-$1$ pendant vertices adjacent to its color-$2$ endpoint. The first triangle-counting call extends these multiplicities to the triangle with vertex colors $2,3,4$. These counts then become the multiplicities of the $\{3,4\}$-colored edges, and a second call extends each partial instance by a color-$5$ vertex adjacent to both endpoints. The resulting count on each $\{4,5\}$-colored edge is therefore the number of colorful diamond-pendant instances containing that edge. The bull and gem corrections are computed analogously, as detailed in Algorithms~\ref{alg:p5-count-bull}--\ref{alg:p5-count-gem}.

\setlength{\floatsep}{4pt plus 2pt minus 2pt}
\setlength{\textfloatsep}{6pt plus 2pt minus 2pt}
\setlength{\intextsep}{4pt plus 2pt minus 2pt}

\begin{algorithm}[H]
\DontPrintSemicolon
\SetAlgorithmName{Algorithm}{alg}{List}
$L_v\gets\emptyset$\;
\For{$t\gets 1$ \KwTo $O(\log n)$}{ \label{step:p5-trial-loop}
  $c_v \gets$ a uniform random color in $\{1,2,3,4,5,\phi\}$\;
  send $c_v$ to every $u\in N(v)$\;

  \If{$c_v\neq\phi$}{
    $\mathcal{P}_e^v \gets 0$, $\mathcal{B}_e^v \gets 0$, $\mathcal{D}_e^v \gets 0$, and $\mathcal{G}_e^v \gets 0$ for every edge $e$ incident to $v$\;
    \ForEach{permutation $\pi$ of $\{1,2,3,4,5\}$}{ \label{step:p5-perm-loop}
      $k\gets\pi(c_v)$\;
      $N_j(v)\gets\{u\in N(v):\pi(c_u)=j\}$ for each $j\in\{1,\dots,5\}$\;
      \FuncSty{CountPath}() \tcp*[l]{Compute $\mathcal{P}$}
      \FuncSty{CountBull}() \tcp*[l]{Compute $\mathcal{B}$}
      \FuncSty{CountDiamond}() \tcp*[l]{Compute $\mathcal{D}$}
      \FuncSty{CountGem}() \tcp*[l]{Compute $\mathcal{G}$}
    }
    \ForEach{edge $e$ incident to $v$}{$s_e^v\gets \mathcal{P}_e^v-\mathcal{B}_e^v-\mathcal{D}_e^v-\mathcal{G}_e^v$\;} \label{step:p5-se}
    \ForEach{incident edge $e$ with $s_e^v\neq 0\pmod 2$}{$L_v\gets L_v\cup\{e\}$\;}
  }
}
\textbf{halt}\;
\caption{Induced $P_5$ End-Edge Membership Detection in Induced-$C_4$-Free Graphs for a Vertex $v$}
\label{alg:restricted-p5-detection}
\end{algorithm}

\begin{algorithm}[H]
\DontPrintSemicolon
\SetAlgorithmName{Algorithm}{alg}{List}
  $\mathrm{in}(u) \gets 0$ for every $u\in N(v)$\;
  \For{$i\gets 1$ \KwTo $5$}{
    \If(\tcp*[h]{$k=\pi(c_v)$ is the current color of $v$}){$k=i$}{
      \eIf{$i=1$}{send the value $1$ to every $w\in N_2(v)$\; \label{step:p5-path-init}}{
        $s_v\gets\sum_{u\in N_{i-1}(v)}\mathrm{in}(u)$\; \label{step:p5-path-sum}
        \lIf{$i\le 4$}{send $s_v$ to every $w\in N_{i+1}(v)$} \label{step:p5-path-fwd}
      }
      \lIf{$i=4$}{$\mathcal{P}_{\{v,u\}}^v\gets \mathcal{P}_{\{v,u\}}^v+s_v$ for every $u\in N_5(v)$} \label{step:p5-path-i4}
      \lIf{$i=5$}{$\mathcal{P}_{\{v,u\}}^v\gets \mathcal{P}_{\{v,u\}}^v+\mathrm{in}(u)$ for every $u\in N_4(v)$} \label{step:p5-path-i5}
    }
    \If{$k=i+1$}{$\mathrm{in}(u)\gets$ value received from $u$, for every $u\in N_i(v)$\;}
  }
\caption{\FuncSty{CountPath}() for a Vertex $v$}
\label{alg:p5-count-path}
\end{algorithm}

\begin{algorithm}[H]
\DontPrintSemicolon
\SetAlgorithmName{Algorithm}{alg}{List}
  set $\mu(e)\gets 1$ for every incident edge $e$\;
  compute $\WT(e,\{2,3,4\})$ for every incident edge $e$ using $\TC(\{2,3,4\},\mu)$ \; \label{step:p5-bull-tc234}
  \If(\tcp*[h]{$k=\pi(c_v)$ is the current color of $v$}){$k=2$}{send $|N_1(v)|$ to every $u\in N_4(v)$\;}
  \If{$k=4$}{
    $\delta_v\gets\sum_{x\in N_2(v)} |N_1(x)|\cdot\WT(\{x,v\},\{2,3,4\})$\;
    $\mathcal{B}_{\{v,u\}}^v\gets \mathcal{B}_{\{v,u\}}^v+\delta_v$ for every $u\in N_5(v)$\;
    send $\delta_v$ to every $u\in N_5(v)$\; \label{step:p5-bull-update}
  }
  \If{$k=5$}{$\mathcal{B}_{\{v,u\}}^v\gets \mathcal{B}_{\{v,u\}}^v+\delta_u$ for every $u\in N_4(v)$\;}
\caption{\FuncSty{CountBull}() for a Vertex $v$}
\label{alg:p5-count-bull}
\end{algorithm}

\begin{algorithm}[H]
\DontPrintSemicolon
\SetAlgorithmName{Algorithm}{alg}{List}
  set $\mu(e)\gets 1$ for every incident edge $e$\;
  \If(\tcp*[h]{$k=\pi(c_v)$ is the current color of $v$}){$k=2$}{send $|N_1(v)|$ to every $u\in N_3(v)$\;}
  \If{$k\in\{2,3\}$}{$\mu(e)\gets|N_1(x)|$ for every incident edge $e$ colored $\{2,3\}$, where $x$ is its color-$2$ endpoint\;} \label{step:p5-diam-mu23}
  run $\TC(\{2,3,4\},\mu)$\; \label{step:p5-diam-tc234}
  \lIf{$k\in\{3,4\}$}{$\mu(e)\gets\WT(e,\{2,3,4\})$ for every incident edge $e$ colored $\{3,4\}$} \label{step:p5-diam-mu34}
  run $\TC(\{3,4,5\},\mu)$\; \label{step:p5-diam-tc345}
  \lIf{$k\in\{4,5\}$}{$\mathcal{D}_e^v\gets \mathcal{D}_e^v+\WT(e,\{3,4,5\})$ for every incident edge $e$ colored $\{4,5\}$} \label{step:p5-diam-update}
\caption{\FuncSty{CountDiamond}() for a Vertex $v$}
\label{alg:p5-count-diamond}
\end{algorithm}

\begin{algorithm}[H]
\DontPrintSemicolon
\SetAlgorithmName{Algorithm}{alg}{List}
  set $\mu(e)\gets 1$ for every incident edge $e$\;
  run $\TC(\{1,2,3\},\mu)$\; \label{step:p5-gem-tc123}
  \lIf{$k\in\{1,3\}$}{$\mu(e)\gets\WT(e,\{1,2,3\})$ for every incident edge $e$ colored $\{1,3\}$} \label{step:p5-gem-mu13}
  run $\TC(\{1,3,4\},\mu)$\; \label{step:p5-gem-tc134}
  \lIf{$k\in\{1,4\}$}{$\mu(e)\gets\WT(e,\{1,3,4\})$ for every incident edge $e$ colored $\{1,4\}$} \label{step:p5-gem-mu14}
  run $\TC(\{1,4,5\},\mu)$\; \label{step:p5-gem-tc145}
  \lIf{$k\in\{4,5\}$}{$\mathcal{G}_e^v\gets \mathcal{G}_e^v+\WT(e,\{1,4,5\})$ for every incident edge $e$ colored $\{4,5\}$} \label{step:p5-gem-update}
\caption{\FuncSty{CountGem}() for a Vertex $v$}
\label{alg:p5-count-gem}
\end{algorithm}

To analyze our algorithm, we need additional definitions. Fix a single iteration of the loop in Line~\ref{step:p5-trial-loop} in Algorithm~\ref{alg:restricted-p5-detection}, which we refer to as a trial, and let $c$ be its corresponding random coloring. Define the subgraphs
\[
\begin{aligned}
E_\mathcal{P} &=\{\{1,2\},\{2,3\},\{3,4\},\{4,5\}\}, \\
E_\mathcal{B} &= E_\mathcal{P}\cup\{\{2,4\}\},\\
E_\mathcal{D} &= E_\mathcal{P}\cup\{\{2,4\},\{3,5\}\}, \text{ and }\\
E_\mathcal{G} &= E_\mathcal{P}\cup\{\{1,3\},\{1,4\},\{1,5\}\}.
\end{aligned}
\]
These are the path subgraph $\mathcal{P}$, the bull subgraph, the diamond-pendant subgraph, and the gem subgraph, respectively. Let $H=\left\{v_1,\ldots,v_5\right\}$ be a colorful subset of 5 vertices such that $c_{v_i}=i$ for every $i\in\{1,\ldots,5\}$, and let $\tau$ be some permutation of $\{1,\ldots,5\}$.  We say that a subgraph $E'$ is present in $H$ under $\tau$ if for every edge $\left\{\tau(i),\tau(j)\right\}\in E'$, the edge $\left\{v_i,v_j\right\}$ exists in $G$.  Let $\chi_\mathcal{P}(H,\tau)$, $\chi_\mathcal{B}(H,\tau)$, $\chi_\mathcal{D}(H,\tau)$, $\chi_\mathcal{G}(H,\tau)$ be 1 if $E_\mathcal{P},E_\mathcal{B},E_\mathcal{D},$ or $E_\mathcal{G}$ are present in $H$ under $\tau$, respectively, and be $0$ otherwise.  Let $\varphi(H,\tau)=\{v_i,v_j\}$, where $i,j\in\{1,\ldots,5\}$ satisfy $\left\{\tau(i),\tau(j)\right\}=\{4,5\}$; that is, $\varphi(H,\tau)$ is the edge of $H$ colored $\{4,5\}$ under the permutation $\tau$.  Define the \emph{contribution of $H$ to an edge $e$} as
\[
C(H,e)=
\sum_{\tau:\ \varphi(H,\tau)=e}
\left(
\chi_\mathcal{P}(H,\tau)-\chi_\mathcal{B}(H,\tau)-\chi_\mathcal{D}(H,\tau)-\chi_\mathcal{G}(H,\tau)
\right).
\]
We now claim that $s_e^v$, as computed by Algorithm~\ref{alg:restricted-p5-detection}, does not depend on which endpoint $v$ of $e$ computes it, which lets us use $s_e$ without a superscript in our analysis.

\begin{lemma}\label{lem:p5-se-symmetric}
For every edge $e=\{u,v\}$, it holds that $s_e^v=s_e^u$ at the end of every trial of Algorithm~\ref{alg:restricted-p5-detection}.
\end{lemma}

\begin{proof}
Since it holds that $s_e^w=\mathcal{P}_e^w-\mathcal{B}_e^w-\mathcal{D}_e^w-\mathcal{G}_e^w$ for $w\in\{u,v\}$, it suffices to show that, in every permutation, both endpoints of $e$ add the same increment to $\mathcal{P}_e^u$ and $\mathcal{P}_e^v$, to $\mathcal{B}_e^u$ and $\mathcal{B}_e^v$, to $\mathcal{D}_e^u$ and $\mathcal{D}_e^v$, and to $\mathcal{G}_e^u$ and $\mathcal{G}_e^v$, respectively. For $\mathcal{P}_e^w$, the increment is computed by one endpoint and sent to the other, which adds the same received value (Lines~\ref{step:p5-path-i4}--\ref{step:p5-path-i5} of Algorithm~\ref{alg:p5-count-path}); the same holds for $\mathcal{B}_e^w$ via the symmetric pair of cases $k=4,5$ in Algorithm~\ref{alg:p5-count-bull}. For $\mathcal{D}_e^w$ and $\mathcal{G}_e^w$, both endpoints add $\WT(e,C)$ for the relevant color triplet $C$ (Line~\ref{step:p5-diam-update} of Algorithm~\ref{alg:p5-count-diamond} and Line~\ref{step:p5-gem-update} of Algorithm~\ref{alg:p5-count-gem}), and by Lemma~\ref{lem:triangle-to-edge-counting} both endpoints of $e$ learn the same value $\WT(e,C)$. Hence $\mathcal{P}_e^w,\mathcal{B}_e^w,\mathcal{D}_e^w,\mathcal{G}_e^w$ agree at $u$ and $v$ after every permutation, and therefore $s_e^w$ does as well.
\end{proof}

Henceforth, we denote by $s_e$ the same value accumulated on both endpoints of $e$. We now state the three lemmas needed for the proof of Theorem~\ref{thm:induced-p5-signal}. The following lemma connects $s_e$ and $C(H,e)$ for every subset $H$ of 5 vertices, essentially showing that the vertices indeed correctly count the required local corrections. 

\begin{restatable}{lemma}{pfiveedgesignalcontribution}\label{lem:p5-edge-signal-contribution}
For each edge $e$, it holds that
\[
    s_e = \sum_{H\in\binom{V}{5}} C(H,e) .
\]
\end{restatable}

Next, we claim the following correctness guarantee that ensures no subgraph that is not an induced $P_5$ can cause false positives. 

\begin{restatable}{lemma}{pfivesubgraphparity}\label{lem:p5-subgraph-parity}
Let $H\subseteq V$ be a colorful set of five vertices that does not include an induced $C_4$.  If $G[H]$ is an induced $P_5$, then for each end edge $e\in E(G[H])$ it holds that $C(H,e)\equiv 1\text{ modulo }2$.  If $G[H]$ is not an induced $P_5$, then it holds that $C(H,e)\equiv 0\text{ modulo }2$ for every $e\in E(G[H])$.
\end{restatable}

We then claim that the random coloring is ``good'' with high probability, as follows.

\begin{lemma}\label{lem:p5-colorings}
If $e$ is an end edge of an induced $P_5$, then there are at least $6^{n-5}$ colorings for which $s_e$ is odd.
\end{lemma}

Let $M$ be the nonempty set of induced copies of $P_5$ for which $e$ is an end edge, and let $M_c$ be its colorful members under a coloring $c$. Lemma~\ref{lem:p5-edge-signal-contribution} expresses $s_e$ as the sum of the contributions $C(H,e)$, while Lemma~\ref{lem:p5-subgraph-parity} implies that, modulo $2$, only the members of $M_c$ contribute $1$. Hence, $s_e\equiv |M_c|\pmod 2$. Applying Lemma~\ref{lem:pk-colorings} with $k=5$ proves the claim. 

Before proving Lemmas~\ref{lem:p5-edge-signal-contribution} and \ref{lem:p5-subgraph-parity}, we prove Theorem~\ref{thm:induced-p5-signal}, given Lemmas~\ref{lem:p5-edge-signal-contribution}, \ref{lem:p5-subgraph-parity}, and~\ref{lem:p5-colorings}. 

\begin{proof}[Proof of Theorem~\ref{thm:induced-p5-signal}]
The vertices run $O(\log n)$ independent trials of Algorithm~\ref{alg:restricted-p5-detection}. In each trial of this algorithm, every vertex samples a color from $\{1,2,3,4,5,\phi\}$, iterates over the $5!$ permutations of the non-$\phi$ colors, computes the number of non-induced $P_5$ that ends with each edge signal on each edge and the three local corrections per edge, and adds each edge $e$ to $L_v$ if it has an odd $s_e$.

By Lemma~\ref{lem:p5-edge-signal-contribution} and Lemma~\ref{lem:p5-subgraph-parity}, if no induced $P_5$ exists, then every set of five vertices contributes an even number to the value $s_e$ of every edge $e$, and thus every vertex halts with an empty $L_v$. Otherwise, let $e$ be an end edge of an induced $P_5$. By Lemma~\ref{lem:p5-colorings}, there are at least $6^{n-5}$ colorings for which $s_e$ is odd, and therefore both endpoints $u,v$ of $e=\{u,v\}$ add it to $L_u, L_v$, respectively, in a single trial of Algorithm~\ref{alg:restricted-p5-detection} with probability at least $6^{-5}$.

Repeating $O(\log n)$ independent trials of Algorithm~\ref{alg:restricted-p5-detection}, the probability that an edge $e$ does not have an odd $s_e$ in any trial of this algorithm is $(1-6^{-5})^{O(\log(n))} \leq 1/n^\gamma$, for any desired constant $\gamma>0$, by choosing the constant in the number of trials of this algorithm sufficiently large. By a union bound, the probability that any edge does not have an odd $s_e$ in any trial of this algorithm is at most $\frac{1}{n^{\gamma-2}}$. By setting $\gamma = \kappa + 2$, we get that the probability that every end edge is detected by its endpoints is at least $1-\frac{1}{n^\kappa}$ for any constant $\kappa$.

For the round complexity, the path signal uses $O(1)$ rounds per permutation.  Each local correction uses a constant number of calls to the triangle counting algorithm of Lemma~\ref{lem:triangle-to-edge-counting} and a constant number of additional local forwarding rounds.  There are $5!$ permutations and $O(\log n)$ trials of Algorithm~\ref{alg:restricted-p5-detection}, and therefore the total round complexity is $\tilde O(n^{1/3})$, concluding the proof.
\end{proof}

We now prove Lemmas~\ref{lem:p5-edge-signal-contribution} and~\ref{lem:p5-subgraph-parity}. 

\pfiveedgesignalcontribution*

\begin{proof}
Let $e\in E$ be an edge, let $\tau$ be a permutation of $\{1,\ldots,5\}$, and let $\mathcal{P}_e^\tau,\mathcal{B}_e^\tau,\mathcal{D}_e^\tau,\mathcal{G}_e^\tau$ be the values that an endpoint of $e$ adds to $\mathcal{P}_e,\mathcal{B}_e,\mathcal{D}_e,\mathcal{G}_e$ during the iteration $\pi=\tau$ of Algorithm~\ref{alg:restricted-p5-detection}, such that
$s_e=\sum_{\tau}(\mathcal{P}_e^\tau-\mathcal{B}_e^\tau-\mathcal{D}_e^\tau-\mathcal{G}_e^\tau)$, summed over all permutations $\tau$ of $\{1,\ldots,5\}$.
Without loss of generality, assume $\tau(i)=i$ for
every $i\in\{1,\ldots,5\}$. We show that
\begin{equation}\label{eq:p5-perm-combined}
    \mathcal{P}_e^\tau-\mathcal{B}_e^\tau-\mathcal{D}_e^\tau-\mathcal{G}_e^\tau
    =\sum_{H\in\binom{V}{5}}\mathbf 1[e=\varphi(H,\tau)]\bigl(\chi_\mathcal{P}(H,\tau)-\chi_\mathcal{B}(H,\tau)-\chi_\mathcal{D}(H,\tau)-\chi_\mathcal{G}(H,\tau)\bigr),
\end{equation}
which, summed over $\tau$, gives $s_e=\sum_H C(H,e)$.

\medskip
\noindent\textbf{The path signal (Algorithm~\ref{alg:p5-count-path}).}
For $i=1$, in Line~\ref{step:p5-path-init}, every vertex $v$
with $c_v=1$ sends the value $1$ to each of its neighbors $u$ with
$c_u=2$. This establishes the base case, since the only such path is the single vertex $v$ itself. For $i\in\{2,3\}$, in Line~\ref{step:p5-path-sum}, every
vertex $v$ with $c_v=i$ sets $s_v$ to the sum of the values it just
received from its neighbors $u$ with $c_u=i-1$, and
in Line~\ref{step:p5-path-fwd}, it forwards $s_v$ to every neighbor $u$ with
$c_u=i+1$.  By induction on $i$, the value received by a vertex
$v_i$ with $c_{v_i}=i$ from a neighbor $v_{i-1}$ with
$c_{v_{i-1}}=i-1$ is the number of colorful ordered paths
$v_1v_2\cdots v_{i}$ with $c_{v_j}=j$ for every $j \le i$.
Consequently, at $i=4$ the quantity $s_v$ computed in
Line~\ref{step:p5-path-sum} for a vertex $v_4$ with $c_{v_4}=4$
equals the number of colorful ordered paths $v_1v_2v_3v_4$ with
$c_{v_j}=j$, and in Line~\ref{step:p5-path-i4}, $v_4$ adds this value to
$\mathcal{P}_{\{v_4,u\}}$ for every neighbor $u$ with $c_u=5$; symmetrically,
in Line~\ref{step:p5-path-i5}, the endpoint with $c_{v_5}=5$ adds
the value received from a neighbor with $c_{v_4}=4$, which by the
same induction equals the number of colorful ordered paths $v_1v_2v_3v_4$
ending at that neighbor.
Hence, for every edge $e=\{v_4,v_5\}$ with $c_{v_4}=4$ and
$c_{v_5}=5$, both endpoints add one unit for every colorful
set of vertices $H=\{v_1,\ldots,v_5\}$ with $c_{v_j}=j$ for every
$j\in\{1,\ldots,5\}$, $\{v_4,v_5\}=e$, which is $\chi_\mathcal{P}(H,\tau)=1$.
Thus
\begin{equation}\label{eq:p5-identity-path}
    \mathcal{P}_e^\tau=\sum_{H'\in\binom{V}{5}}
    \mathbf 1\!\left[e=\varphi(H',\tau)\right]\cdot\chi_\mathcal{P}(H',\tau).
\end{equation}

\smallskip\noindent\textbf{Gem correction (Algorithm~\ref{alg:p5-count-gem}).}
For $j\in\{1,\ldots,5\}$, write $V_j:=\{u\in V:c_u=j\}$ for the set of vertices colored $j$.
In Line~\ref{step:p5-gem-mu13}, $v_1$ and $v_3$ set, for
edge $e_{13}=\{v_1,v_3\}\in V_1\times V_3$,
\[
    \mu(e_{13})=\WT(e_{13},\{1,2,3\})
    =\bigl|\{v_2\in V_2:\{v_1,v_2\},\{v_3,v_2\}\in E\}\bigr|.
\]
Then, in Line~\ref{step:p5-gem-mu14}, $v_1$ and $v_4$ set, for edge $e_{14}=\{v_1,v_4\}\in V_1\times V_4$,
\begin{align*}
    \mu(e_{14}) &= \WT(e_{14},\{1,3,4\})\\
    &= \sum_{\substack{v_3\in V_3\\ \{v_1,v_3\}\in E\\ \{v_3,v_4\}\in E}}\mu(\{v_1,v_3\})\\
    &= \bigl|\{(v_3,v_2)\in V_3\times V_2:\{v_1,v_3\},\{v_3,v_4\},\{v_1,v_2\},\{v_3,v_2\}\in E\}\bigr|.
\end{align*}
Finally, in Line~\ref{step:p5-gem-update}, $v_4$ and $v_5$ set, for edge $e=\{v_4,v_5\} \in V_4\times V_5$,
\begin{align}
    \mathcal{G}_e^\tau&=\WT(e,\{1,4,5\})
    =\sum_{\substack{v_1\in V_1\\\{v_1,v_4\}\in E\\\{v_1,v_5\}\in E}}\mu(\{v_1,v_4\}) \nonumber\\
    &=\sum_{\substack{v_1\in V_1\\v_3\in V_3\\v_2\in V_2}}\mathbf 1\bigl[\{v_1,v_4\},\{v_1,v_5\},\{v_1,v_3\},\{v_3,v_4\},\{v_1,v_2\},\{v_3,v_2\}\in E\bigr] \nonumber\\
    &= \sum_{H'\in\binom{V}{5}}\mathbf 1[e=\varphi(H',\tau)]\chi_\mathcal{G}(H',\tau).
    \label{eq:p5-identity-gem}
\end{align}
The bull and diamond-pendant correction proofs are analogous (Algorithms~\ref{alg:p5-count-bull} and~\ref{alg:p5-count-diamond}, respectively), leading to
\begin{align}
    \mathcal{B}_e^\tau&=\sum_{H'}\mathbf 1[e=\varphi(H',\tau)]\chi_\mathcal{B}(H',\tau),
    \label{eq:p5-identity-bull}\\
    \mathcal{D}_e^\tau&=\sum_{H'}\mathbf 1[e=\varphi(H',\tau)]\chi_\mathcal{D}(H',\tau).
    \label{eq:p5-identity-diamond}
\end{align}
\smallskip\noindent
In Line~\ref{step:p5-se} of Algorithm~\ref{alg:restricted-p5-detection}, $v$ sets $s_e^v=\mathcal{P}_e^v-\mathcal{B}_e^v-\mathcal{D}_e^v-\mathcal{G}_e^v$.
Substituting Equations~\ref{eq:p5-identity-path}, \ref{eq:p5-identity-gem}, \ref{eq:p5-identity-bull}, and~\ref{eq:p5-identity-diamond} into Equation~\ref{eq:p5-perm-combined} and summing over $\tau$ gives $s_e=\sum_{H\in\binom{V}{5}}C(H,e)$.
\end{proof}

Next, we prove the claimed correctness guarantee.

\pfivesubgraphparity*

\begin{figure}[htbp]
    \centering
    \begin{tikzpicture}[x=1cm,y=1cm]
    \begin{scope}[shift={(0.00,0.00)}]
    \draw[paper highlighted edge] (-1.48,-0.70)--(-0.97,0.16);
    \draw[paper edge] (-0.97,0.16)--(0.00,0.70);
    \draw[paper edge] (0.00,0.70)--(0.97,0.16);
    \draw[paper highlighted edge] (0.97,0.16)--(1.48,-0.70);
    \node[paper highlighted label] at (-1.47,-0.27) {$1$};
    \node[paper edge label] at (-0.65,0.63) {$0$};
    \node[paper edge label] at (0.65,0.63) {$0$};
    \node[paper highlighted label] at (1.47,-0.27) {$1$};
    \node[paper vertex] at (-1.48,-0.70) {};
    \node[paper vertex] at (-0.97,0.16) {};
    \node[paper vertex] at (0.00,0.70) {};
    \node[paper vertex] at (0.97,0.16) {};
    \node[paper vertex] at (1.48,-0.70) {};
    \end{scope}
    \begin{scope}[shift={(4.70,0.00)}]
    \draw[paper edge] (-1.48,-0.70)--(-0.97,0.16);
    \draw[paper edge] (-0.97,0.16)--(0.00,0.70);
    \draw[paper edge] (-0.97,0.16)--(0.97,0.16);
    \draw[paper edge] (0.00,0.70)--(0.97,0.16);
    \draw[paper edge] (0.97,0.16)--(1.48,-0.70);
    \node[paper edge label] at (-1.47,-0.27) {$0$};
    \node[paper edge label] at (-0.72,0.40) {$0$};
    \node[paper edge label] at (0.00,0.3) {$0$};
    \node[paper edge label] at (0.72,0.40) {$0$};
    \node[paper edge label] at (1.47,-0.27) {$0$};
    \node[paper vertex] at (-1.48,-0.70) {};
    \node[paper vertex] at (-0.97,0.16) {};
    \node[paper vertex] at (0.00,0.70) {};
    \node[paper vertex] at (0.97,0.16) {};
    \node[paper vertex] at (1.48,-0.70) {};
    \end{scope}
    \begin{scope}[shift={(9.40,0.00)}]
    \draw[paper edge] (-1.48,-0.70)--(-0.97,0.16);
    \draw[paper edge] (-1.48,-0.70)--(0.00,0.70);
    \draw[paper edge] (-0.97,0.16)--(0.00,0.70);
    \draw[paper edge] (0.00,0.70)--(0.97,0.16);
    \draw[paper edge] (0.97,0.16)--(1.48,-0.70);
    \node[paper edge label] at (-1.47,-0.27) {$2$};
    \node[paper edge label] at (-0.56,-0.20) {$0$};
    \node[paper edge label] at (-0.65,0.63) {$0$};
    \node[paper edge label] at (0.65,0.63) {$0$};
    \node[paper edge label] at (1.47,-0.27) {$2$};
    \node[paper vertex] at (-1.48,-0.70) {};
    \node[paper vertex] at (-0.97,0.16) {};
    \node[paper vertex] at (0.00,0.70) {};
    \node[paper vertex] at (0.97,0.16) {};
    \node[paper vertex] at (1.48,-0.70) {};
    \end{scope}
    \begin{scope}[shift={(0.00,-2.70)}]
    \draw[paper edge] (-1.48,-0.70)--(-0.97,0.16);
    \draw[paper edge] (-1.48,-0.70)--(1.48,-0.70);
    \draw[paper edge] (-0.97,0.16)--(0.00,0.70);
    \draw[paper edge] (0.00,0.70)--(0.97,0.16);
    \draw[paper edge] (0.97,0.16)--(1.48,-0.70);
    \node[paper edge label] at (-1.47,-0.27) {$2$};
    \node[paper edge label] at (0.00,-0.96) {$2$};
    \node[paper edge label] at (-0.65,0.63) {$2$};
    \node[paper edge label] at (0.65,0.63) {$2$};
    \node[paper edge label] at (1.47,-0.27) {$2$};
    \node[paper vertex] at (-1.48,-0.70) {};
    \node[paper vertex] at (-0.97,0.16) {};
    \node[paper vertex] at (0.00,0.70) {};
    \node[paper vertex] at (0.97,0.16) {};
    \node[paper vertex] at (1.48,-0.70) {};
    \end{scope}
    \begin{scope}[shift={(4.70,-2.70)}]
    \draw[paper edge] (-1.48,-0.70)--(-0.97,0.16);
    \draw[paper edge] (-1.48,-0.70)--(0.97,0.16);
    \draw[paper edge] (-0.97,0.16)--(0.00,0.70);
    \draw[paper edge] (-0.97,0.16)--(0.97,0.16);
    \draw[paper edge] (0.00,0.70)--(0.97,0.16);
    \draw[paper edge] (0.97,0.16)--(1.48,-0.70);
    \node[paper edge label] at (-1.47,-0.27) {$0$};
    \node[paper edge label] at (-0.40,-0.51) {$0$};
    \node[paper edge label] at (-0.65,0.63) {$0$};
    \node[paper edge label] at (0.00,-0.02) {$0$};
    \node[paper edge label] at (0.65,0.63) {$0$};
    \node[paper edge label] at (1.47,-0.27) {$0$};
    \node[paper vertex] at (-1.48,-0.70) {};
    \node[paper vertex] at (-0.97,0.16) {};
    \node[paper vertex] at (0.00,0.70) {};
    \node[paper vertex] at (0.97,0.16) {};
    \node[paper vertex] at (1.48,-0.70) {};
    \end{scope}
    \begin{scope}[shift={(9.40,-2.70)}]
    \draw[paper edge] (-1.48,-0.70)--(-0.97,0.16);
    \draw[paper edge] (-1.48,-0.70)--(0.00,0.70);
    \draw[paper edge] (-0.97,0.16)--(0.00,0.70);
    \draw[paper edge] (0.00,0.70)--(0.97,0.16);
    \draw[paper edge] (0.00,0.70)--(1.48,-0.70);
    \draw[paper edge] (0.97,0.16)--(1.48,-0.70);
    \node[paper edge label] at (-1.47,-0.27) {$4$};
    \node[paper edge label] at (-0.56,-0.20) {$0$};
    \node[paper edge label] at (-0.65,0.63) {$0$};
    \node[paper edge label] at (0.65,0.63) {$0$};
    \node[paper edge label] at (0.56,-0.20) {$0$};
    \node[paper edge label] at (1.47,-0.27) {$4$};
    \node[paper vertex] at (-1.48,-0.70) {};
    \node[paper vertex] at (-0.97,0.16) {};
    \node[paper vertex] at (0.00,0.70) {};
    \node[paper vertex] at (0.97,0.16) {};
    \node[paper vertex] at (1.48,-0.70) {};
    \end{scope}
    \begin{scope}[shift={(0.00,-5.40)}]
    \draw[paper edge] (-1.48,-0.70)--(-0.97,0.16);
    \draw[paper edge] (-1.48,-0.70)--(0.00,0.70);
    \draw[paper edge] (-1.48,-0.70)--(0.97,0.16);
    \draw[paper edge] (-0.97,0.16)--(0.00,0.70);
    \draw[paper edge] (0.00,0.70)--(0.97,0.16);
    \draw[paper edge] (0.97,0.16)--(1.48,-0.70);
    \node[paper edge label] at (-1.47,-0.27) {$0$};
    \node[paper edge label] at (-0.56,-0.20) {$0$};
    \node[paper edge label] at (-0.40,-0.51) {$0$};
    \node[paper edge label] at (-0.65,0.63) {$0$};
    \node[paper edge label] at (0.65,0.63) {$0$};
    \node[paper edge label] at (1.47,-0.27) {$2$};
    \node[paper vertex] at (-1.48,-0.70) {};
    \node[paper vertex] at (-0.97,0.16) {};
    \node[paper vertex] at (0.00,0.70) {};
    \node[paper vertex] at (0.97,0.16) {};
    \node[paper vertex] at (1.48,-0.70) {};
    \end{scope}
    \begin{scope}[shift={(4.70,-5.40)}]
    \draw[paper edge] (-1.48,-0.70)--(-0.97,0.16);
    \draw[paper edge] (-1.48,-0.70)--(0.97,0.16);
    \draw[paper edge] (-0.97,0.16)--(0.00,0.70);
    \draw[paper edge] (-0.97,0.16)--(0.97,0.16);
    \draw[paper edge] (-0.97,0.16)--(1.48,-0.70);
    \draw[paper edge] (0.00,0.70)--(0.97,0.16);
    \draw[paper edge] (0.97,0.16)--(1.48,-0.70);
    \node[paper edge label] at (-1.47,-0.27) {$0$};
    \node[paper edge label] at (-0.40,-0.51) {$0$};
    \node[paper edge label] at (-0.65,0.63) {$0$};
    \node[paper edge label] at (0.00,-0.02) {$0$};
    \node[paper edge label] at (0.40,-0.51) {$0$};
    \node[paper edge label] at (0.65,0.63) {$0$};
    \node[paper edge label] at (1.47,-0.27) {$0$};
    \node[paper vertex] at (-1.48,-0.70) {};
    \node[paper vertex] at (-0.97,0.16) {};
    \node[paper vertex] at (0.00,0.70) {};
    \node[paper vertex] at (0.97,0.16) {};
    \node[paper vertex] at (1.48,-0.70) {};
    \end{scope}
    \begin{scope}[shift={(9.40,-5.40)}]
    \draw[paper edge] (-1.48,-0.70)--(-0.97,0.16);
    \draw[paper edge] (-1.48,-0.70)--(0.00,0.70);
    \draw[paper edge] (-1.48,-0.70)--(0.97,0.16);
    \draw[paper edge] (-0.97,0.16)--(0.00,0.70);
    \draw[paper edge] (-0.97,0.16)--(0.97,0.16);
    \draw[paper edge] (0.00,0.70)--(0.97,0.16);
    \draw[paper edge] (0.97,0.16)--(1.48,-0.70);
    \node[paper edge label] at (-1.47,-0.27) {$-2$};
    \node[paper edge label] at (-0.56,-0.20) {$-2$};
    \node[paper edge label] at (-0.40,-0.51) {$0$};
    \node[paper edge label] at (-0.65,0.63) {$-2$};
    \node[paper edge label] at (0.00,-0.02) {$0$};
    \node[paper edge label] at (0.65,0.63) {$0$};
    \node[paper edge label] at (1.47,-0.27) {$0$};
    \node[paper vertex] at (-1.48,-0.70) {};
    \node[paper vertex] at (-0.97,0.16) {};
    \node[paper vertex] at (0.00,0.70) {};
    \node[paper vertex] at (0.97,0.16) {};
    \node[paper vertex] at (1.48,-0.70) {};
    \end{scope}
    \begin{scope}[shift={(0.00,-8.10)}]
    \draw[paper edge] (-1.48,-0.70)--(-0.97,0.16);
    \draw[paper edge] (-1.48,-0.70)--(0.00,0.70);
    \draw[paper edge] (-1.48,-0.70)--(0.97,0.16);
    \draw[paper edge] (-1.48,-0.70)--(1.48,-0.70);
    \draw[paper edge] (-0.97,0.16)--(0.00,0.70);
    \draw[paper edge] (0.00,0.70)--(0.97,0.16);
    \draw[paper edge] (0.97,0.16)--(1.48,-0.70);
    \node[paper edge label] at (-1.47,-0.27) {$0$};
    \node[paper edge label] at (-0.56,-0.20) {$0$};
    \node[paper edge label] at (-0.40,-0.51) {$0$};
    \node[paper edge label] at (0.00,-0.96) {$0$};
    \node[paper edge label] at (-0.65,0.63) {$2$};
    \node[paper edge label] at (0.65,0.63) {$0$};
    \node[paper edge label] at (1.47,-0.27) {$2$};
    \node[paper vertex] at (-1.48,-0.70) {};
    \node[paper vertex] at (-0.97,0.16) {};
    \node[paper vertex] at (0.00,0.70) {};
    \node[paper vertex] at (0.97,0.16) {};
    \node[paper vertex] at (1.48,-0.70) {};
    \end{scope}
    \begin{scope}[shift={(4.70,-8.10)}]
    \draw[paper edge] (-1.48,-0.70)--(-0.97,0.16);
    \draw[paper edge] (-1.48,-0.70)--(0.00,0.70);
    \draw[paper edge] (-1.48,-0.70)--(0.97,0.16);
    \draw[paper edge] (-1.48,-0.70)--(1.48,-0.70);
    \draw[paper edge] (-0.97,0.16)--(0.00,0.70);
    \draw[paper edge] (-0.97,0.16)--(0.97,0.16);
    \draw[paper edge] (0.00,0.70)--(0.97,0.16);
    \draw[paper edge] (0.97,0.16)--(1.48,-0.70);
    \node[paper edge label] at (-1.47,-0.27) {$-2$};
    \node[paper edge label] at (-0.56,-0.20) {$-2$};
    \node[paper edge label] at (-0.40,-0.51) {$0$};
    \node[paper edge label] at (0.00,-0.96) {$-2$};
    \node[paper edge label] at (-0.65,0.63) {$-4$};
    \node[paper edge label] at (0.00,-0.02) {$-2$};
    \node[paper edge label] at (0.65,0.63) {$-2$};
    \node[paper edge label] at (1.47,-0.27) {$-2$};
    \node[paper vertex] at (-1.48,-0.70) {};
    \node[paper vertex] at (-0.97,0.16) {};
    \node[paper vertex] at (0.00,0.70) {};
    \node[paper vertex] at (0.97,0.16) {};
    \node[paper vertex] at (1.48,-0.70) {};
    \end{scope}
    \begin{scope}[shift={(9.40,-8.10)}]
    \draw[paper edge] (-1.48,-0.70)--(-0.97,0.16);
    \draw[paper edge] (-1.48,-0.70)--(0.00,0.70);
    \draw[paper edge] (-1.48,-0.70)--(0.97,0.16);
    \draw[paper edge] (-1.48,-0.70)--(1.48,-0.70);
    \draw[paper edge] (-0.97,0.16)--(0.00,0.70);
    \draw[paper edge] (-0.97,0.16)--(0.97,0.16);
    \draw[paper edge] (-0.97,0.16)--(1.48,-0.70);
    \draw[paper edge] (0.00,0.70)--(0.97,0.16);
    \draw[paper edge] (0.97,0.16)--(1.48,-0.70);
    \node[paper edge label] at (-1.47,-0.27) {$-4$};
    \node[paper edge label] at (-0.56,-0.20) {$-10$};
    \node[paper edge label] at (-0.40,-0.51) {$-4$};
    \node[paper edge label] at (0.00,-0.96) {$-10$};
    \node[paper edge label] at (-0.65,0.63) {$-10$};
    \node[paper edge label] at (0.00,-0.015) {$-4$};
    \node[paper edge label] at (0.40,-0.51) {$-10$};
    \node[paper edge label] at (0.65,0.63) {$-10$};
    \node[paper edge label] at (1.47,-0.27) {$-10$};
    \node[paper vertex] at (-1.48,-0.70) {};
    \node[paper vertex] at (-0.97,0.16) {};
    \node[paper vertex] at (0.00,0.70) {};
    \node[paper vertex] at (0.97,0.16) {};
    \node[paper vertex] at (1.48,-0.70) {};
    \end{scope}
    \begin{scope}[shift={(4.70,-10.80)}]
    \draw[paper edge] (-1.48,-0.70)--(-0.97,0.16);
    \draw[paper edge] (-1.48,-0.70)--(0.00,0.70);
    \draw[paper edge] (-1.48,-0.70)--(0.97,0.16);
    \draw[paper edge] (-1.48,-0.70)--(1.48,-0.70);
    \draw[paper edge] (-0.97,0.16)--(0.00,0.70);
    \draw[paper edge] (-0.97,0.16)--(0.97,0.16);
    \draw[paper edge] (-0.97,0.16)--(1.48,-0.70);
    \draw[paper edge] (0.00,0.70)--(0.97,0.16);
    \draw[paper edge] (0.00,0.70)--(1.48,-0.70);
    \draw[paper edge] (0.97,0.16)--(1.48,-0.70);
    \node[paper edge label] at (0,-1.02) {$\forall e:\ C(H,e) = 24$};
    \node[paper vertex] at (-1.48,-0.70) {};
    \node[paper vertex] at (-0.97,0.16) {};
    \node[paper vertex] at (0.00,0.70) {};
    \node[paper vertex] at (0.97,0.16) {};
    \node[paper vertex] at (1.48,-0.70) {};
    \end{scope}
    \end{tikzpicture}
    \caption{The final edge values under the full set of local corrections,
    evaluated on every induced-$C_4$-free five-vertex graph that contains a
    Hamiltonian path (each graph is drawn with its path
    along the outline).  Each edge $e$ is annotated with its value $C(H,e)$; bold
    red edges are those carrying an odd $C(H,e)$.  Only the induced $P_5$
    (top-left) has edges with an odd $C(H,e)$, while every other subgraph yields an even $C(H,e)$ on each of its edges.}
    \label{fig:p5-full-local-corrections}
    \end{figure}
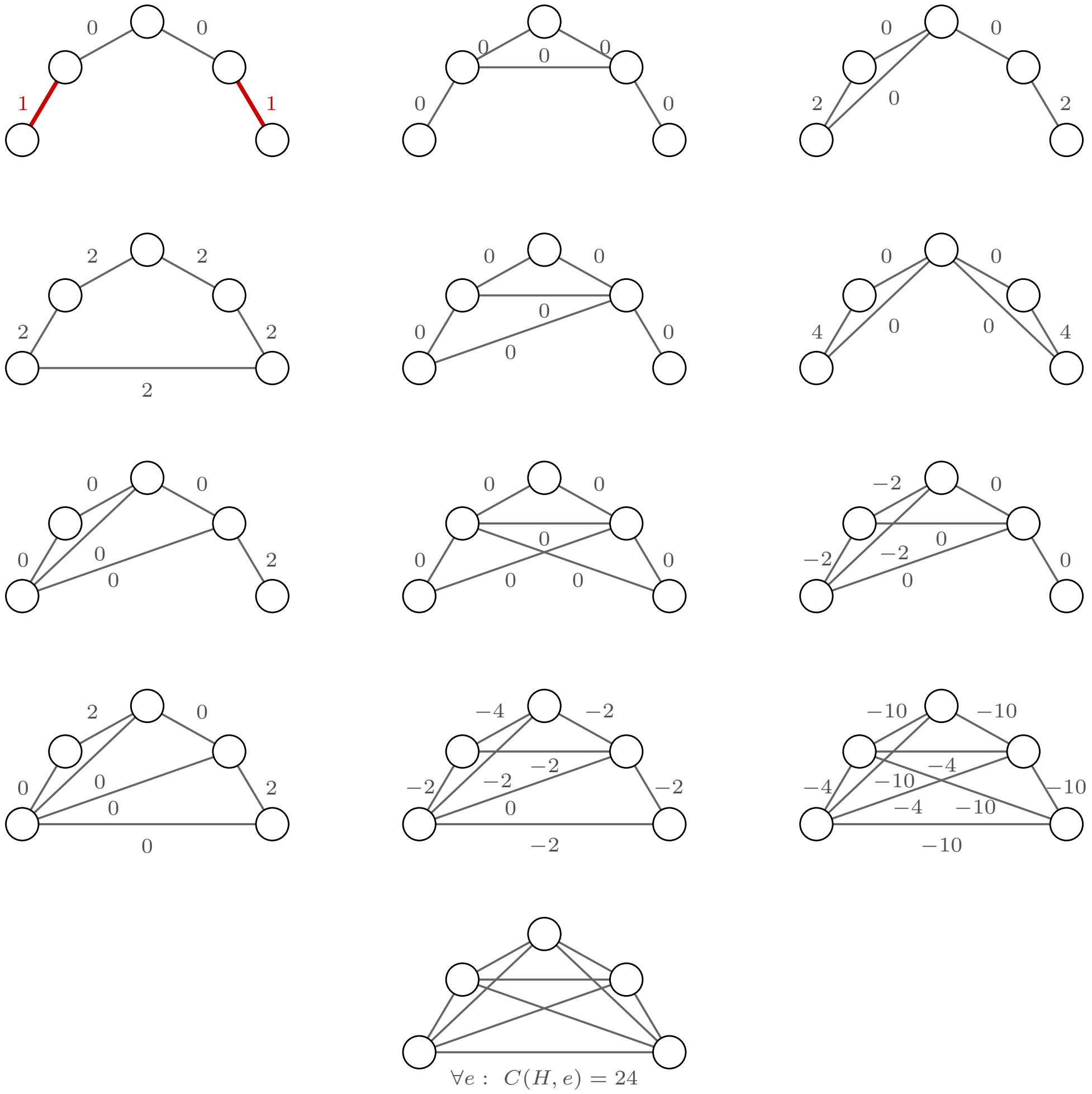

\begin{proof}
The proof is a finite subgraph check, which we verify using the SageMath script
of Figure~\ref{lst:p5-sage-check}: for every five-vertex graph $H$ that does
not contain an induced $C_4$, it computes $C(H,e)$ on
every edge $e$ and checks that some $C(H,e)$ is odd if and only if $H$ is an
induced $P_5$.  See Figure~\ref{fig:p5-full-local-corrections} for the value
of $C(H,e)$ for each edge $e$ in each induced-$C_4$-free subgraph $H$.
\end{proof}

\begin{figure}[H]
\begin{lstlisting}[style=sagepy]
V=range(5); P,C=graphs.PathGraph(5),graphs.CycleGraph(4)
T=[(((0,1),(1,2),(2,3),(3,4)),1),(((0,1),(1,2),(2,3),(3,4),(1,3)),-1),(((0,1),(1,2),(2,3),(3,4),(1,3),(2,4)),-1),(((0,1),(1,2),(2,3),(3,4),(0,2),(0,3),(0,4)),-1)]
for G in graphs.nauty_geng("5"):
    s={tuple(sorted(e)):0 for e in G.edges(labels=False)}
    if not any(G.subgraph(q).is_isomorphic(C) for q in combinations(V,4)):
        for p in permutations(V):
            for H,a in T:
                if all(G.has_edge(p[i],p[j]) for i,j in H): s[tuple(sorted((p[3],p[4])))] += a
    if not G.is_isomorphic(P) and any(v%2 for v in s.values()):
        print("reject", G.graph6_string(), s); break
else: print("accept")
\end{lstlisting}
\caption{The SageMath script verifying Lemma~\ref{lem:p5-subgraph-parity} by exhaustive search over all five-vertex graphs.}
\label{lst:p5-sage-check}
\end{figure}

Finally, we prove Lemma~\ref{lem:triangle-to-edge-counting}, which shows how to compute $\WT(e,C)$ for a given coloring $C$ and multiplicities $\mu$, for each incident edge $e$. 

Recall that for a $C$-colored triangle $T$, its weight is $W_T=\prod_{f\in E(T)}\mu(f)$, and the weighted triangle count $\WT(e,C)$ is the sum of $W_T$ over all $C$-colored triangles $T$ containing~$e$. 

\triangleedgecounting*

We use the following implicit result of \cite{CPSZ21}, which we restate here.

\begin{theorem}[Implicit in Theorem 1 in \cite{CPSZ21}]\label{thm:triangle-listing}
There exists a randomized $\tilde{O}(n^{1/3})$-round algorithm that, given an $n$-vertex graph $G=(V,E)$, lists all the triangles of $G$ with high probability in \congest. Moreover, each triangle $T$ of $G$ is listed \emph{once} by a unique representative vertex $\rep(T)\in V$.
\end{theorem}

At a high level, the algorithm given by Theorem \ref{thm:triangle-listing} is as follows. The vertices recursively compute an expander decomposition and partition the edges into cluster edges and inter-cluster edges. Inter-cluster edges are deferred to later iterations in the recursion. In each cluster, the vertices route the identifiers explicitly to collect the set of edges for every triangle $T$ in the cluster at a unique vertex $\rep(T)$. The vertex $\rep(T)$ then has all the information to list $T$. Since the expander decomposition guarantees the number of deferred inter-cluster edges is at most a constant fraction of the total edges, after $O(\log n)$ iterations no inter-cluster edges remain and the vertices terminate. This takes $R=\tilde O(n^{1/3})$ rounds. 

We use this algorithm to prove Lemma \ref{lem:triangle-to-edge-counting}.

\begin{proof}
We run the algorithm of Theorem \ref{thm:triangle-listing}. After the expander decomposition, whenever a vertex sends a message containing the identifier of an edge $e$, we add $O(\log n)$ bits that carry the multiplicity $\mu(e)$ and the colors of its endpoints. During this part, which we refer to as the \emph{forward propagation}, vertices locally build a tree $\mathcal{F}_e$ as follows. 

\paragraph{Forward propagation.}
Recall that $R = \tilde O(n^{1/3})$ denotes the number of rounds in the algorithm of Theorem \ref{thm:triangle-listing}.
If a vertex $v$ sends the identifier of an edge $e$ to a neighbor $w$ in round~$t\in\{1,\dots,R\}$, we say that $w$ \emph{learns $e$ from $v$ at round~$t$}.
Define the set of edges seen by $v$ during the forward propagation by
\[
    \mathcal{L}(v)
    \;=\;
    \bigl\{e\in E : \text{the identifier of }e\text{ reaches }v\bigr\}.
\]
For each $e\in \mathcal{L}(v)$, let $t(v,e)$ be the round at which $v$ first learns~$e$, and let the \emph{predecessor}
\[
    \pred(v,e)
    \;=\;
    \arg\min\bigl\{\id(u) : u\in N(v),\ v\text{ learned }e\text{ from }u \text { at } t(v,e)\bigr\},
\]
That is, $\pred(v,e)$ is the first vertex $u$ that $v$ learned $e$ from at round $t(v,e)$, breaking ties by $\id(u)$.
We may imagine that in the first round the identifier of $e$ is sent from the endpoint of $e$ with the \emph{minimum} ID to the endpoint with the \emph{maximum} ID; this min-ID endpoint is then the unique source of $\mathrm{ID}(e)$, and for it we set $\pred(v,e)=\bot$ and $t(v,e)=0$.
Because every vertex that learns $e$ has a chain of predecessors ending at its single min-ID source, the pairs $(v,\pred(v,e))$ with $\pred(v,e)\neq\bot$ form a single rooted \emph{tree} on a subset of $V$, which we denote $\mathcal F_e$.

After the forward propagation, the representative vertex $\rep(T)$ knows all three edges of $T$ and their multiplicities and colors, and therefore computes~$W_T$.

Next, for every triangle $T$ with $e\in E(T)$, we send the value $W_T$ back to the root of $\mathcal F_e$, thereby allowing it to compute $\WT(e,C)$. We call this the \emph{backward propagation}.

\paragraph{Backward propagation.}
Let $C=\{\alpha,\beta,\gamma\}$ be a color triplet, and let $e$ be an edge.
We aggregate triangle weights on~$e$ by running the forward propagation \emph{in reverse}.
This takes another $R$ rounds. 
For every vertex $v$, once the forward propagation is complete, it initializes
\[
    \omega_v(e) = \sum_{\substack{T:\ v=\rep(T)\\ e\in E(T)\\ T \text{ is }C\text{-colored}}} W_T 
\]
At round $t'(v,e)$, $v$ sends $\omega_v(e)$ to its $\pred(v,e)$, where $t'(v,e)$ is defined as
\[
    t'(v,e)
    \;=\;
    2R - t(v,e).
\]

It holds that $\omega_v(e)$ and $\omega_v(e')$ for different edges $e,e'$ are not sent on the same edge since they could not be sent to $v$ during the same round. At every round, each vertex $u\in V$ adds the values it receives for each edge $e$ to the corresponding $\omega_u(e)$. 

By Theorem \ref{thm:triangle-listing}, each triangle $T$ is listed by a single representative vertex $\rep(T)\in \mathcal{F}_e$ for every $e\in E(T)$, since the representative $\rep(T)$ must learn $e$ in order to list $T$. We now prove by induction that upon sending the value to its predecessor, the value $\omega_v(e)$ is the sum of all triangle weights $W_T$ for every $C$-colored triangle $T$ whose $\rep(T)$ is in the subtree of $v$ in $\mathcal{F}_e$. The base case is the leaves of $\mathcal{F}_e$, for which the claim holds by definition. Let $v$ be some internal vertex in $\mathcal{F}_e$, and let the claim be correct for every child $u$ of $v$ in $\mathcal{F}_e$. Since the vertex $v$ must learn $e$ before sending it to any child vertex $u$, it holds that $t(v,e) < t(u,e)$, and hence $t'(v,e) > t'(u,e)$. Therefore, the vertex $v$ sums $\omega_u(e)$ for every child $u$ of $v$ in $\mathcal{F}_e$ to $\omega_v(e)$. Furthermore, by the induction hypothesis this sums $W_T$ for every $C$-colored triangle $T$ whose $\rep(T)$ is in the subtree of any child $u$ of $v$. Adding $W_T$ for every $C$-colored triangle $T$ such that $\rep(T) = v$ to this sum proves the claim. Applying this claim to the root $r$ of $\mathcal{F}_e$ shows that the value $\omega_r(e)$ equals $\WT(e,C)$.

To ensure both endpoints of $e$ learn the correct value of $\WT(e,C)$, the root of $\mathcal F_e$ sends the value to the other endpoint of $e$ at the cost of one additional round.
\end{proof}

\setlength{\floatsep}{12pt plus 2pt minus 2pt}
\setlength{\textfloatsep}{20pt plus 2pt minus 4pt}
\setlength{\intextsep}{12pt plus 2pt minus 2pt}

\section{Induced $P_6$ Detection}
\label{sec:induced-p6-detection}

In this section, we prove the following result. 

\inducedpsixdetection*

The vertices begin by building a rooted BFS tree of depth at most 5. This is done in a standard way by starting parallel BFS trees from all vertices, where a message for a tree from root $r$ is forwarded by $v$ only if $r$ has the smallest ID among all vertices whose BFS trees reach $v$ so far. After $5$ rounds, any vertex that has a neighbor with a different BFS root, deduces that the diameter of the graph is more than 5 and therefore rejects, since this implies the existence of an induced $P_6$. Otherwise, denote by $T$ the BFS tree rooted at $v_0$. Next, all vertices send their neighbor list $N(v)$ to all of their neighbors. 

Then, the vertices run $O(\log n)$ trials of Algorithm~\ref{alg:induced-p6-detection}. All vertices send the colors of their neighbors to all of their neighbors. Each such trial counts the number of (non-induced) $P_6$ that end with each edge using the procedure \FuncSty{CountPath}, and a set of local corrections that can be computed using the two-hop neighborhood they learned in the previous step, as we explain soon. The vertices then use the procedure \FuncSty{Aggregate} in Algorithm~\ref{alg:induced-p6-detection} to aggregate their signals into a global signal $Z:=Z_{v_0}$ which we rely on to detect an induced $P_6$, using the BFS tree $T$. This signal is the sum of the local corrections and non-induced $P_6$ counts, computed by all vertices in the graph. As we shall prove, this value is equal to twice the number of colorful induced paths modulo 4. The vertices reject if at least one trial of this algorithm rejects.

To describe the local corrections for this algorithm, we define \emph{6-vertex patterns}, as presented in Table~\ref{tab:p6-corrections} and illustrated in Figure~\ref{fig:p6-patterns}. Each pattern is a tuple of 4 elements: a set of required edges, a set of required non-edges, a representative color, and a coefficient. Our local corrections consist of the sum of the coefficients of each pattern, multiplied by the number of instances of this pattern. The coefficients are chosen so that for each subgraph on 6 vertices, the weighted sum of the number of instances of each pattern is nonzero modulo 4 only for an induced $P_6$, which ensures other subgraphs do not introduce false positives. For a subset of vertices to be an instance of a pattern, the required edges must be present in the induced subgraph on it, and the required non-edges must be absent. Other edges may or may not be present in the induced subgraph. The representative color specifies which vertex of an instance of a pattern is the one to include it in its count.

\begin{table}[htbp]
\centering
\renewcommand{\arraystretch}{1.25}
\begin{tabular}{c l l c c}
\hline
pattern $p$ & \multicolumn{1}{c}{required edges $E(p)\setminus E(S)$} & \multicolumn{1}{c}{required non-edges $\hat E(p)$} & rep.\ $\Lambda(p)$ & coeff.\ $\alpha(p)$\\
\hline
$S$   & ---                                    & ---              & $6$   & $+1$\\
$F_1$ & $\{2,4\}$                               & ---              & $4$   & $-2$\\
$F_2$ & $\{2,5\}$                               & $\{2,4\},\{3,5\}$  & $2$   & $-1$\\
$F_3$ & $\{1,3\},\{1,5\}$                         & ---              & $1$   & $+2$\\
$F_4$ & $\{1,3\},\{1,4\},\{1,6\}$                   & ---              & $1$   & $+2$\\
$F_5$ & $\{1,3\},\{1,5\},\{3,5\}$                   & ---              & $1$   & $+2$\\
$F_6$ & $\{1,3\},\{1,4\},\{1,5\},\{3,6\}$             & ---              & $1$   & $+2$\\
$F_7$ & $\{1,3\},\{1,4\},\{1,5\},\{4,6\}$             & ---              & $1$   & $+1$\\
$F_8$ & $\{1,3\},\{1,4\},\{1,5\},\{2,6\},\{3,6\}$       & ---              & $1$   & $+1$\\
\hline
\end{tabular}
\caption{The set of all patterns, along with their required edges, required non-edges, representative color, and coefficient. The required edges column represents $E(p) \setminus E(S)$ where $E(S) = \{\{1,2\},\{2,3\},\{3,4\},\{4,5\},\{5,6\}\}$ is the set of edges of the path $S$. }
\label{tab:p6-corrections}
\end{table}

\newcommand{\pSixCoords}{%
    \coordinate (v1) at (-1.35,-0.45);
    \coordinate (v2) at (-0.95,0.25);
    \coordinate (v3) at (-0.35,0.60);
    \coordinate (v4) at (0.35,0.60);
    \coordinate (v5) at (0.95,0.25);
    \coordinate (v6) at (1.35,-0.45);
}
\newcommand{\pSixBase}{%
    \draw[base] (v1)--(v2)--(v3)--(v4)--(v5)--(v6);
}
\newcommand{\pSixNodes}{%
    \node[vertex] at (v1) {$1$};
    \node[vertex] at (v2) {$2$};
    \node[vertex] at (v3) {$3$};
    \node[vertex] at (v4) {$4$};
    \node[vertex] at (v5) {$5$};
    \node[vertex] at (v6) {$6$};
}

\begin{figure}[htbp]
\centering
\begin{tikzpicture}[
    x=0.88cm,
    y=0.9cm,
    base/.style={paper path},
    extra/.style={paper extra edge},
    forb/.style={paper forbidden edge},
    vertex/.style={paper vertex},
    rep/.style={paper representative},
    panel/.style={paper panel}
]

\begin{scope}[shift={(0,0)}]\pSixCoords\pSixBase\pSixNodes
\node[rep] at (v6) {$6$};
\node[panel] at (0,-1.15) {path $S$ ($+1$)};\end{scope}

\begin{scope}[shift={(3.3,0)}]\pSixCoords\pSixBase
\draw[extra] (v2)--(v4);
\pSixNodes\node[rep] at (v4) {$4$};
\node[panel] at (0,-1.15) {$F_1$ ($-2$)};\end{scope}

\begin{scope}[shift={(6.6,0)}]\pSixCoords\pSixBase
\draw[extra] (v2)--(v5);
\draw[forb] (v2)--(v4);\draw[forb] (v3)--(v5);
\pSixNodes\node[rep] at (v2) {$2$};
\node[panel] at (0,-1.15) {$F_2$ ($-1$)};\end{scope}

\begin{scope}[shift={(9.9,0)}]\pSixCoords\pSixBase
\draw[extra] (v1)--(v3);\draw[extra] (v1)--(v5);
\pSixNodes\node[rep] at (v1) {$1$};
\node[panel] at (0,-1.15) {$F_3$ ($+2$)};\end{scope}

\begin{scope}[shift={(13.2,0)}]\pSixCoords\pSixBase
\draw[extra] (v1)--(v3);\draw[extra] (v1)--(v4);\draw[extra] (v1)--(v6);
\pSixNodes\node[rep] at (v1) {$1$};
\node[panel] at (0,-1.15) {$F_4$ ($+2$)};\end{scope}

\begin{scope}[shift={(0,-2.6)}]\pSixCoords\pSixBase
\draw[extra] (v1)--(v3);\draw[extra] (v1)--(v5);\draw[extra] (v3)--(v5);
\pSixNodes\node[rep] at (v1) {$1$};
\node[panel] at (0,-1.15) {$F_5$ ($+2$)};\end{scope}

\begin{scope}[shift={(3.3,-2.6)}]\pSixCoords\pSixBase
\draw[extra] (v1)--(v3);\draw[extra] (v1)--(v4);\draw[extra] (v1)--(v5);\draw[extra] (v3)--(v6);
\pSixNodes\node[rep] at (v1) {$1$};
\node[panel] at (0,-1.15) {$F_6$ ($+2$)};\end{scope}

\begin{scope}[shift={(6.6,-2.6)}]\pSixCoords\pSixBase
\draw[extra] (v1)--(v3);\draw[extra] (v1)--(v4);\draw[extra] (v1)--(v5);\draw[extra] (v4)--(v6);
\pSixNodes\node[rep] at (v1) {$1$};
\node[panel] at (0,-1.15) {$F_7$ ($+1$)};\end{scope}

\begin{scope}[shift={(9.9,-2.6)}]\pSixCoords\pSixBase
\draw[extra] (v1)--(v3);\draw[extra] (v1)--(v4);\draw[extra] (v1)--(v5);
\draw[extra] (v2)--(v6);\draw[extra] (v3)--(v6);
\pSixNodes\node[rep] at (v1) {$1$};
\node[panel] at (0,-1.15) {$F_8$ ($+1$)};\end{scope}

\end{tikzpicture}
\caption{The nine patterns $S,F_1,\dots,F_8$ of the
induced $P_6$ detection algorithm (Algorithm~\ref{alg:induced-p6-detection}).  Every pattern shares the ordered path
$1\text{-}2\text{-}3\text{-}4\text{-}5\text{-}6$, drawn in \emph{bold}; the
\emph{required} chords of each pattern are dashed blue, and the \emph{required
non-edges} are drawn as red
dashed lines (needed only for $F_2$).  The representative color $\Lambda(p)$ of each pattern is drawn
in red. The coefficient of each pattern is shown in parentheses.}
\label{fig:p6-patterns}
\end{figure}
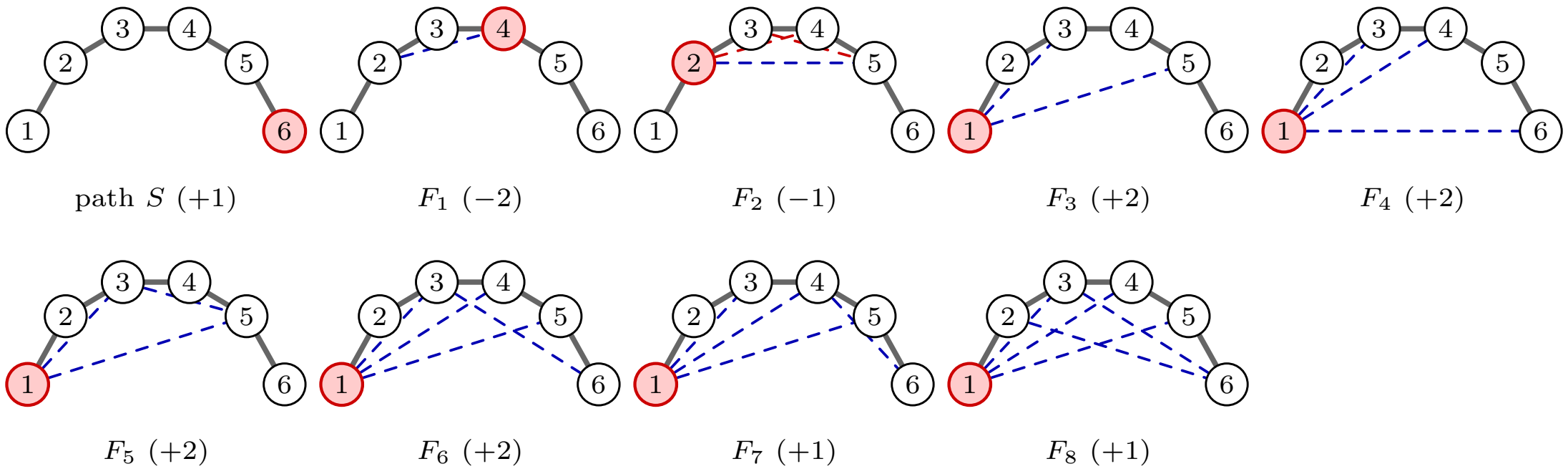

Formally, let $\mathcal{P}$ be the set of patterns in Table~\ref{tab:p6-corrections}. For every pattern $p\in\mathcal P$, let $E(p)$ be its set of required edges, let $\hat E(p)$ be its set of required non-edges, let $\Lambda(p)$ be its representative color, and let $\alpha(p)$ be its coefficient. Given a coloring $c$ and a permutation $\pi$ of $\{1,\ldots,6\}$, we say that a colorful subset of 6 vertices $H$ \emph{fits} $p$ under $\pi(c)$ if the following two conditions hold.
\begin{enumerate}
    \item For every edge $\{i,j\} \in E(p)$, the vertices of $H$ colored $i$ and $j$ under $\pi(c)$ are adjacent in $G$.
    \item For every non-edge $\{i,j\} \in \hat{E}(p)$, the vertices of $H$ colored $i$ and $j$ under $\pi(c)$ are non-adjacent in $G$.
\end{enumerate}

Fix one trial and permutation $\pi$. Each vertex $v$ sets $k=\pi(c_v)$ and partitions its neighbors by color. The procedure \FuncSty{CountPath} counts colorful ordered paths and adds their contribution to $\omega_v$. For each pattern $p$ with $\Lambda(p)=k$, vertex $v$ enumerates connected colorful six-vertex sets containing it and checks the required edges and non-edges using the neighbor lists learned earlier. The choices of the representative colors ensure that $v$ knows every adjacency needed for this check. For each fitting set, $v$ adds $\alpha(p)$ to $\omega_v$. All counters are maintained modulo $4$ because an induced $P_6$ adds $2$ modulo $4$ to $Z$, whereas every other six-vertex graph adds $0$ modulo $4$ to $Z$. Finally, \FuncSty{Aggregate} sums all $\omega_v$ values modulo $4$ over the BFS tree, producing $Z$ at the root.

\begin{algorithm}[H]
\DontPrintSemicolon
\SetAlgorithmName{Algorithm}{alg}{List}
\BlankLine
$c_v \gets$ uniform random color in $\{1,2,3,4,5,6,\phi\}$\;
send $c_v$ to every $u\in N(v)$\;
send $\{c_{u'}~|~{u'} \in N(v)\}$ to every $u\in N(v)$\; 
$\omega_v \gets 0$\;
\ForEach{permutation $\pi$ of $\{1,2,3,4,5,6\}$, if $c_v\neq\phi$}{   
  $k\gets\pi(c_v)$\;
  $N_j(v)\gets\{u\in N(v):\pi(c_u)=j\}$ for each $j\in\{1,\dots,6\}$\;
  \FuncSty{CountPath}() \tcp*[f]{Add path count to $\omega_v$}\;
  \ForEach{pattern $p\in\{F_1,\dots,F_8\}$ with $\Lambda(p)=k$}{
    \ForEach{colorful set of 6 vertices $H\ni v$ that fits $p$ under $\pi$}{
      $\omega_v\gets \omega_v+\alpha(p) \pmod 4$\;
    }
  }
}
\FuncSty{Aggregate}() \tcp*[f]{Compute $Z_v$ over $T$}\;
\eIf{$v=v_0$ and $Z_v\not\equiv 0\pmod 4$}{\textbf{reject}\;}{\textbf{accept}\;}
\caption{Single Trial for Induced $P_6$ Detection for a Vertex $v$}
\label{alg:induced-p6-detection}
\end{algorithm}

\begin{algorithm}[H]
\DontPrintSemicolon
\SetAlgorithmName{Algorithm}{alg}{List}
  $\mathrm{in}(u) \gets 0$ for every $u\in N(v)$\;
  \For{$i\gets 1$ \KwTo $6$}{
    \If{$k=i$}{
      \eIf{$i=1$}{send the value $1$ to every $u\in N_2(v)$\;}{
        $s_v\gets\sum_{u\in N_{i-1}(v)}\mathrm{in}(u)\pmod 4$\;
        \eIf{$i\le 5$}{send $s_v$ to every $u\in N_{i+1}(v)$\;}{$\omega_v\gets \omega_v+\alpha(S)\cdot s_v \pmod 4$\;}
      }
    }
    \If{$k=i+1$}{$\mathrm{in}(u)\gets$ value received from $u$, for every $u\in N_i(v)$\;}
  }
\caption{\FuncSty{CountPath}() for a Vertex $v$}
\label{alg:p6-count-path}
\end{algorithm}

\begin{algorithm}[H]
\DontPrintSemicolon
\SetAlgorithmName{Algorithm}{alg}{List}
  $Z_v\gets \omega_v$\;
  \ForEach{child $u$ of $v$ in $T$}{receive $Z_u$ and set $Z_v\gets Z_v+Z_u \pmod 4$\;}
  \If{$\mathrm{parent}(v)$ exists}{send $Z_v$ to $\mathrm{parent}(v)$\;}
\caption{\FuncSty{Aggregate}() for a Vertex $v$}
\label{alg:p6-aggregate}
\end{algorithm}

This completes the description of the algorithm for Theorem~\ref{thm:induced-p6-signal}. We now move to proving its correctness. 
Fix a single trial of Algorithm~\ref{alg:induced-p6-detection} and its random coloring. Let $H$ be a colorful subset of 6 vertices, and let $\tau$ be a permutation of $\{1,\dots,6\}$.
For a pattern $p\in\mathcal P$, let
$\chi_p(H,\tau)\in\{0,1\}$ characterize whether $H$ fits $p$ under
$\tau$, i.e., $\chi_p(H,\tau)=1$ if and only $H$ fits $p$ under
$\tau$.  Define the contribution of $H$ as
\[
    C(H)=\sum_{\tau}\ \sum_{p\in\mathcal P}\alpha(p)\,\chi_p(H,\tau),
\]
summed over all $6!$ permutations $\tau$.

We begin by connecting this definition to the quantities of Algorithm~\ref{alg:induced-p6-detection}. Let $Z=Z_{v_0}$.
\begin{restatable}{lemma}{psixedgesignalcontribution}\label{lem:p6-edge-signal-contribution}
By the end of a trial of Algorithm~\ref{alg:induced-p6-detection} it holds that
$Z\equiv\sum_{H\in\binom{V}{6}}C(H)\pmod 4$.
\end{restatable}

Then, we prove that an induced $P_6$ is the only graph to have a nonzero contribution modulo $4$. This, combined with Lemma~\ref{lem:p6-edge-signal-contribution}, implies that the aggregate $Z$ is equal modulo $4$ to twice the number of copies of colorful induced $P_6$ in a single trial of Algorithm~\ref{alg:induced-p6-detection}.

\begin{restatable}{lemma}{psixpatternparity}\label{lem:p6-pattern-parity}
Let $H\subseteq V$ be a colorful set of six vertices.  If $G[H]$ is an
induced $P_6$, then it holds that $C(H)\equiv 2\text{ modulo }4$; otherwise, it holds that $C(H)\equiv 0\text{ modulo }4$.
\end{restatable}

We now state the following lemma that proves there are sufficiently many ``good'' colorings, and hence the algorithm rejects correctly with high probability.

\begin{lemma}\label{lem:p6-colorings}
If $G$ contains an induced $P_6$, then there are at least $7^{n-6}$ colorings for which $Z\not\equiv 0\pmod 4$.
\end{lemma}

Let $M$ be the set of induced copies of $P_6$, and let $M_c$ be the set of colorful members of $M$ under the current coloring $c$. Lemma~\ref{lem:p6-edge-signal-contribution} expresses $Z$ as the sum of the contributions $C(H)$ over all six-vertex sets $H$, while Lemma~\ref{lem:p6-pattern-parity} shows that each member of $M_c$ contributes $2$ modulo $4$ and every other set contributes $0$. Hence, $Z\equiv 2|M_c|\pmod 4$. Lemma~\ref{lem:pk-colorings} guarantees at least $7^{n-6}$ colorings for which $|M_c|$ is odd; for every such coloring, $Z\equiv 2\pmod 4$ and is therefore nonzero. This proves Lemma~\ref{lem:p6-colorings}.

We first prove Theorem~\ref{thm:induced-p6-signal} assuming Lemmas~\ref{lem:p6-edge-signal-contribution}, \ref{lem:p6-pattern-parity}, and~\ref{lem:p6-colorings}, and later we prove the first two lemmas.

\begin{proof}[Proof of Theorem~\ref{thm:induced-p6-signal}]
The vertices first construct a BFS tree $T$ with depth at most $5$ or some vertex rejects. The vertices
then run
$O(\log n)$ independent trials of Algorithm~\ref{alg:induced-p6-detection}. Fix a single trial of this algorithm and its corresponding coloring. 

By Lemma~\ref{lem:p6-edge-signal-contribution}, $Z$ is equal modulo $4$ to $\sum_{H}C(H)$, and by
Lemma~\ref{lem:p6-pattern-parity} every colorful induced $P_6$ contributes
$C(H)\equiv 2\text{ modulo }4$ and every other colorful $H$ contributes $0\text{ modulo }4$; letting $M := \{H\in\binom{V}{6} : G[H]\equiv P_6\}$, this gives
\[
    Z
    \;\equiv\;
    2\cdot |M_c|
    \pmod 4 ,
\]
where $M_c$ is the set of colorful members of $M$ under the current coloring $c$. If no induced $P_6$ exists, the set $M$ is empty; hence, the aggregate is $Z\ \equiv 0 \pmod 4$ and the root $v_0$ accepts. If
an induced $P_6$ exists, then by Lemma~\ref{lem:p6-colorings}, there are at least $7^{\,n-6}$ colorings for which
the aggregate $Z$ is not equal to $0\text{ modulo }4$; hence the root $v_0$
rejects with probability at least $7^{-6}$. Repeating
$O(\log n)$ independent trials of Algorithm~\ref{alg:induced-p6-detection} gives a success probability of
$1-1/n^{\gamma}$ for any constant $\gamma>0$, by choosing the constant in the number of trials accordingly.

For the round complexity, the breadth-first search from $v_0$ uses
$O(1)$ rounds. It takes $O(\Delta)$ rounds for every vertex to send its list of neighbors to all of its neighbors, and another  $O(\Delta)$ rounds to send their colors throughout all trials, since we can compress the $O(1)$-bit colors. In a single trial of Algorithm~\ref{alg:induced-p6-detection}, the vertices iterate over $6!$ permutations, and for each permutation they use \FuncSty{CountPath} for a constant number of rounds to forward $s_v$ along the path, while computing the corrections for $F_1,\ldots,F_8$ locally from the already-sent two-hop neighborhoods. Since we run trials of this algorithm in parallel, all trials of this algorithm require $O(1)$ rounds. The vertices perform the convergecast in \FuncSty{Aggregate} in $O(1)$ rounds across all trials of this algorithm due to the bounded depth of the BFS tree. Hence, the total round complexity is $O(\Delta)$ rounds.
\end{proof}

We now move to proving Lemma \ref{lem:p6-edge-signal-contribution}.

\psixedgesignalcontribution*

\begin{proof}
Fix a permutation $\tau$ and consider the iteration of
Algorithm~\ref{alg:induced-p6-detection} in which $\pi=\tau$. We consider each pattern separately.

\smallskip\noindent\textbf{Path pattern $S$.}
The procedure \FuncSty{CountPath} propagates signals along the color order.
We show by induction on $i$ that, for every vertex $v$ with $\tau(c_v)=i$,
the value sent (if $i=1$) or computed as $s_v$ (if $i\ge 2$) equals the
number of colorful ordered paths $v_1,\ldots,v_i$ such that $v_i=v$ and $\tau(c_{v_j})=j$
for every $j\le i$.  For $i=1$ this is the value $1$ sent by $v$,
since the only path of length 1 is simply the vertex $v$ itself.  For $i\ge2$, $s_v=\sum_{u\in
N_{i-1}(v)}\mathrm{in}(u)$, where $\mathrm{in}(u)$ is the value $u$ sent in
the previous step; by the induction hypothesis this value is the number of
colorful ordered paths $v_1,\ldots,v_{i-1}$ such that $v_{i-1}=u$, and summing over
$u\in N_{i-1}(v)$ (which requires $\{u,v\}\in E$) counts the
colorful ordered paths $v_1,\ldots,v_{i-1},v$.  At $i=6$, $v$ sets
$\omega_v\mathrel{+}=\alpha(S)\cdot s_v$; thus, $v$ adds the following term to $\omega_v$:

\begin{align} \label{eq:p6-s-contribution}
    \alpha(S)\sum_{H\ni v}\mathbf 1[\tau(c_v)=\Lambda(S)] \cdot \chi_S(H,\tau)
\end{align}

\smallskip\noindent\textbf{Patterns $F_1,\ldots,F_8$.}
We claim that for every pattern $p \in \{F_1,\ldots,F_8\}$, the vertex colored $\Lambda(p)$ learns all the colors and edges (and non-edges) needed to list all six-vertex subsets that fit $p$ under $\tau$. This is because it learns about all vertices in its two-hop neighborhood and the required edges and non-edges. Therefore, over all patterns $p\in\{F_1,\ldots,F_8\}$, the representative vertices add the following total term to their counters $\omega_v$:
\begin{align} \label{eq:p6-f1-f8-contribution}
\sum_{p=F_1}^{F_8}\alpha(p)\sum_{H\ni v}\mathbf 1[\tau(c_v)=\Lambda(p)]\,\chi_p(H,\tau).
\end{align}

\smallskip
Summing Equations~\ref{eq:p6-s-contribution} and~\ref{eq:p6-f1-f8-contribution} over every $v$ and every permutation
$\tau$ gives 
\begin{equation*}
    \sum_{v}\omega_v\equiv\sum_{H}C(H)\pmod 4.
\end{equation*}

By the convergecast, it holds that
$Z=Z_{v_0}\equiv\sum_{v}\omega_v\pmod 4$, concluding the proof.
\end{proof}

Finally, we prove Lemma~\ref{lem:p6-pattern-parity} by computing the value of $C(H)$ for all possible graphs on 6 vertices.

\begin{figure}[H]
\begin{lstlisting}[style=sagepy]
from itertools import permutations

V = range(6)
PATH = graphs.PathGraph(6).edges(labels=False)
k = lambda u, v: tuple(sorted((u, v)))

T = [([],[],1),([(1,3)],[],-2),([(1,4)],[(1,3),(2,4)],-1),([(0,2),(0,4)],[],2),([(0,2),(0,3),(0,5)],[],2),([(0,2),(0,4),(2,4)],[],2),([(0,2),(0,3),(0,4),(2,5)],[],2),([(0,2),(0,3),(0,4),(3,5)],[],1),([(0,2),(0,3),(0,4),(1,5),(2,5)],[],1)]

def C(E):
    total = 0
    for t in permutations(V):
        for req, non, a in T:
            has_req = all(k(t[i],t[j]) in E for i,j in PATH+req)
            has_non = any(k(t[i],t[j]) in E for i,j in non)
            if has_req and not has_non:
                total += a
    return total

for G in graphs.nauty_geng("6"):
    E = set(G.edges(labels=False))
    is_path = G.is_isomorphic(graphs.PathGraph(6))
    assert C(E) % 4 == (2 if is_path else 0)

print("Verified on all isomorphism types of six-vertex graphs.")
\end{lstlisting}
\caption{The SageMath script verifying Lemma~\ref{lem:p6-pattern-parity} by exhaustive search over all six-vertex graphs.}
\label{lst:p6-python-check}
\end{figure}

\psixpatternparity*

\begin{proof}
The proof is by exhaustively computing $C(H)$ for all colorful subsets of 6 vertices, which we do using the SageMath script in Figure~\ref{lst:p6-python-check}.
\end{proof}

\bigskip
\paragraph{Acknowledgments.}
The research is supported in part by the Israel Science Foundation (grant 529/23).

%\bigskip
\paragraph*{Use of AI Tools.}
AI tools were used to compute the contributions and the number of paths including each edge and vertex in all subgraphs on 4,5, and 6 vertices. We also used AI to run wide searches for the set of local corrections, write scripts and visualizations, and verify the counting results. We further used AI to search for relevant papers and sources. AI was also used to improve the readability of the paper.  

\bibliographystyle{alpha}
\bibliography{refs}

@inproceedings{AbboudSW26,
  author       = {Amir Abboud and
                  Ron Safier and
                  Nathan Wallheimer},
  editor       = {Shubhangi Saraf},
  title        = {Triangle Detection in H-Free Graphs},
  booktitle    = {17th Innovations in Theoretical Computer Science Conference, {ITCS}
                  2026, Bocconi University, Milan, Italy, January 27-30, 2026},
  series       = {LIPIcs},
  volume       = {362},
  pages        = {1:1--1:19},
  publisher    = {Schloss Dagstuhl - Leibniz-Zentrum f{\"{u}}r Informatik},
  year         = {2026},
  url          = {https://doi.org/10.4230/LIPIcs.ITCS.2026.1},
  doi          = {10.4230/LIPICS.ITCS.2026.1},
  bibsource    = {dblp computer science bibliography, https://dblp.org}
}

@article{AFS12,
  author  = {Omid Amini and Fedor V. Fomin and Saket Saurabh},
  title   = {Counting Subgraphs via Homomorphisms},
  journal = {{SIAM} Journal on Discrete Mathematics},
  volume  = {26},
  number  = {2},
  pages   = {695--717},
  year    = {2012},
  url     = {https://doi.org/10.1137/100789403},
  doi     = {10.1137/100789403}
}

@incollection{BCLSV06,
  author    = {Christian Borgs and Jennifer Chayes and L{\'a}szl{\'o} Lov{\'a}sz and Vera T. S{\'o}s and Katalin Vesztergombi},
  title     = {Counting Graph Homomorphisms},
  booktitle = {Topics in Discrete Mathematics},
  series    = {Algorithms and Combinatorics},
  volume    = {26},
  pages     = {315--371},
  publisher = {Springer},
  year      = {2006},
  url       = {https://doi.org/10.1007/3-540-33700-8_18},
  doi       = {10.1007/3-540-33700-8_18}
}

@inproceedings{CurticapeanDM17,
  author    = {Radu Curticapean and Holger Dell and D{\'a}niel Marx},
  title     = {Homomorphisms Are a Good Basis for Counting Small Subgraphs},
  booktitle = {Proceedings of the 49th Annual {ACM} {SIGACT} Symposium on Theory of Computing},
  pages     = {210--223},
  publisher = {{ACM}},
  year      = {2017},
  url       = {https://doi.org/10.1145/3055399.3055502},
  doi       = {10.1145/3055399.3055502}
}

@article{KLL13,
  author  = {Miros{\l}aw Kowaluk and Andrzej Lingas and Eva{-}Marta Lundell},
  title   = {Counting and Detecting Small Subgraphs via Equations},
  journal = {{SIAM} Journal on Discrete Mathematics},
  volume  = {27},
  number  = {2},
  pages   = {892--909},
  year    = {2013},
  url     = {https://doi.org/10.1137/110859798},
  doi     = {10.1137/110859798}
}

@inproceedings{DruckerKO13,
  author       = {Andrew Drucker and
                  Fabian Kuhn and
                  Rotem Oshman},
  editor       = {Magn{\'{u}}s M. Halld{\'{o}}rsson and
                  Shlomi Dolev},
  title        = {On the power of the congested clique model},
  booktitle    = {{ACM} Symposium on Principles of Distributed Computing, {PODC} '14,
                  Paris, France, July 15-18, 2014},
  pages        = {367--376},
  publisher    = {{ACM}},
  year         = {2014},
  url          = {https://doi.org/10.1145/2611462.2611493},
  doi          = {10.1145/2611462.2611493},
  bibsource    = {dblp computer science bibliography, https://dblp.org}
}

@inproceedings{Pandurangan0S18,
  author       = {Gopal Pandurangan and
                  Peter Robinson and
                  Michele Scquizzato},
  editor       = {Christian Scheideler and
                  Jeremy T. Fineman},
  title        = {On the Distributed Complexity of Large-Scale Graph Computations},
  booktitle    = {Proceedings of the 30th on Symposium on Parallelism in Algorithms
                  and Architectures, {SPAA} 2018, Vienna, Austria, July 16-18, 2018},
  pages        = {405--414},
  publisher    = {{ACM}},
  year         = {2018},
  url          = {https://doi.org/10.1145/3210377.3210409},
  doi          = {10.1145/3210377.3210409},
  bibsource    = {dblp computer science bibliography, https://dblp.org}
}

@inproceedings{IzumiG17,
  author       = {Taisuke Izumi and
                  Fran{\c{c}}ois Le Gall},
  editor       = {Elad Michael Schiller and
                  Alexander A. Schwarzmann},
  title        = {Triangle Finding and Listing in {CONGEST} Networks},
  booktitle    = {Proceedings of the {ACM} Symposium on Principles of Distributed Computing,
                  {PODC} 2017, Washington, DC, USA, July 25-27, 2017},
  pages        = {381--389},
  publisher    = {{ACM}},
  year         = {2017},
  url          = {https://doi.org/10.1145/3087801.3087811},
  doi          = {10.1145/3087801.3087811},
  bibsource    = {dblp computer science bibliography, https://dblp.org}
}

@article{Censor-HillelKK19,
  author       = {Keren Censor{-}Hillel and
                  Petteri Kaski and
                  Janne H. Korhonen and
                  Christoph Lenzen and
                  Ami Paz and
                  Jukka Suomela},
  title        = {Algebraic methods in the congested clique},
  journal      = {Distributed Comput.},
  volume       = {32},
  number       = {6},
  pages        = {461--478},
  year         = {2019},
  url          = {https://doi.org/10.1007/s00446-016-0270-2},
  doi          = {10.1007/S00446-016-0270-2},
  bibsource    = {dblp computer science bibliography, https://dblp.org}
}

@inproceedings{FischerGKO18,
  author       = {Orr Fischer and
                  Tzlil Gonen and
                  Fabian Kuhn and
                  Rotem Oshman},
  editor       = {Christian Scheideler and
                  Jeremy T. Fineman},
  title        = {Possibilities and Impossibilities for Distributed Subgraph Detection},
  booktitle    = {Proceedings of the 30th on Symposium on Parallelism in Algorithms
                  and Architectures, {SPAA} 2018, Vienna, Austria, July 16-18, 2018},
  pages        = {153--162},
  publisher    = {{ACM}},
  year         = {2018},
  url          = {https://doi.org/10.1145/3210377.3210401},
  doi          = {10.1145/3210377.3210401},
  bibsource    = {dblp computer science bibliography, https://dblp.org}
}

@inproceedings{AssadiS25,
  author       = {Sepehr Assadi and
                  Janani Sundaresan},
  title        = {Distributed Triangle Detection is Hard in Few Rounds},
  booktitle    = {66th {IEEE} Annual Symposium on Foundations of Computer Science, {FOCS}
                  2025, Sydney, Australia, December 14-17, 2025},
  pages        = {755--766},
  publisher    = {{IEEE}},
  year         = {2025},
  url          = {https://doi.org/10.1109/FOCS63196.2025.00040},
  doi          = {10.1109/FOCS63196.2025.00040},
  bibsource    = {dblp computer science bibliography, https://dblp.org}
}

@article{AbboudCKL20,
  author       = {Amir Abboud and
                  Keren Censor{-}Hillel and
                  Seri Khoury and
                  Christoph Lenzen},
  title        = {Fooling views: a new lower bound technique for distributed computations
                  under congestion},
  journal      = {Distributed Comput.},
  volume       = {33},
  number       = {6},
  pages        = {545--559},
  year         = {2020},
  url          = {https://doi.org/10.1007/s00446-020-00373-4},
  doi          = {10.1007/S00446-020-00373-4},
  bibsource    = {dblp computer science bibliography, https://dblp.org}
}

@article{CensorHillelLV24,
  author       = {Keren Censor{-}Hillel and
                  Dean Leitersdorf and
                  David Vulakh},
  title        = {Deterministic near-optimal distributed listing of cliques},
  journal      = {Distributed Comput.},
  volume       = {37},
  number       = {4},
  pages        = {363--385},
  year         = {2024},
  url          = {https://doi.org/10.1007/s00446-024-00470-8},
  doi          = {10.1007/S00446-024-00470-8},
  bibsource    = {dblp computer science bibliography, https://dblp.org}
}

@article{FraigniaudLT24,
  author       = {Pierre Fraigniaud and
                  Ma{\"{e}}l Luce and
                  Ioan Todinca},
  title        = {On the power of threshold-based algorithms for detecting cycles in
                  the {CONGEST} model},
  journal      = {Theor. Comput. Sci.},
  volume       = {996},
  pages        = {114500},
  year         = {2024},
  url          = {https://doi.org/10.1016/j.tcs.2024.114500},
  doi          = {10.1016/J.TCS.2024.114500},
  bibsource    = {dblp computer science bibliography, https://dblp.org}
}

@inproceedings{FraigniaudLMT24,
  author       = {Pierre Fraigniaud and
                  Ma{\"{e}}l Luce and
                  Fr{\'{e}}d{\'{e}}ric Magniez and
                  Ioan Todinca},
  editor       = {Ran Gelles and
                  Dennis Olivetti and
                  Petr Kuznetsov},
  title        = {Even-Cycle Detection in the Randomized and Quantum {CONGEST} Model},
  booktitle    = {Proceedings of the 43rd {ACM} Symposium on Principles of Distributed
                  Computing, {PODC} 2024, Nantes, France, June 17-21, 2024},
  pages        = {209--219},
  publisher    = {{ACM}},
  year         = {2024},
  url          = {https://doi.org/10.1145/3662158.3662767},
  doi          = {10.1145/3662158.3662767},
  bibsource    = {dblp computer science bibliography, https://dblp.org}
}

@inproceedings{FraigniaudLMT25,
  author       = {Pierre Fraigniaud and
                  Ma{\"{e}}l Luce and
                  Fr{\'{e}}d{\'{e}}ric Magniez and
                  Ioan Todinca},
  editor       = {Keren Censor{-}Hillel and
                  Fabrizio Grandoni and
                  Jo{\"{e}}l Ouaknine and
                  Gabriele Puppis},
  title        = {Deterministic Even-Cycle Detection in Broadcast {CONGEST}},
  booktitle    = {52nd International Colloquium on Automata, Languages, and Programming,
                  {ICALP} 2025, Aarhus, Denmark, July 8-11, 2025},
  series       = {LIPIcs},
  volume       = {334},
  pages        = {80:1--80:19},
  publisher    = {Schloss Dagstuhl - Leibniz-Zentrum f{\"{u}}r Informatik},
  year         = {2025},
  url          = {https://doi.org/10.4230/LIPIcs.ICALP.2025.80},
  doi          = {10.4230/LIPICS.ICALP.2025.80},
  bibsource    = {dblp computer science bibliography, https://dblp.org}
}

@inproceedings{KorhonenR17,
  author       = {Janne H. Korhonen and
                  Joel Rybicki},
  editor       = {James Aspnes and
                  Alysson Bessani and
                  Pascal Felber and
                  Jo{\~{a}}o Leit{\~{a}}o},
  title        = {Deterministic Subgraph Detection in Broadcast {CONGEST}},
  booktitle    = {21st International Conference on Principles of Distributed Systems,
                  {OPODIS} 2017, Lisbon, Portugal, December 18-20, 2017},
  series       = {LIPIcs},
  volume       = {95},
  pages        = {4:1--4:16},
  publisher    = {Schloss Dagstuhl - Leibniz-Zentrum f{\"{u}}r Informatik},
  year         = {2017},
  url          = {https://doi.org/10.4230/LIPIcs.OPODIS.2017.4},
  doi          = {10.4230/LIPICS.OPODIS.2017.4},
  bibsource    = {dblp computer science bibliography, https://dblp.org}
}

@article{ChungPS17,
  author       = {Kai{-}Min Chung and
                  Seth Pettie and
                  Hsin{-}Hao Su},
  title        = {Distributed algorithms for the Lov{\'{a}}sz local lemma and graph
                  coloring},
  journal      = {Distributed Comput.},
  volume       = {30},
  number       = {4},
  pages        = {261--280},
  year         = {2017},
  url          = {https://doi.org/10.1007/s00446-016-0287-6},
  doi          = {10.1007/S00446-016-0287-6},
  bibsource    = {dblp computer science bibliography, https://dblp.org}
}

@article{PettieS15,
  author       = {Seth Pettie and
                  Hsin{-}Hao Su},
  title        = {Distributed coloring algorithms for triangle-free graphs},
  journal      = {Inf. Comput.},
  volume       = {243},
  pages        = {263--280},
  year         = {2015},
  url          = {https://doi.org/10.1016/j.ic.2014.12.018},
  doi          = {10.1016/J.IC.2014.12.018},
  bibsource    = {dblp computer science bibliography, https://dblp.org}
}

@inproceedings{Censor-HillelFG20,
  author       = {Keren Censor{-}Hillel and
                  Orr Fischer and
                  Tzlil Gonen and
                  Fran{\c{c}}ois Le Gall and
                  Dean Leitersdorf and
                  Rotem Oshman},
  editor       = {Hagit Attiya},
  title        = {Fast Distributed Algorithms for Girth, Cycles and Small Subgraphs},
  booktitle    = {34th International Symposium on Distributed Computing, {DISC} 2020,
                  Virtual Conference, October 12-16, 2020},
  series       = {LIPIcs},
  volume       = {179},
  pages        = {33:1--33:17},
  publisher    = {Schloss Dagstuhl - Leibniz-Zentrum f{\"{u}}r Informatik},
  year         = {2020},
  url          = {https://doi.org/10.4230/LIPIcs.DISC.2020.33},
  doi          = {10.4230/LIPICS.DISC.2020.33},
  bibsource    = {dblp computer science bibliography, https://dblp.org}
}

@inproceedings{CH21Survey,
  author        = {Keren Censor{-}Hillel},
  title         = {Distributed Subgraph Finding: Progress and Challenges (Invited Talk)},
  booktitle     = {48th International Colloquium on Automata, Languages, and Programming (ICALP 2021)},
  series        = {Leibniz International Proceedings in Informatics (LIPIcs)},
  volume        = {198},
  pages         = {3:1--3:14},
  publisher     = {Schloss Dagstuhl -- Leibniz-Zentrum f{\"u}r Informatik},
  year          = {2021},
  doi           = {10.4230/LIPIcs.ICALP.2021.3}
}

@article{EFFKO19,
  author        = {Talya Eden and Nimrod Fiat and Orr Fischer and Fabian Kuhn and Rotem Oshman},
  title         = {Sublinear-time distributed algorithms for detecting small cliques and even cycles},
  journal       = {Distributed Computing},
  volume        = {35},
  number        = {3},
  pages         = {207--234},
  year          = {2022},
  url           = {https://doi.org/10.1007/s00446-021-00409-3},
  doi           = {10.1007/s00446-021-00409-3}
}

@inproceedings{GM21,
  author        = {Fran{\c{c}}ois {Le Gall} and Masayuki Miyamoto},
  title         = {Lower bounds for induced cycle detection in distributed computing},
  booktitle     = {32nd International Symposium on Algorithms and Computation (ISAAC 2021)},
  series        = {Leibniz International Proceedings in Informatics (LIPIcs)},
  volume        = {212},
  pages         = {58:1--58:19},
  publisher     = {Schloss Dagstuhl -- Leibniz-Zentrum f{\"u}r Informatik},
  year          = {2021},
  doi           = {10.4230/LIPIcs.ISAAC.2021.58}
}

@inproceedings{WWWY15,
  author        = {Virginia {Vassilevska Williams} and Joshua R. Wang and Richard Ryan Williams and Huacheng Yu},
  title         = {Finding four-node subgraphs in triangle time},
  booktitle     = {Proceedings of the 26th Annual ACM-SIAM Symposium on Discrete Algorithms (SODA 2015)},
  pages         = {1671--1680},
  publisher     = {SIAM},
  year          = {2015},
  doi           = {10.1137/1.9781611973730.111}
}

@article{HRSS17,
  author        = {Juho Hirvonen and Joel Rybicki and Stefan Schmid and Jukka Suomela},
  title         = {Large Cuts with Local Algorithms on Triangle-Free Graphs},
  journal       = {The Electronic Journal of Combinatorics},
  volume        = {24},
  number        = {4},
  pages         = {P4.21},
  year          = {2017},
  doi           = {10.37236/6862}
}

@misc{HMP24,
  author        = {Magn{\'u}s M. Halld{\'o}rsson and Yannic Maus and Saku Peltonen},
  title         = {Distributed {Lov{\'a}sz} Local Lemma under Bandwidth Limitations},
  howpublished  = {arXiv:2405.07353 [cs.DS]},
  year          = {2024},
  eprint        = {2405.07353},
  archivePrefix = {arXiv},
  primaryClass  = {cs.DS},
  doi           = {10.48550/arXiv.2405.07353}
}

@inproceedings{KS25,
  author        = {Seri Khoury and Aaron Schild},
  title         = {Breaking Barriers for Distributed {MIS} by Faster Degree Reduction},
  booktitle     = {Proceedings of the 58th Annual ACM Symposium on Theory of Computing (STOC 2026)},
  pages         = {1037--1048},
  publisher     = {ACM},
  year          = {2026},
  doi           = {10.1145/3798129.3800816}
}

@inproceedings{NK21,
  author        = {Amir Nikabadi and Janne H. Korhonen},
  title         = {Beyond Distributed Subgraph Detection: Induced Subgraphs, Multicolored Problems and Graph Parameters},
  booktitle     = {25th International Conference on Principles of Distributed Systems (OPODIS 2021)},
  series        = {Leibniz International Proceedings in Informatics (LIPIcs)},
  volume        = {217},
  pages         = {15:1--15:18},
  publisher     = {Schloss Dagstuhl -- Leibniz-Zentrum f{\"u}r Informatik},
  year          = {2022},
  doi           = {10.4230/LIPIcs.OPODIS.2021.15}
}

@article{CPSZ21,
  author        = {Yi-Jun Chang and Seth Pettie and Thatchaphol Saranurak and Hengjie Zhang},
  title         = {Near-optimal Distributed Triangle Enumeration via Expander Decompositions},
  journal       = {Journal of the ACM},
  volume        = {68},
  number        = {3},
  pages         = {21:1--21:36},
  year          = {2021},
  doi           = {10.1145/3446330}
}

@inproceedings{EvenEtAl17,
  author        = {Guy Even and Orr Fischer and Pierre Fraigniaud and Tzlil Gonen and Reut Levi and Moti Medina and Pedro Montealegre and Dennis Olivetti and Rotem Oshman and Ivan Rapaport and Ioan Todinca},
  title         = {Three Notes on Distributed Property Testing},
  booktitle     = {31st International Symposium on Distributed Computing (DISC 2017)},
  series        = {Leibniz International Proceedings in Informatics (LIPIcs)},
  volume        = {91},
  pages         = {15:1--15:30},
  publisher     = {Schloss Dagstuhl -- Leibniz-Zentrum f{\"u}r Informatik},
  year          = {2017},
  doi           = {10.4230/LIPIcs.DISC.2017.15},
  url           = {https://doi.org/10.4230/LIPIcs.DISC.2017.15}
}

@article{FLLL15,
  author        = {Peter Floderus and Miros{\l}aw Kowaluk and Andrzej Lingas and Eva-Marta Lundell},
  title         = {Induced subgraph isomorphism: Are some patterns substantially easier than others?},
  journal       = {Theoretical Computer Science},
  volume        = {605},
  pages         = {119--128},
  year          = {2015},
  doi           = {10.1016/j.tcs.2015.09.001}
}

@inproceedings{BKS18,
  author        = {Markus Bl{\"a}ser and Balagopal Komarath and Karteek Sreenivasaiah},
  title         = {Graph Pattern Polynomials},
  booktitle     = {38th {IARCS} Annual Conference on Foundations of Software Technology and Theoretical Computer Science (FSTTCS 2018)},
  series        = {Leibniz International Proceedings in Informatics (LIPIcs)},
  volume        = {122},
  pages         = {18:1--18:13},
  publisher     = {Schloss Dagstuhl -- Leibniz-Zentrum f{\"u}r Informatik},
  year          = {2018},
  doi           = {10.4230/LIPIcs.FSTTCS.2018.18}
}

@inproceedings{miyamoto2025pk,
  author        = {Masayuki Miyamoto},
  title         = {Distributed Complexity of {$P_k$}-Freeness: Decision and Certification},
  booktitle     = {36th International Symposium on Algorithms and Computation (ISAAC 2025)},
  series        = {Leibniz International Proceedings in Informatics (LIPIcs)},
  volume        = {359},
  pages         = {51:1--51:21},
  publisher     = {Schloss Dagstuhl -- Leibniz-Zentrum f{\"u}r Informatik},
  year          = {2025},
  doi           = {10.4230/LIPIcs.ISAAC.2025.51}
}

@article{AYZ95,
  author        = {Noga Alon and Raphael Yuster and Uri Zwick},
  title         = {Color-coding},
  journal       = {Journal of the ACM},
  volume        = {42},
  number        = {4},
  pages         = {844--856},
  year          = {1995},
  doi           = {10.1145/210332.210337}
}

@article{CLB81,
  author        = {Derek G. Corneil and Helmut Lerchs and Lorna {Stewart Burlingham}},
  title         = {Complement reducible graphs},
  journal       = {Discrete Applied Mathematics},
  volume        = {3},
  number        = {3},
  pages         = {163--174},
  year          = {1981},
  doi           = {10.1016/0166-218X(81)90013-5}
}

@inproceedings{KariMRS15,
  author        = {Jarkko Kari and Mart{\'\i}n Matamala and Ivan Rapaport and Ville Salo},
  title         = {Solving the Induced Subgraph Problem in the Randomized Multiparty Simultaneous Messages Model},
  booktitle     = {Structural Information and Communication Complexity (SIROCCO 2015)},
  series        = {Lecture Notes in Computer Science},
  volume        = {9439},
  pages         = {370--384},
  publisher     = {Springer},
  year          = {2015},
  doi           = {10.1007/978-3-319-25258-2_26}
}

@article{Bressan21,
  author        = {Marco Bressan},
  title         = {Faster Algorithms for Counting Subgraphs in Sparse Graphs},
  journal       = {Algorithmica},
  volume        = {83},
  number        = {8},
  pages         = {2578--2605},
  year          = {2021},
  doi           = {10.1007/s00453-021-00811-0}
}

@article{KKM00,
  author        = {Ton Kloks and Dieter Kratsch and Haiko M{\"u}ller},
  title         = {Finding and counting small induced subgraphs efficiently},
  journal       = {Information Processing Letters},
  volume        = {74},
  number        = {3--4},
  pages         = {115--121},
  year          = {2000},
  doi           = {10.1016/S0020-0190(00)00047-8}
}

@article{FLRSR12,
  author        = {Fedor V. Fomin and Daniel Lokshtanov and Venkatesh Raman and Saket Saurabh and B. V. Raghavendra Rao},
  title         = {Faster algorithms for finding and counting subgraphs},
  journal       = {Journal of Computer and System Sciences},
  volume        = {78},
  number        = {3},
  pages         = {698--706},
  year          = {2012},
  doi           = {10.1016/j.jcss.2011.10.001}
}

@article{FKLL15,
  author        = {Peter Floderus and Miros{\l}aw Kowaluk and Andrzej Lingas and Eva-Marta Lundell},
  title         = {Detecting and Counting Small Pattern Graphs},
  journal       = {SIAM Journal on Discrete Mathematics},
  volume        = {29},
  number        = {3},
  pages         = {1322--1339},
  year          = {2015},
  doi           = {10.1137/140978211}
}

@inproceedings{KKMS23,
  author        = {Balagopal Komarath and Anant Kumar and Suchismita Mishra and Aditi Sethia},
  title         = {Finding and Counting Patterns in Sparse Graphs},
  booktitle     = {40th International Symposium on Theoretical Aspects of Computer Science (STACS 2023)},
  series        = {Leibniz International Proceedings in Informatics (LIPIcs)},
  volume        = {254},
  pages         = {40:1--40:20},
  publisher     = {Schloss Dagstuhl -- Leibniz-Zentrum f{\"u}r Informatik},
  year          = {2023},
  doi           = {10.4230/LIPIcs.STACS.2023.40}
}

\end{document}